\documentstyle[amstex,amssymb,epsf,11pt]{article}

\def\hybrid{\topmargin 0pt      \oddsidemargin 0pt
        \headheight 0pt \headsep 0pt
        \textwidth 6.25in       
        \textheight 9.5in       
        \marginparwidth 0.0in
        \parskip 5pt plus 1pt   \jot = 1.5ex}
\catcode`\@=11
\def\marginnote#1{}

\newcount\hour
\newcount\minute
\newtoks\amorpm
\hour=\time\divide\hour by60
\minute=\time{\multiply\hour by60 \global\advance\minute by-\hour}
\edef\standardtime{{\ifnum\hour<12 \global\amorpm={am}%
        \else\global\amorpm={pm}\advance\hour by-12 \fi
        \ifnum\hour=0 \hour=12 \fi
        \number\hour:\ifnum\minute<10 0\fi\number\minute\the\amorpm}}
\edef\militarytime{\number\hour:\ifnum\minute<10 0\fi\number\minute}

\def\draftlabel#1{{\@bsphack\if@filesw {\let\thepage\relax
   \xdef\@gtempa{\write\@auxout{\string
      \newlabel{#1}{{\@currentlabel}{\thepage}}}}}\@gtempa
   \if@nobreak \ifvmode\nobreak\fi\fi\fi\@esphack}
        \gdef\@eqnlabel{#1}}
\def\@eqnlabel{}
\def\@vacuum{}
\def\draftmarginnote#1{\marginpar{\raggedright\scriptsize\tt#1}}

\def\draftlabel#1{{\@bsphack\if@filesw {\let\thepage\relax
   \xdef\@gtempa{\write\@auxout{\string
      \newlabel{#1}{{\@currentlabel}{\thepage}}}}}\@gtempa
   \if@nobreak \ifvmode\nobreak\fi\fi\fi\@esphack}
        \gdef\@eqnlabel{#1}}
\def\@eqnlabel{}
\def\@vacuum{}
\def\draftmarginnote#1{\marginpar{\raggedright\scriptsize\tt#1}}

\def\draft{\oddsidemargin -.5truein
        \def\@oddfoot{\sl preliminary draft \hfil
        \rm\thepage\hfil\sl\today\quad\militarytime}
        \let\@evenfoot\@oddfoot \overfullrule 3pt
        \let\label=\draftlabel
        \let\marginnote=\draftmarginnote
   \def\@eqnnum{(\theequation)\rlap{\kern\marginparsep\tt\@eqnlabel}%
\global\let\@eqnlabel\@vacuum}  }

\def\numberbysection{\@addtoreset{equation}{section}
        \def\theequation{\thesection.\arabic{equation}}}

\def\underline#1{\relax\ifmmode\@@underline#1\else
        $\@@underline{\hbox{#1}}$\relax\fi}

\def\titlepage{\@restonecolfalse\if@twocolumn\@restonecoltrue\onecolumn
     \else \newpage \fi \thispagestyle{empty}\c@page\z@
        \def\thefootnote{\fnsymbol{footnote}} }

\def\endtitlepage{\if@restonecol\twocolumn \else  \fi
        \def\thefootnote{\arabic{footnote}}
        \setcounter{footnote}{0}}  
\relax

\numberbysection
\hybrid

\newenvironment{Proof}{\noindent {\it Proof:\ }}%
{\hfill {\mbox{$\Box$}} \par  \vspace{1.5ex}}

\newenvironment{Rem}{\noindent {\it Remark.\ }}%

\newtheorem{Th}{Theorem}[section]
\newtheorem{Prop}{Proposition}[section]
\newtheorem{Lem}{Lemma}[section]
\newtheorem{Cor}{Corollary}[section]
\newtheorem{Def}{Definition}[section]

\newcommand{\ZZ}{{\Bbb Z}}

\newcommand{\RR}{{\Bbb R}}
\newcommand{\LL}{{\Bbb L}}
\newcommand{\PP}{{\Bbb P}}
\newcommand{\CC}{{\Bbb C}}

\newcommand{\WW}{{\Bbb W}}
\newcommand{\VV}{{\Bbb V}}
\newcommand{\FF}{{\Bbb F}}
\newcommand{\II}{{\Bbb I}}
\newcommand{\EE}{{\Bbb E}}
\newcommand{\calC}{{\cal C}}
\newcommand{\cD}{{\cal D}}
\newcommand{\cT}{{\cal T}}
\newcommand{\cF}{{\cal F}}
\newcommand{\cL}{{\cal L}}
\newcommand{\cP}{{\cal P}}

\newcommand{\D}{\Delta}

\newcommand{\nin}{\noindent}

\newcommand{\bx}{{\boldsymbol x}}
\newcommand{\bu}{{\boldsymbol u}}
\newcommand{\bz}{{\boldsymbol z}}

\newcommand{\by}{{\boldsymbol y}}
\newcommand{\bq}{{\boldsymbol q}}
\newcommand{\bt}{{\boldsymbol t}}
\newcommand{\bv}{{\boldsymbol v}}
\newcommand{\bw}{{\boldsymbol w}}
\newcommand{\bn}{{\boldsymbol n}}
\newcommand{\bp}{{\boldsymbol p}}
\newcommand{\gl}{{\frak l}}
\newcommand{\gr}{{\frak r}}

\newcommand{\gt}{{\frak t}}

\newcommand{\bX}{{\boldsymbol X}}

\newcommand{\bY}{{\boldsymbol Y}}
\newcommand{\bnu}{{\boldsymbol \nu}}
\newcommand{\bOm}{{\boldsymbol \Omega}}
\newcommand{\bphi}{{\boldsymbol \phi}}
\newcommand{\eps}{{\varepsilon}}

\newcommand{\ra}{\longrightarrow}

\begin{document}

\begin{center}

{\sc \Large Transformations of Quadrilateral Lattices}

\bigskip

{\large Adam Doliwa$^{1,2,\dagger}$, Paolo Maria Santini$^{1,3,\S}$
and Manuel Ma\~nas$^{4,5,\ddagger}$}

\medskip

{\it $^1$Istituto Nazionale di Fisica Nucleare, Sezione di Roma\\
P-le Aldo Moro 2, I--00185 Roma, Italy


$^2$Instytut Fizyki Teoretycznej, Uniwersytet Warszawski \\
ul. Ho\.{z}a 69, 00-681 Warszawa, Poland


$^3$Dipartimento di Fisica, Universit\`a di Catania \\
Corso Italia 57, I-95129 Catania, Italy

$^4$Departamento de Matem\'{a}tica Aplicada y Estad\'{\i}stica,\\
 Escuela
Universitaria de Ingenieria T\'{e}cnica Areona\'{u}tica, \\ Universidad
Polit\'{e}cnica de Madrid\\ E28040-Madrid, Spain

$^5$Departamento de F\'\i sica Te\'orica, Universidad Complutense\\
E28040-Madrid, Spain}

\smallskip

$^\dagger$e-mail: {\tt doliwa@roma1.infn.it,
doliwa@fuw.edu.pl}

$^{\S}$e-mail:  {\tt santini@catania.infn.it,
santini@roma1.infn.it}

$^\ddagger$e-mail: {\tt manuel@dromos.fis.ucm.es}

\end{center}

\begin{abstract}
Motivated by the classical studies on transformations of conjugate nets, we
develop the general geometric theory of transformations of their discrete
analogues: the multidimensional quadrilateral lattices, i.e. lattices $\bx:
\ZZ^N \ra \RR^M$, $N\leq M$, whose elementary quadrilaterals are planar.
Our investigation is based on the discrete analogue of the theory
of the rectilinear congruences, which we also present in detail. We study,
in particular, the discrete analogues of the Laplace, Combescure,
L\'evy, radial and fundamental transformations and their interrelations.
The composition of these transformations and their permutability is
also investigated from a geometric point of view. The deep
connections between ``transformations" and ``discretizations" is
also investigated for quadrilateral lattices. We finally interpret 
these results within the $\bar{\partial}$ formalism.

\end{abstract}

\section{Introduction}
An interesting topic developed by distinguished geometers of the
turn of the last century is the theory of submanifolds equipped
with conjugate systems of coordinates (conjugate nets)
\cite{DarbouxOS,Eisenhart-TS}, i.e., mappings $\bx:\RR^N\ra\RR^M$,
$N\leq M$, satisfying the Laplace equations
\begin{equation} \frac{\partial^2\bx}{\partial u_i \partial u_j} =
\frac{1}{H_i}\frac{\partial H_i}{\partial u_j}  
\frac{\partial\bx}{\partial u_i} +
\frac{1}{H_j}\frac{\partial H_j}{\partial u_i} 
\frac{\partial\bx}{\partial u_j} \; , 
\quad i,j=1,...,N, \; i\ne j
\end{equation}
whose compatibility for $N>2$ gives the Darboux equations
\begin{equation} \label{eq:Darboux-H}
 \frac{\partial^2 H_k}{\partial u_i\partial u_j} = 
\frac{1}{H_i}\frac{\partial H_i}{\partial u_j}  
\frac{\partial H_k}{\partial u_i} +
\frac{1}{H_j}\frac{\partial H_j}{\partial u_i} 
\frac{\partial H_k}{\partial u_j}  \; ,
  \quad i\ne j\ne k \ne j.
\end{equation}
Imposing suitable geometric constraints on the conjugate nets, one obtains
significant reductions like the orthogonal systems of 
coordinates \cite{DarbouxOS}. 
It was recently shown by Zakharov and Manakov \cite{ZakMa} that the
Darboux equations can be solved using the $\bar{\partial}$ method
and that a suitable constraint on the associated
$\bar{\partial}$ datum allows one to solve its orthogonality
reduction \cite{Zakharov,ManakovZakharov}. These examples show once
more the deep connections between geometry and integrability, which
was observed in the past in other cases \cite{Sym,Bobenko2b2}. Actually,
basically all the known ``integrable geometries" or ``soliton
surfaces" and their integrability schemes can be viewed as special
reductions of the conjugate nets and of their integrability
properties.

During the last years some of these results have been generalized to
a discrete level~\cite{BP1,DS-AL,BP2,BS}. 
Based on a result by Sauer, which introduced the
proper discrete analogue of a conjugate net on a surface
\cite{Sauer}, Doliwa and Santini introduced the notion of
``Multidimensional Quadrilateral Lattice" (MQL), i.e. a lattice
$\bx: \ZZ^N \ra \RR^M$, $N\leq M$, 
with all its elementary quadrilaterals planar,
which is the discrete analogue of a multidimensional conjugate net
\cite{MQL}. 
Futhermore they showed that the planarity constraint (which is a linear 
constraint) provides a way to construct the lattice uniquely, once 
a suitable set of initial data is given. Therefore this lattice, generated
by a set of linear constraints is ``geometrically integrable".
They also found that the discrete nonlinear equations
characterizing the MQL had been already introduced, using the $\bar{\partial}$
formalism, by
Bogdanov and Konopelchenko \cite{BoKo} as a natural integrable
discrete analogue of the Darboux equations.

Also the orthogonality constraint has been successfully
discretized. This discretization consists in imposing that the
elementary quadrilaterals of the MQL are inscribed in circles. This
notion was first proposed in \cite{2dcl1,2dcl2} for $N=2$, $M=3$, as a 
discrete analogue of surfaces parametrized by curvature lines (see
also~\cite{BP2}); later
by Bobenko for $N=M=3$ \cite{Bobenko} and,
finally for arbitrary $N\leq M$ by Cie\'{s}li\'{n}ski, Doliwa
and Santini \cite{CDS}. These lattices are
now called ``Multidimensional Circular Lattices" (MCL) or 
discrete orthogonal lattices. In \cite{CDS} it was also shown that
the geometric integrability scheme for MQLs is consistent with the
circularity reduction, thus proving the integrability
of the MCL in pure geometric terms. Soon after that, Doliwa, Manakov and 
Santini have proven the (analytic)
integrability of the MCL generalizing to a discrete level the method of
solution, proposed in \cite{ManakovZakharov}, for the Lam\'e system and for
other reductions of the Darboux equations.
More recently, Konopelchenko and Schief have obtained a
convenient set of equations characterizing the circular lattices in $\EE^3$
\cite{KoSchief2}.

An extensive literature exists on the classes of
transformations of the conjugate nets, which provide an effective way to
construct new (and more
complicated) conjugate nets from given (simple) ones. The basic classes of 
transformations
of conjugate nets, listed for instance in \cite{Eisenhart-TS},
include the so-called Laplace, Combescure, L\'evy, radial and
fundamental transformations. The transformations preserving 
additional geometric constraints were also extensively investigated;
in particular, the reduction of the fundamental transformation compatible with
the orthogonality constraint is called the Ribaucour
transformation~\cite{Bianchi}.
We finally remark that the classical transformations 
of conjugate nets provide an
interesting geometric interpretation to the basic operations associated with
the multicomponent KP hierarchy~\cite{DMMMS}.

Guided by Sauer's definition of 2 dimensional discrete conjugate
net~\cite{Sauer} and by
the studies of Darboux on the Laplace transformations of
2 dimensional conjugate nets \cite{DarbouxIV,Eisenhart-TS}, Doliwa
has found in \cite{DCN} the discrete analogue of the Laplace
transform of a 2D quadrilateral lattice, which provides the
geometric interpretation of the Hirota equation \cite{Hirota}
(discrete 2D Toda system). Motivated by the general theory of
transformations of conjugate nets, in this paper we make a detailed
study of the geometric and analytic properties of the classes of
transformations of MQLs. These transformations turn out to be
particular cases of a general algebraic formulation recently
proposed by us in \cite{MDS}.

In order to construct the geometric theory of transformations of
MQLs, one has first to develop the discrete analogue of the theory
of rectilinear congruences, which we present in Section~\ref{sec:congruences}.
In Section~\ref{sec:congruences} we also define two basic relations between 
quadrilateral 
lattices and congruences: {\em focal lattices of a congruence} and
{\em lattices conjugate to a congruence}.
In the subsequent Sections \ref{sec:Laplace}--\ref{sec:fundamental}
we construct and study (the discrete analogues of) the Laplace,
Combescure, L\'evy, adjoint L\'evy, radial and fundamental
transformations of MQLs, emphasizing the geometric significance of
all the ingredients of these transformations and explaining the
geometric steps involved in the construction of a new MQL from a
given one. These transformations are the natural analogues of the corresponding
transformations of the conjugate nets and their definitions can be obtained
from the corresponding definitions, replacing the
expressions "focal net" and "net conjugate to a congruence" by "focal lattice" 
and "lattice conjugate to a congruence", respectively.
In Section \ref{sec:fundamental}, in addition, we also
give the geometric meaning of the composition of fundamental
transformations. The interpretation of the L\'evy, adjoint L\'evy and
Laplace transformations as geometrically distinguished limits of
the fundamental transformation is also used to describe
analytically these limits (Section \ref{sec:limits}). In Section
\ref{sec:vect-fund} we show how all these transformations
are particular cases of the general vectorial transformation
obtained in \cite{MDS}. A very successful, but empirical rule used
in the literature \cite{LeBen} to build integrable
discrete analogues of integrable differential equations consists in
finding the finite transformations of the differential systems and
in interpreting them as integrable discretizations; the validity of this rule 
is confirmed as a consequence of our theory. Section \ref{sec:D-bar} is
dedicated to the formulation of the geometric results of the paper
within the $\bar{\partial}$ formalism.

We remark that some aspects of the theory of the Combescure and fundamental
transformations of quadrilateral lattices have been already considered
independently by Konopelchenko and Schief in~\cite{KoSchief2} (see
Sections~\ref{sec:Combescure} and \ref{sec:fundamental} of the present paper); 
in that work they also found the discrete analogue of the Ribaucour 
transformation.

In the rest of this introductory Section, we recall the necessary results on
MQLs. For details, see \cite{MQL} and \cite{MDS}.

Consider a MQL; i. e., a mapping $\bx :\ZZ^N \ra \RR^M$, $N\leq M$,
with all elementary quadrilaterals planar \cite{MQL}. The planarity
condition can be formulated in terms of the Laplace equations
\begin{equation}  \label{eq:Laplace}
\D_i\D_j\bx=(T_{i} A_{ij})\D_i\bx+
(T_j A_{ji})\D_j\bx,\;\; i\not= j, \; \; \;  i,j=1 ,\dots, N,
\end{equation}
where the coefficients $A_{ij}$ satisfy the MQL equation
\begin{equation} \label{eq:MQL-A}
\D_k A_{ij} =
 (T_jA_{jk})A_{ij} +(T_k A_{kj})A_{ik} - (T_kA_{ij})A_{ik},
\;\; i\neq j\neq k\neq i.
\end{equation}
It is often convenient to reformulate  equations (\ref{eq:Laplace}) as
first order systems \cite{MQL}. The suitably scaled tangent vectors
$\bX_i$, $i=1,...,N$,
\begin{equation}  \label{def:HX}
\D_i\bx = (T_iH_i) \bX_i,
\end{equation}
satisfy the linear system
\begin{equation} \label{eq:lin-X}
\D_j\bX_i = (T_j Q_{ij})\bX_j,    \; \; \; i\ne j \; ,
\end{equation}
whose compatibility condition gives the following new form of the MQL equations
\begin{equation} \label{eq:MQL-Q}
\D_kQ_{ij} = (T_kQ_{ik})Q_{kj}\;\; i\neq j\neq k\neq i.
\end{equation}
The scaling factors $H_i$,
called the Lam\'e coefficients, solve the linear equations
\begin{equation} \label{eq:lin-H}
\D_iH_j = (T_iH_i) Q_{ij}, \; \; \; i\ne j \; ,
\end{equation}
 whose compatibility gives equations
(\ref{eq:MQL-Q}) again; moreover
\begin{equation*}   \label{def:A-H}
A_{ij}= \frac{\D_j H_i}{H_i} \; , \; \; i\ne j \; .
\end{equation*}
The Laplace equations (\ref{eq:Laplace})
and the MQL equations (\ref{eq:MQL-A})
read
\begin{align}\label{eq:Laplace-H}
\Delta_i\Delta_j\bx &= T_i\left( (\D_jH_i)H_i^{-1}\right)\D_i\bx +
T_j\left( (\D_iH_j)H_j^{-1}\right)\D_j \bx, \; i\not=j,\\
\label{eq:MQL-H}
\Delta_i\Delta_jH_k &= T_i\left( (\D_jH_i)H_i^{-1}\right)\D_iH_k +
T_j\left( (\D_iH_j)H_j^{-1}\right)\D_jH_k, \; i\neq j\neq k
\neq i,
\end{align}
in terms of the Lam\'e coefficients.

In the recent paper \cite{MDS} we proved the following
basic results:
\begin{Th} \label{th:vect-Darb}
Let $Q_{ij}$, $i,j=1,\dots,N$, $i\neq j$, be a solution of the MQL equations
(\ref{eq:MQL-Q}) and $\bY_i$ and $\bY^*_i$, $i=1,\dots N$, be solutions of the
associated linear systems (\ref{eq:lin-X}) and (\ref{eq:lin-H})
taking values in a linear space $\WW$ and in its adjoint $\WW^*$ respectively.
Let $\bOm[\bY,\bY^*]\in L(\WW)$ be a linear operator in $\WW$
defined by the compatible equations
\begin{equation} \label{eq:omega}
\D_i\bOm[\bY,\bY^*] =\bY_i\otimes (T_i\bY^*_i), \;\;\; i=1,\dots,N.
\end{equation}
If the potential $\bOm$ is invertible,  $\bOm[\bY,\bY^*]\in GL(\WW)$,
then the functions
\begin{equation}
\hat{Q}_{ij} = Q_{ij}-\langle \bY^*_j|\bOm^{-1}|\bY_i \rangle,
\;\;\; i,j=1,\dots,N, \;\;i\neq j, \label{eq:hatQ}
\end{equation}
are new solutions of the equation (\ref{eq:MQL-Q}), and
\begin{align}
\hat{\bY}_i &=\bOm^{-1}\bY_i, \;\; \; i=1,...,N,
\label{eq:hatY}\\
\hat{\bY}^*_i&=\bY_i^*\bOm^{-1},\;\;\; i=1,...,N,\label{eq:hatY*}
\end{align}
are corresponding new solutions of the equations (\ref{eq:lin-X}),
(\ref{eq:lin-H}). In addition,
\begin{equation} \label{eq:hatOm}
\bOm[\hat{\bY},\hat{\bY}^*] = C - \bOm[\bY,\bY^*]^{-1},
\end{equation}
where $C$ is a constant operator.
\end{Th}
\begin{Prop} \label{prop:vect-Darb-latt}
Consider a constant vector $\bw \in \WW$
and the projection operator
$P$ on an $M$ dimensional subspace $\VV$ of $\WW$,
then the vector function $\bx: \ZZ^N \ra \VV\equiv \RR^M$, defined by
\begin{equation} \bx = P(\bOm [\bY,\bY^*]\bw) \; ,
\end{equation}
defines an $N$-dimensional quadrilateral lattice whose  Lam\'e
coefficients and scaled tangent vectors are of the form
\begin{align}
H_i &= \langle \bY^*_i | \bw \rangle,  \\
\bX_i &= P(\bY_i).
\end{align}
\end{Prop}

As we shall see in the following Sections, the vectorial transformations
obtained in Theorem \ref{th:vect-Darb} contain all the transformations studied
in this paper as particular and/or limiting cases.

\section{Rectilinear congruences and quadrilateral lattices}
It is well known that rectilinear congruences play a fundamental role in
the theory of transformations of multiconjugate systems
\cite{Eisenhart-TS}. In this Section we discretize the theory of
congruences whose importance in the theory of transformations of MQLs
will be evident in the following Sections.

Study of families of lines was motivated by the theory of
optics, and mathematicians like Monge, Malus and Hamilton initiated
the general theory of rays. However it was Pl\"ucker, who first
considered straight lines in $\RR^3$ as points of some space; he also 
found a convenient way to parametrize that
space.  In the second half of the XIX-th century this subject was 
very popular and studied, after Pl\"ucker, 
by many distinguished geometers; to mention
Klein, Lie, Bianchi and Darboux
only~\cite{Rowe,Bianchi,DarbouxIV,Eisenhart-TS,Lane}. 

It turns out (see Chapter XII of~\cite{Eisenhart-TCS} for more details) 
that, for a generic
two-parameter family of lines in $\RR^3$ (called {\em rectilinear congruence}),
there exist, roughly speaking, two surfaces 
(called {\em focal surfaces of the 
congruence}) characterized by the property that every line of the
family is tangent to both surfaces. This fact does not hold for 
bigger dimensions of the ambient space and, by definition, {\em 
a two-parameter family of straight lines in $\RR^M$ is called (rectilinear)
congruence iff it has focal surfaces}. 
One-parameter families of straight lines
tangent to a curve are called {\em developable surfaces}; one can consider
developable surfaces as one dimensional congruences. A three-parameter family
of lines in $\RR^3$ is sometimes also called line-complex.

Our goal is to construct the theory of {\em $N$ dimensional}
congruences of straight
lines within the {\em discrete geometry approach}. In doing this we 
use the idea of the {\em constructability} of the discrete integrable
geometries presented in \cite{MQL,CDS,Dol-RC}.

\label{sec:congruences}

\subsection{Congruences and their focal lattices}

\label{subs:cong-foc}

\begin{Def} \label{def:int-congr}
An $N$-dimensional rectilinear congruence
(or, simply, congruence) is a mapping $\gl : \ZZ^N \ra \LL(M)$ 
from the integer lattice to the
space of lines in $\RR^M$ such that every two neighboring lines
$\gl$ and $T_i \gl$, $i=1,...,N$, are coplanar.
\end{Def}

Let us make a trivial, but important remark: the planarity of two
neighboring lines of the congruence allows for their intersection.
When the lines are parallel, we consider their intersection in the
hyperplane at infinity. In fact, as it was observed in
\cite{DCN,MQL}, the quadrilateral lattices should be considered
within the projective geometry approach; i.e. the ambient space
should be the $M$-dimensional projective space $\PP^M$.
Accordingly, the space of lines in the affine space modelled on
$\RR^M$ should be then replaced by the space of lines in $\PP^M$;
that is to say, by the Grassmannian  $\text{Gr}(2,M+1)$.

One can associate with any $N$-dimensional congruence in a canonical way $N$
lattices defined as follows.

\begin{Def}
The $i$-th focal lattice $\by_i(\gl)$ of a congruence $\gl$
is the lattice constructed out of the intersection points of the lines $
\gl$ with $T_i^{-1}\gl$.
\end{Def}
In our paper we study the interplay between congruences of lines and
quadrilateral lattices, and we shall show that the focal lattices
of a ``generic" congruence are indeed quadrilateral.
To explain what a generic congruence is, let us consider any four lines
\[ \gl, \; \; T_i\gl, \; \; T_j\gl,
\; \; T_k\gl, \; \; i\ne j \ne k \ne i \; ;
\]
the congruence is {\em generic} if the linear space $\VV_{ijk}(\gl)$ generated
by these lines is of the maximal possible dimension: $\dim \VV_{ijk}(\gl) = 4$.
The congruence is called {\em weakly generic} if the linear space 
$\VV_{ij}(\gl)$ 
generated by any three lines $\gl$, $T_i\gl$, $T_j\gl$, $i\ne j$, is of
maximal possible dimension: $\dim \VV_{ij}(\gl) = 3$. 

Obviously, any generic congruence is also a weakly generic one. In our studies 
we may violate the genericity assumption but {\em we always assume we deal with 
weakly generic congruences}.  

\begin{Th} \label{th:foc-int}
Focal lattices  of a generic congruence are quadrilateral lattices.
\end{Th}
\begin{Proof} The proof splits naturally into two parts. In the first part,
illustrated on Fig.~1, we
show the planarity of the elementary quadrilaterals with vertices
$\by_i$, $T_i\by_i$,
$T_j\by_i$, $T_iT_j\by_i$, where $j\ne i$. In the second part, illustrated on 
Fig.~2, we prove the same for the elementary quadrilaterals with vertices
$\by_i$, $T_j\by_i$,
$T_k\by_i$, $T_jT_k\by_i$, where $j,k\ne i$, $j\ne k$. 
\begin{center}
\leavevmode\epsfxsize=8cm\epsffile{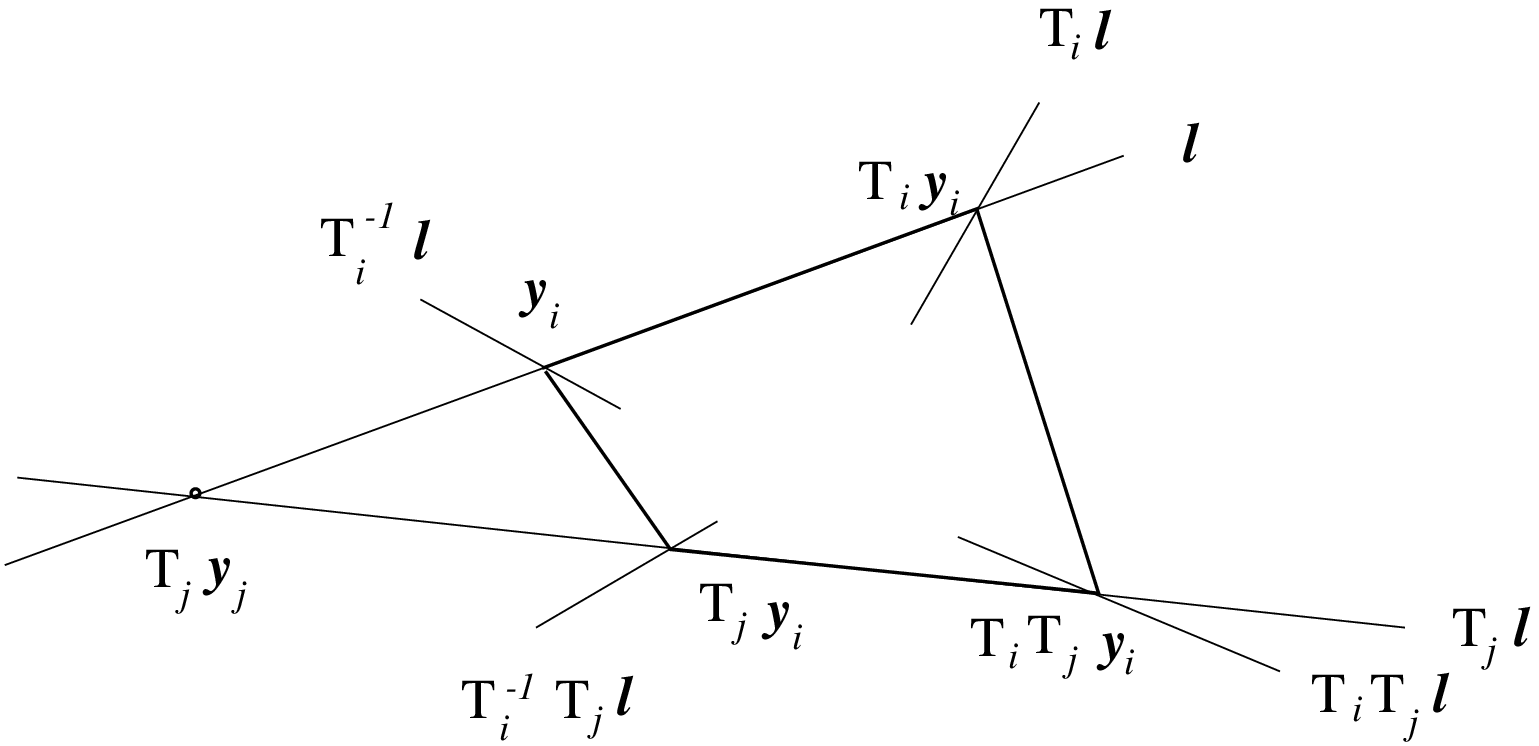}
\end{center}
Figure 1.

\nin i) Let us observe that the vertices $\by_i$ and $T_i\by_i$ are points of
the
line $\gl$. Similarly, the vertices $T_j\by_i$ and $T_iT_j\by_i$ belong to the
line $T_j\gl$. But the lines $\gl$ and $T_j\gl$ are coplanar, which concludes
the first part of the proof.
\begin{center}
\leavevmode\epsfxsize=8cm\epsffile{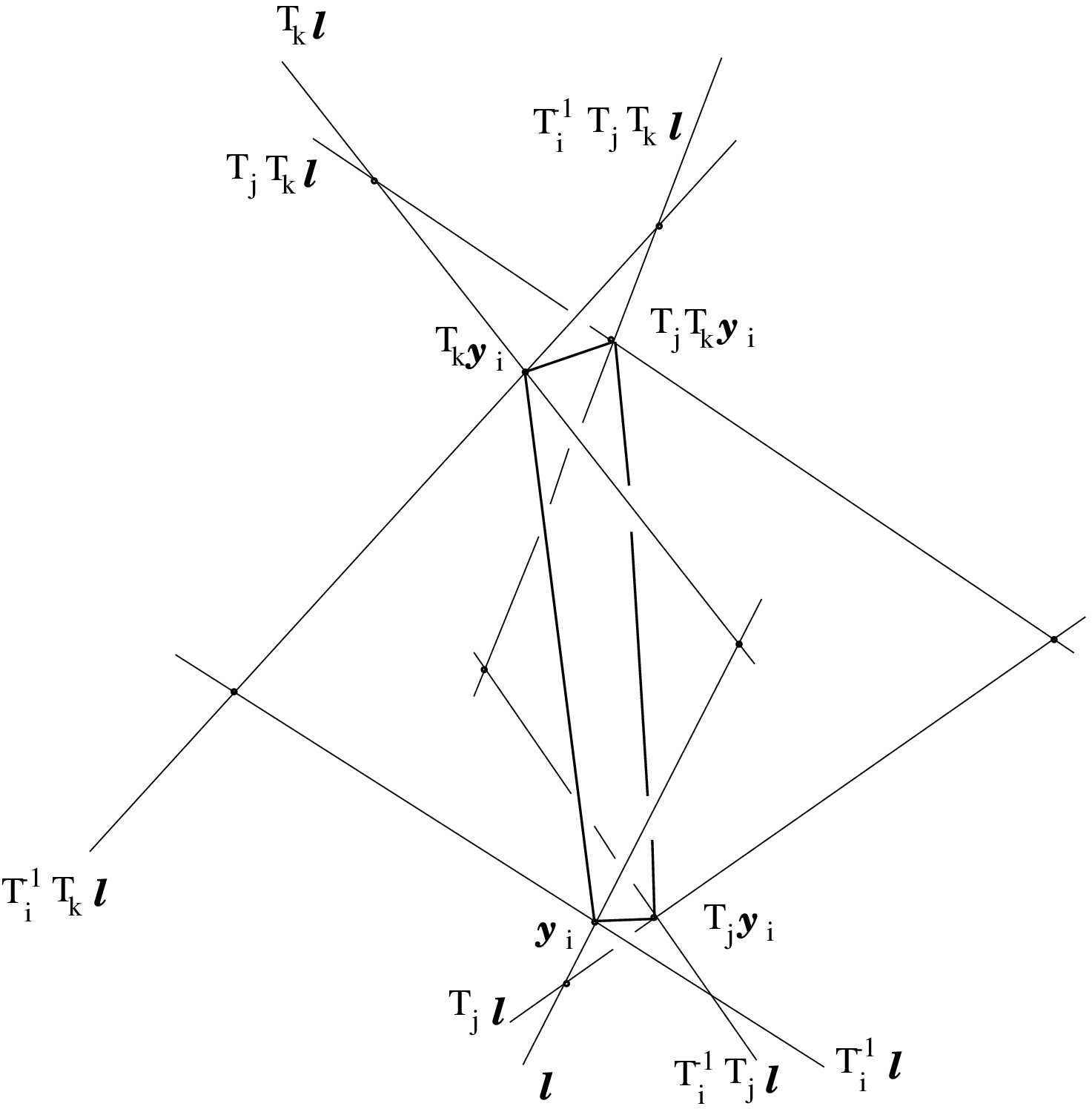}
\end{center}
Figure 2.

\nin ii) Consider the configuration of the four lines
\[ \gl, \; \; T_j\gl,
\; \; T_k\gl, \; \; T_jT_k\gl,
\]
contained in the three dimensional space $\VV_{jk}(\gl)$,
and the similar configuration of four lines
\[ T_i^{-1}\gl, \; \; T_i^{-1}T_j\gl,
\;\; T_i^{-1}T_k\gl, \; \; T_i^{-1}T_jT_k\gl,
\]
contained in a three dimensional subspace
$\VV_{jk}(T_i^{-1}\gl)$. We remark that $\VV_{ijk}(T_i^{-1}\gl) = 
\VV_{jk}(T_i^{-1}\gl) + \VV_{jk}(\gl)$.

Let us notice that corresponding lines of the two configurations have
one point in common
\begin{gather*}
\by_i = (T_i^{-1}\gl)\: \cap \;\gl,
\quad  T_j\by_i = (T_i^{-1}T_j \gl )\: \cap \;(T_j\gl), \\
T_k\by_i = (T_i^{-1}T_k\gl)\: \cap \;(T_k\gl),
\quad T_jT_k\by_i = (T_i^{-1}T_jT_k \gl )\: \cap \;(T_jT_k\gl) ;
\end{gather*}
these points are vertices of the quadrilateral whose planarity we would
like to show. The points $\by_i$, $T_j\by_i$,
$T_k\by_i$ define a plane $\VV_{jk}(\by_i)$, which 
is contained in both subspaces
$\VV_{jk}(\gl)$ and $\VV_{jk}(T_i^{-1}\gl)$. Since, for a generic congruence,
\[ \dim (\VV_{jk}(\gl) \cap \VV_{jk}(T_i^{-1}\gl)) = 
\dim \VV_{jk}(T_i^{-1}\gl) + \dim \VV_{jk}(\gl) -
\dim \VV_{ijk}(T_i^{-1}\gl) =2 \; ,
\] 
then 
\[ \VV_{jk}(\by_i) = \VV_{jk}(\gl) \cap \VV_{jk}(T_i^{-1}\gl)
\]
and, therefore, also $T_jT_k\by_i\in\VV_{jk}(\by_i)$; this proves the 
planarity of 
the quadrilateral under consideration.
\end{Proof}

It turns out that 
even in the non generic case, if one of the focal lattices is quadrilateral,
then all the others are quadrilateral as well; to show it we need the following
simple but basic fact:
\begin{Lem} \label{lem:ab}
Consider, in the three dimensional space, two different 
coplanar lines $a$ and $b$ and two different
planes $\pi_a$ and $\pi_b$ which contain the lines $a$ and $b$,
correspondingly:
$a\subset\pi_a$, $b\subset\pi_b$. Then the common line (it exists and is unique)
of the two planes contains the intersection point $\bp$ of the two lines:
$\bp = (a\cap b) \in \pi_a\cap \pi_b$.
\end{Lem}
\begin{center}\leavevmode\epsfxsize=9.3cm\epsffile{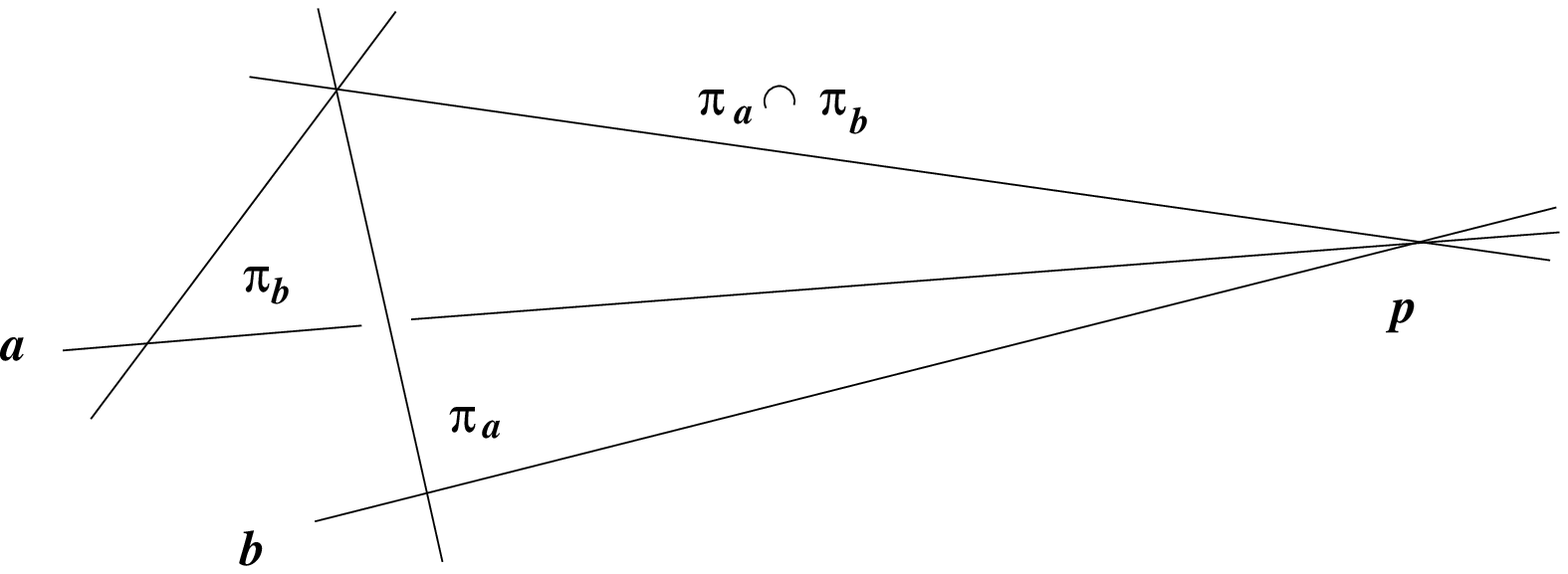}\end{center}
Figure 3.

\begin{Prop} \label{prop:foc-quad-ng}
If one of the focal lattices of the congruence is quadrilateral then the other
focal lattices are quadrilateral as well.
\end{Prop}
\begin{Proof} 
Let us assume that the $i$-th focal lattice be planar. Therefore
the lines $a=\langle T_j\by_i , T_jT_k\by_i \rangle$ and 
$b=\langle \by_i , T_k\by_i \rangle$ intersect at $\bp$. From 
Lemma~\ref{lem:ab}, the intersection line $\langle T_j\by_j , T_jT_k\by_j
\rangle $ of the planes $\pi_a = \langle T_j \gl , T_jT_k \gl \rangle $ and
$\pi_b = \langle \gl , T_k \gl \rangle $ passes through the point $\bp$.
Analogously, also the line $\langle T_i^{-1}T_j\by_j , T_i^{-1}T_jT_k\by_j 
\rangle $
passes through $\bp$. This proves the planarity of the quadrilateral
$T_i^{-1}T_j\{ \by_j, T_i\by_j, T_k\by_j, T_iT_k\by_j \}$.
\end{Proof}
\begin{Cor} \label{cor:foc-ident-Lapl}
The intersection points of the pairs of lines $\langle T_i\by_i ,
T_iT_k\by_i \rangle $ with $\langle T_iT_j\by_i$, $T_iT_jT_k\by_i
\rangle $ and $\langle T_j\by_j , T_jT_k\by_j \rangle$ with
$\langle T_iT_j\by_j ,T_iT_jT_k\by_j \rangle$ coincide.
\end{Cor}

\subsection{Constructability of congruences}

In this Section we look at the congruences from the point of view
of their constructability.
We recall that, in the case of quadrilateral lattices \cite{MQL},
given the points $\bx$, $T_i\bx$, $T_j\bx$, $T_k\bx$ in general position, and
points 
$T_iT_j\bx\in\VV_{ij}(\bx)$,
$T_iT_k\bx\in\VV_{ik}(\bx)$
and $T_jT_k\bx\in\VV_{jk}(\bx)$, then the point $T_iT_jT_k\bx$ is uniquely
determined
as the intersection point of the three planes $\VV_{jk}(T_i\bx)$,
$\VV_{ik}(T_j\bx)$ and $\VV_{ij}(T_k\bx)$
in the three dimensional
space $\VV_{ijk}(\bx)$.

A similar procedure is valid also for congruences. Given the lines $\gl$,
$T_i\gl$ and $T_j\gl$, the admissible lines $T_iT_j\gl$ form a
two-parameter space (any pair of points of $T_i\gl$ and $T_j \gl$
may be connected by a line), like for the lattice case. 
This is actually another reason why one can view congruences of lines as dual
objects to quadrilateral lattices. 

In a generic situation, the ``initial" lines $\gl$, $T_i\gl$, $T_j\gl$,
$T_k\gl$, $T_iT_j\gl$, $T_iT_k\gl$ and $T_jT_k\gl$  are contained
in the four dimensional space $\VV_{ijk}(\gl)$. The line
$T_iT_jT_k\gl$ is therefore the {\em unique} line which intersects the
three lines $T_iT_j\gl$, $T_iT_k\gl$ and $T_jT_k\gl$ (or, equivalently, the
intersection line of the three spaces $\VV_{ij}(T_k\gl)$,
$\VV_{ik}(T_j\gl)$, and $\VV_{jk}(T_i\gl)$). Therefore genericity of the
congruence and uniqueness of the construction are sinonimous, implying that the
focal lattices are quadrilateral.

In the non-generic case, when the lines $T_iT_j\gl$, $T_iT_k\gl$
and $T_jT_k\gl$  are contained  in a three dimensional space, there
exists a one-parameter family of lines intersecting the three given 
lines and the construction is not unique. We remark that, in this situation,
for any point of the line $T_jT_k\gl$, say, there exists a unique line passing
through the other two lines $T_iT_k\gl$ and $T_iT_j\gl$; such family of lines 
forms a one-sheeted hyperboloid. Any element of this family is admissible, but 
may not give rise to quadrilateral focal lattices.

However, in this non-generic case, we may single-out the line
$T_iT_jT_k\gl$ from the above one-parameter family of lines 
by requiring that the
intersection point $T_iT_jT_k\by_i$ of $T_iT_jT_k \gl$ with the line 
$T_jT_k\gl$ belong to the plane $\VV_{jk}(T_i\by_i)=
\langle T_i\by_i , T_iT_j\by_i,$ $T_iT_k\by_i \rangle$ or, equivalently,
that {\em the focal lattice $\by_i$ be quadrilateral}.
We remark that this procedure does not depend on the focal
lattice we consider (from Proposition~\ref{prop:foc-quad-ng}).

We have seen that, given an $N$-dimensional
congruence, one can associate with it $N$ focal (quadrilateral, in general) 
lattices. There
is of course a dual picture, and one can associate with a lattice which is
quadrilateral $N$ (tangent) congruences.

\begin{Def} \label{def:tan-congruence}
Given an $N$-dimensional quadrilateral lattice $\bx$,
its $i$-th tangent congruence $\gt_i(\bx)$ consists of the lines
passing through the points $\bx$ of the lattice and directed along
the tangent vectors $\D_i\bx$.
\end{Def}
We remark that the planarity of the elementary quadrilaterals of $\bx$
implies that the tangent congruence is a congruence of lines in the sense of
Definition \ref{def:int-congr}.
Obviously, excluding degenerations,
any congruence $\gl$ can be viewed as the $i$-th tangent congruence of
its $i$-th focal lattice $\by_i(\gl)$.

In the previous Section we have shown that, for non-generic congruences,
the focal lattices may not be quadrilateral. However, for tangent
congruences, due to Proposition~\ref{prop:foc-quad-ng},
we have the following
\begin{Th}
Focal lattices of tangent congruences are quadrilateral lattices.
\end{Th}

\subsection{Conjugacy of quadrilateral lattices and rectilinear congruences}

The following mutual relation between a congruence and a quadrilateral
lattice is of particular importance in our theory.
\begin{Def} \label{def:conj-congr-latt}
An $N$-dimensional quadrilateral lattice $\bx$ and an $N$-dimensional
congruence  $\gl$
are called conjugate if
$\bx(\bn)\in\gl(\bn)$, for all $\bn\in\ZZ^N$ .
\end{Def}
In the definition of conjugate net (on a surface) conjugate to a congruence,  
first given by Guichard~\cite{Eisenhart-TS}, the developables of the congruence intersect the net in 
conjugate-parameter lines; the focal nets of the congruence were excluded a 
priori from the definition.

In our approach, instead, we include focal
lattices (and focal manifolds) in a natural way as special limiting cases 
of generic lattices (manifolds)
conjugate to the congruence; this observation will be used in Section
\ref{sec:limits}.

We will show now that a quadrilateral lattice conjugate to a congruence may be 
conveniently used to improve the construction of the congruence itself 
making it unique in the non-generic case.

We first show that, for a generic congruence $\gl$, the construction of a 
quadrilateral lattice $\bx$ conjugate to the congruence is compatible with the 
construction of the congruence itself. We assume, for simplicity, that the 
points of the lattice are not the focal ones. We observe that, given three 
points $\bx$, $T_i\bx$ and $T_j\bx$, $i\ne j$, marked on the lines $\gl$, 
$T_i\gl$ and $T_j\gl$, the point $T_iT_j\bx$
is then uniquely determined as the intersection point of the plane
$\VV_{ij}(\bx)=\langle
\bx , T_i\bx , T_j\bx \rangle$ with the line $T_iT_j\gl$ in the three
dimensional space $\VV_{ij}(\gl)$. In the dual picture, given the point
$T_iT_j\bx$, then the line $T_iT_j\gl$ is the intersection line of
the planes $\langle T_i\gl, T_iT_j\bx \rangle$ and $\langle T_j\gl,
T_iT_j\bx \rangle$. 

If we also give the point $T_k\bx$ on $T_k\gl$, then the lines $T_iT_k\gl$
and $T_jT_k\gl$ allow to find the points $T_iT_k\bx$ and $T_jT_k\bx$, and vice 
versa.

Now we can use the standard construction of the MQL lattice to find
the eight point $T_iT_jT_k \bx$ from the seven points $\bx$,..., $T_jT_k\bx$, 
and we can use the above presented construction of the non-degenerate congruence 
to find the line $T_iT_jT_k\gl$ from the seven lines $\gl$,.., $T_jT_k\gl$.
At this point a natural and important question arises: {\em does the point 
$T_iT_jT_k\bx$ belong to the line $T_iT_jT_k\gl$?\/}If it doesn't, then the 
notion of quadrilateral lattice conjugate to congruence would not be a very 
relevant one. 

To show that the answer is positive let us proceed as follows.  
Denote by $\bz$ the {\em unique} intersection point of the line
$T_iT_jT_k\gl$ with the three dimensional subspace $\VV_{ijk}(\bx)=
\langle \bx , T_i\bx , T_j\bx, T_k \bx \rangle $ (our congruence is a
generic one). Since $\VV_{jk}(T_i\bx)\subset \VV_{ijk}(\bx)$ and
$\VV_{jk}(T_i\bx) \cap T_iT_jT_k\gl \not= \emptyset$, then $\bz \in
\VV_{jk}(T_i\bx)$. Similarly, $\bz \in \VV_{ik}(T_j\bx)$ and $\bz \in
\VV_{ij}(T_k\bx)$; which implies that $\bz = T_iT_jT_k\bx$.

\begin{Rem}
The above construction properties imply that, for a given generic congruence,
a quadrilateral lattice conjugate to it is uniquely defined assigning
its initial curves.
\end{Rem}

In the non-generic case we may again single-out the  line
$T_iT_jT_k\gl$ from the one-parameter family of lines by the
following requirement, which has been proved to hold in the generic
situation:\\ 
{\it i) the line passes through the point
$T_iT_jT_k\bx$ and meets the lines $T_iT_j\gl$, $T_iT_k\gl$ and
$T_jT_k\gl$.}\\
If such a line exists, for the construction to be the canonical one
we would like also two additional conditions to be satisfied:\\
{\it ii) the line does not depend on the particular positions of
the initial points $\bx$, $T_i\bx$, $T_j\bx$ and $T_k\bx$;\\ 
iii) the new construction gives the same result as the
previous one; i. e., the focal lattices are quadrilateral.} 

To check
that the above construction is the canonical one, we first show that
there exists a
unique line which satisfies conditions i) and iii); due to the
uniqueness of the line satisfying condition iii), the condition ii)
will be also proven.

Assume we have points $\bx,...T_jT_k\bx$ satisfying the planarity conditions
and belonging to the corresponding lines $\gl,...,T_jT_k\gl$.
Using the standard MQL construction we find the point $T_iT_jT_k\bx$;
the point $T_iT_jT_k\by_i$ is the intersection point of the plane 
$\VV_{jk}(T_i\by_i)$ with the line $T_jT_k\gl$.

\begin{center}
\leavevmode\epsfxsize=12cm\epsffile{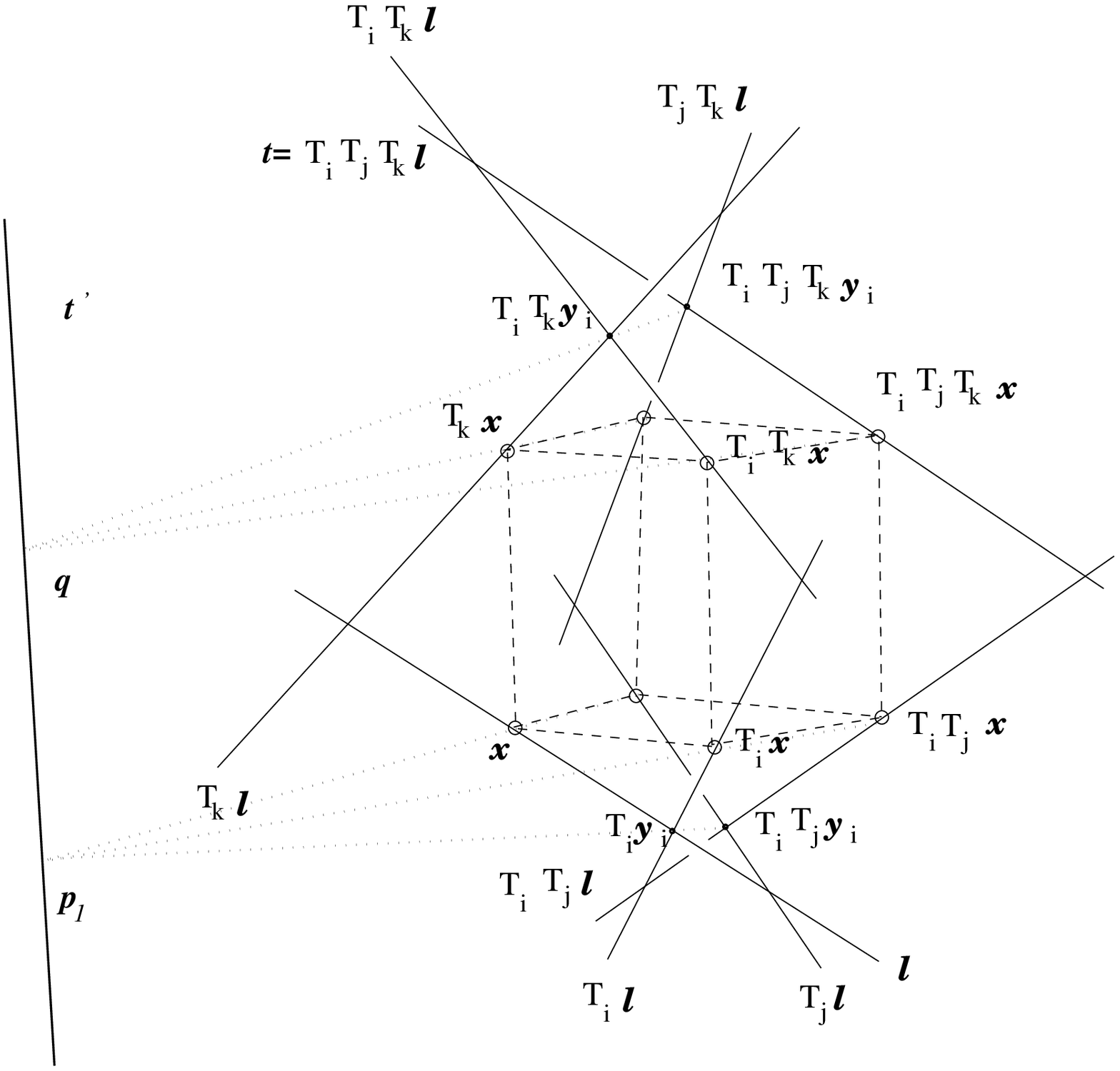}
\end{center}
Figure 4.

Denote by $\gt$ the line passing through $T_iT_jT_k\bx$ and
$T_iT_jT_k\by_i$ (see Fig. 4.). Our goal is to demonstrate that the
quadrilaterals $\{ T_iT_k\by_i$, $T_iT_jT_k\by_i, T_iT_k\bx,
T_iT_jT_k\bx \}$ and $\{ T_iT_j\by_i$, $T_iT_jT_k\by_i$, $T_iT_j\bx,$
$T_iT_jT_k\bx \}$ are planar; this would show that the line $\gt$
meets lines $T_iT_k\gl$ and $T_iT_j\gl$, which would imply that the
line $T_iT_jT_k\gl = \gt$ satisfying condition i) does exist.

Denote by $\gt^\prime$ the intersection line of the planes $\VV_{jk}(\bx)$ 
and $\VV_{jk}(T_i\bx)$. Obviously, the points
$\bp_1=\langle \bx, T_j\bx \rangle\cap \langle T_i\bx, T_iT_j\bx \rangle=$ 
$\VV_{j}(\bx)\cap\VV_j(T_i\bx)$
and
$\bp_2=\VV_{k}(\bx)\cap\VV_k(T_i\bx)$ belong to $\gt^\prime$.
Application of Lemma~\ref{lem:ab} gives $\bp_1\in\VV_j(T_i\by_i)$
and $\bp_2\in\VV_k(T_i\by_i)$,
which implies that the line
$\gt^\prime$ is contained in the plane $\VV_{jk}(T_i\by_i)$.

Since the quadrilateral $\{ T_k\bx , T_kT_j\bx , T_iT_k\by_i , T_iT_jT_k\by_i
\}$ is planar then the lines $\VV_j(T_iT_k\by_i)$ and $\VV_j(T_k\bx)$ intersect
in a point $\bq$, which, accordingly to the reasoning above, must belong to 
the line $\gt^\prime$.
Since the lines $\VV_j(T_k\bx)$ and $\VV_j(T_iT_k\bx)$ intersect
also in a point of $\gt^\prime$, then the point $\bq$ is the intersection
point of all the three lines. This implies that 
the quadrilateral 
$\{T_iT_k\by_i$, $T_iT_jT_k\by_i$, $T_iT_k\bx$, $T_iT_jT_k\bx\}$ is
planar. Similar reasonings show that the quadrilateral 
$\{T_iT_j\by_i$, $T_iT_jT_k\by_i$, $T_iT_j\bx$, $T_iT_jT_k\bx\}$ is
planar as well, which shows that the new construction of the congruence
is indeed the canonical one.

The above reasoning allows to formulate the following
\begin{Prop}
If, for a non-generic congruence, there exists a quadrilateral lattice conjugate 
to it, then the focal lattices of the congruence are quadrilateral.
\end{Prop}
This result, together with Proposition~\ref{prop:foc-quad-ng} implies the 
following important
\begin{Cor}
Focal lattices of congruences conjugate to quadrilateral lattices
are quadrilateral lattices.
\end{Cor}

In the sequel we will need also the following result.
\begin{Prop} \label{prop:conj-foc-conj}
Given two congruences $\gl_1$, $\gl_2$ conjugate to the same quadrilateral
lattice $\bx$,
then the lines defined by joining corresponding points of two
focal lattices $\by_i(\gl_1)$ and $\by_i(\gl_2)$
form a congruence $\gt_i$ conjugate to both focal
lattices.
\end{Prop}
\begin{Proof}
In the Fig. 5 below two congruences $\gl_1$ and $\gl_2$ are represented, 
respectively, by dotted and dashed lines.
We have to prove that the lines $\gt_i$ form a congruence.
The lines $\gt_i$ and $T_i^{-1}\gt_i$ are coplanar because
they belong to the plane of the two intersecting (in $T_i^{-1}\bx$)
lines $T_i^{-1}\gl_1$ and $T_i^{-1}\gl_2$.

To show that the lines $\gt_i$ and $T_j^{-1}\gt_i$, $j\ne i$, are coplanar,
let us consider the quadrilateral with vertices $\by_i(\gl_1)$, $\by_i(\gl_2)$,
$T_j^{-1}\by_i(\gl_1)$ and $T_j^{-1}\by_i(\gl_2)$. Due to Lemma~\ref{lem:ab}
the lines $\langle \by_i(\gl_1) , T_j^{-1}\by_i(\gl_1) \rangle $
and $\langle \by_i(\gl_2) , T_j^{-1}\by_i(\gl_2) \rangle $ intersect in the 
point $\langle \bx , T_j^{-1}\bx \rangle \cup \langle T_i^{-1}\bx , 
T_i^{-1} T_j^{-1}\bx \rangle $, which proves the planarity of the 
quadrilateral and, therefore, the coplanarity of the lines
$\gt_i= \langle \by_i(\gl_2) , \by_i(\gl_2) \rangle $ and 
$T_j^{-1}\gt_i = \langle T_j^{-1}\by_i(\gl_2) , T_j^{-1}\by_i(\gl_2) \rangle $.

\begin{center}\leavevmode\epsfxsize=8cm\epsffile{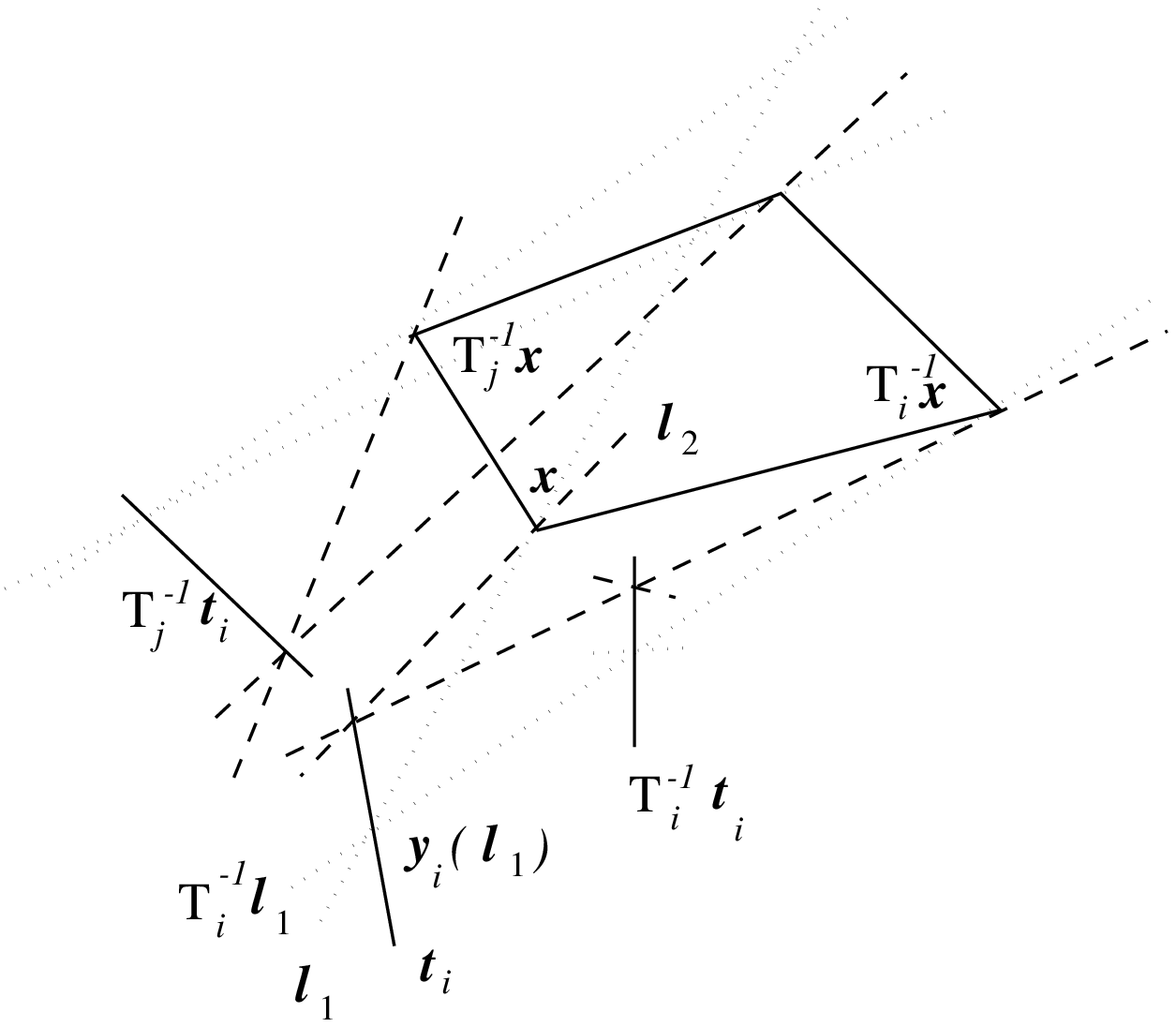}\end{center}
Figure 5.

\end{Proof}

\section{Laplace transformations}
\label{sec:Laplace}

In Section \ref{sec:congruences} we considered congruences of lines and
their focal lattices. In this Section we
are interested, in particular, in the relations between two focal
lattices of the same congruence;these relations are described by the Laplace
transformations.

The Laplace transformations of conjugate nets were introduced by Darboux 
(see \cite{DarbouxIV,Eisenhart-TS,Ferapontov}). For $N=2$ this transformation 
provides the geometric meaning of the transformation (known already to Laplace) 
connecting solutions of two Laplace equations.

\begin{Def}
The Laplace transform $\cL_{ij}(\bx)$ of the quadrilateral lattice $\bx$
is the $j$-th focal lattice of its $i$-th tangent congruence
\begin{equation}
 \cL_{ij}(\bx) = \by_j (\gt_i (\bx )).
\end{equation}
\end{Def}
In simple terms, $\cL_{ij}(\bx)$ is the intersection point of the line
passing through $T_j^{-1}\bx$ and $T_j^{-1}T_i\bx$ with the line
passing through $\bx$ and $T_i\bx$ \cite{DCN}.

\begin{center}\leavevmode\epsfxsize=9cm\epsffile{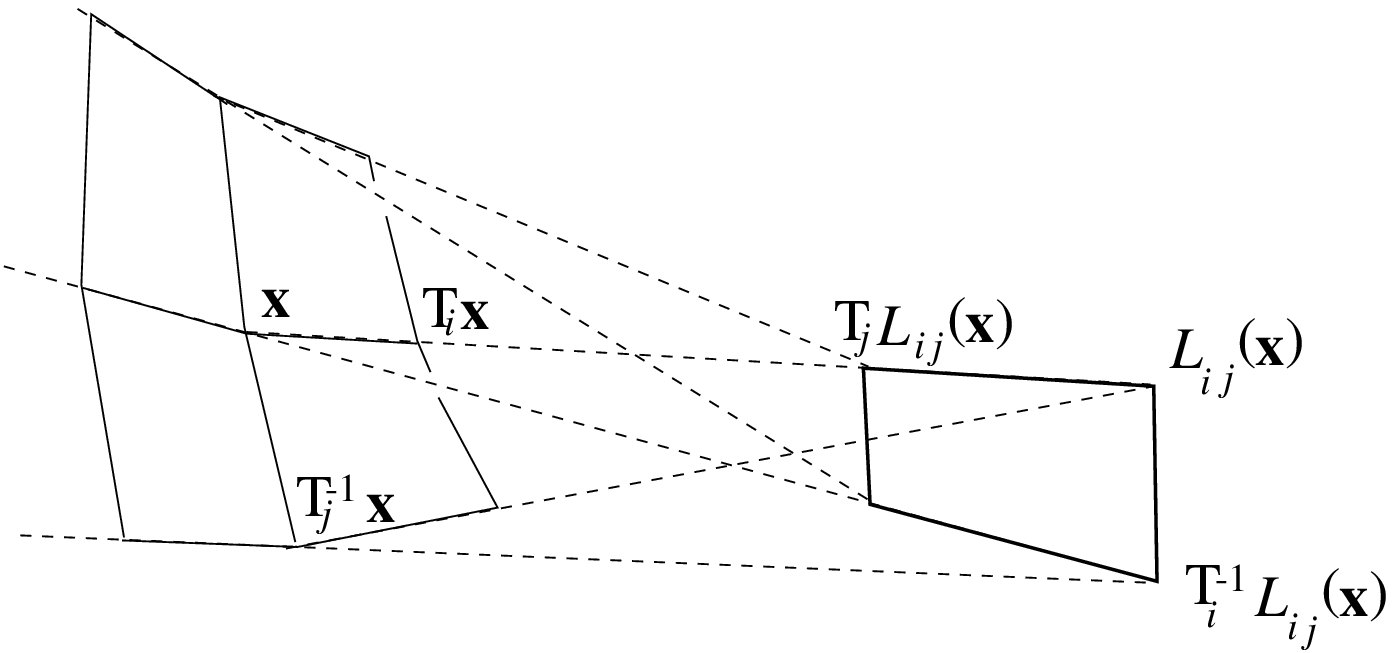} \end{center}
Figure 6.

The points of the first line are of the form
\begin{equation}
\bp(t)=T_j^{-1}\bx + t T_j^{-1}\bX_i,
\end{equation}
which can be transformed, using (\ref{def:HX}) and (\ref{eq:lin-X}), into
\begin{equation}
\bp(t) = \bx + t\bX_i - (H_j + t Q_{ij}) T_j^{-1}\bX_j;
\end{equation}
the intersection point of the two lines is therefore given by
\begin{equation}
 t=-\frac{H_j}{Q_{ij}}.
\end{equation}
Therefore we have the following
\begin{Prop} \label{prop:Laplace}
The Laplace transformation of the quadrilateral lattice $\bx$ is given by
\begin{equation} \label{def:Lij}
\cL_{ij}(\bx) = \bx -\frac{H_j}{Q_{ij}}\bX_i =
\bx - \frac{1}{A_{ji}} \D_i\bx.
\end{equation}
\end{Prop}
By direct calculations, one has the
\begin{Cor}
i) The Laplace transformed $A$--coefficients are of the form
\begin{align}
\cL_{ij}(A_{ij}) &=\frac{A_{ji}}{T_jA_{ji}} (T_iA_{ij} +1) -1 \; ,\\
\cL_{ij}(A_{jk})  & =  T_j^{-1}\left(
\frac{T_k\cL_{ij}(A_{ij})}{\cL_{ij}(A_{ij})} \left( A_{jk} + 1\right)\right) 
-1 , \; \;  \\
\cL_{ij}(A_{ik}) & =  A_{jk}T_k\left( 1 - \frac{A_{ki}}{A_{ji}}\right), 
\quad k \ne i,j,\\
\cL_{ij}(A_{kl})  & = (A_{kl} + 1)
\frac{T_k\left(1 -A_{ki}/A_{ji}\right)}
{\left(1 -A_{ki}/A_{ji} \right)} -1 \quad k\neq j,i \quad l\neq k.
\label{eq:Lij-kl}
\end{align}
ii) The Lam\'e coefficients of the transformed
lattice read
\begin{align}
\cL_{ij}(H_i) &= \frac{T_iH_i}{A_{ji}} = \frac{H_j}{Q_{ij}} 
\; , \label{eq:L-Hi} \\
\cL_{ij}(H_j) &= T_j^{-1}\left( H_j \cL_{ij}(A_{ij}) \right) = 
T_j^{-1}\left( Q_{ij}\D_j \left(\frac{H_j}{Q_{ij}}  \right) \right)
\; , \label{eq:L-Hj} \\
\cL_{ij}(H_k) &= H_k \left( 1 - \frac{A_{ki}}{A_{ji}}\right) 
= H_k - \frac{Q_{ik}}{Q_{ij}} H_j 
\; , \; \; k\ne i,j, \label{eq:L-Hk} 
\end{align}
iii) The tangent vectors of the new lattice read
\begin{align}
\cL_{ij}(\bX_i) & = - \D_i\bX_i + \frac{\D_iQ_{ij}}{Q_{ij}} \bX_i 
\; ,\label{eq:L-Xi} \\
\cL_{ij}(\bX_j) & = -\frac{1}{Q_{ij}}\bX_i \; ,
\label{eq:L-Xj} \\
\cL_{ij}(\bX_k) & = \bX_k - \frac{Q_{kj}}{Q_{ij}} \bX_i 
\; ,\; \; k\ne i,j, \label{eq:L-Xk} \; .
\end{align}
\end{Cor}

Finally we remark that, apart from the identity
\begin{equation} \label{eq:Lid-ij}
\cL_{ij} \circ \cL_{ji}  = \text{id} \; ,
\end{equation}
which follows just from the definition of the Laplace transformation
(see also \cite{DCN}), there are two other identities
\begin{align}
\cL_{jk} \circ \cL_{ij} &= \cL_{ik}, \\
\cL_{ki} \circ \cL_{ij} &= \cL_{kj}; \label{eq:Lid-ijk}
\end{align}
which follow from the Corollary \ref{cor:foc-ident-Lapl}, or may be verified
directly from the above equations.

Notice that, to construct a line of the new lattice, one needs a
quadrilateral strip of the old lattice (see Fig.~7). 
Similarly, one $(N-1)$ dimensional level of
the new lattice can be constructed out of
two $(N-1)$ dimensional levels of the original lattice (i.e., out of
a quadrilateral strip with an $(N-1)$ dimensional basis). In fact, we may 
define the
Laplace transform of a quadrilateral strip; this last observation will be used in the next Sections.

\begin{center}
\leavevmode
\epsfxsize=5cm
\epsffile{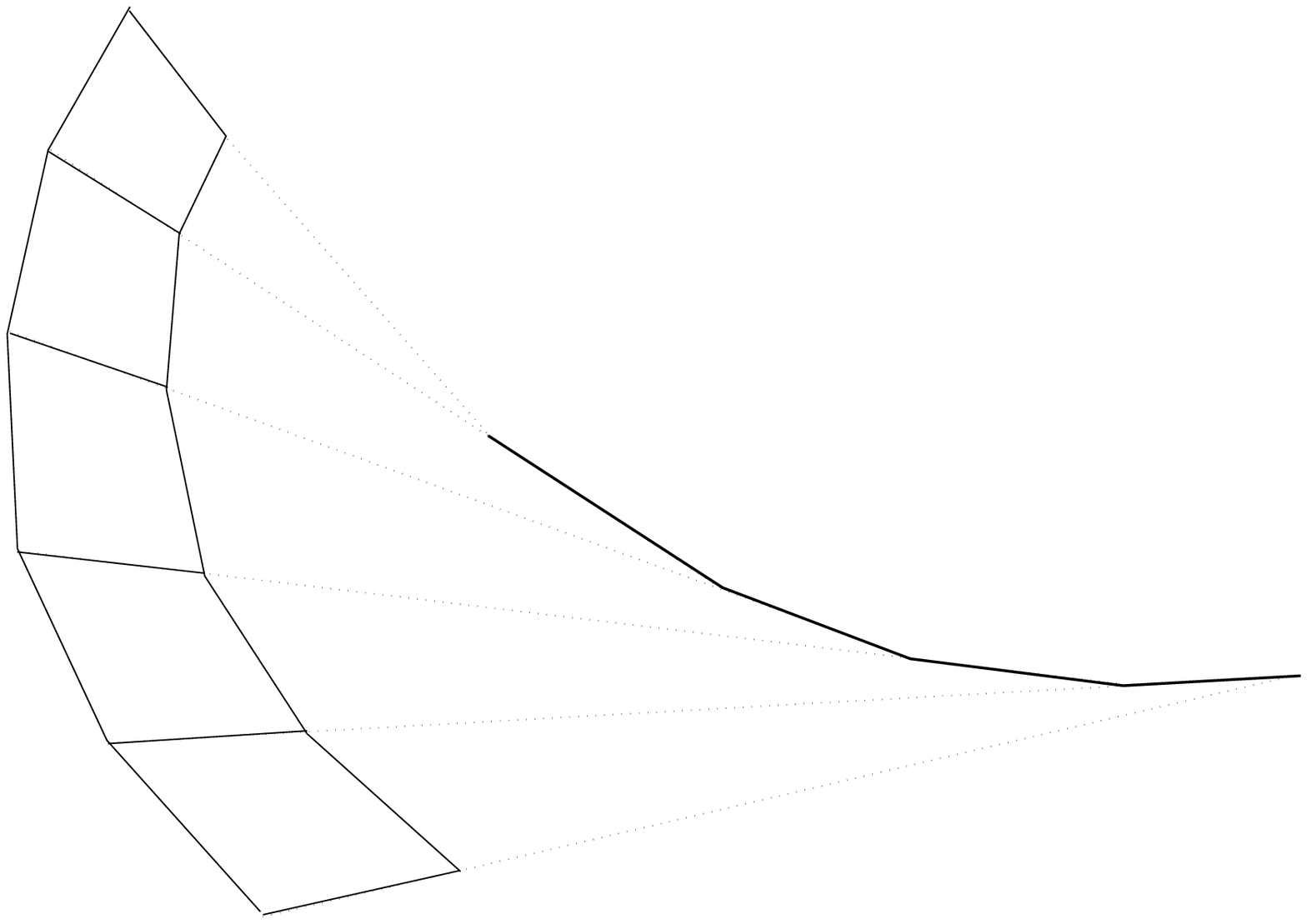}
\end{center}
Figure 7.

\section{Combescure transformations}

\label{sec:Combescure}

In this Section we study quadrilateral lattices related by parallelism of the 
tangent vectors. Basically, we generalize to a discrete level the results
about the Combescure transformations of the conjugate nets, as presented
in the monograph~\cite{Eisenhart-TS}. The Definition~\ref{def:Combescure}
and Proposition~\ref{prop:parallel} of  
Section~\ref{sec:subCombescure} is also contained  in~\cite{KoSchief2}.

\subsection{Combescure transformations of quadrilateral lattices}

\label{sec:subCombescure}

\begin{Def} \label{def:Combescure}
A lattice ${\cal C}(\bx):\ZZ^N\ra\RR^M$
is called  Combescure transform of (or parallel to) the
quadrilateral lattice $\bx:\ZZ^N\ra\RR^M$ if the tangent vectors of
both lattices in the corresponding points are proportional:
\begin{equation}
 \D_i{\cal C}(\bx) = (T_iC_i)\D_i\bx, \; \; i=1,...,N.
\end{equation}
\end{Def}

We mention that the definition of the Combescure transformation
makes use of the notion of parallelism, which has an  affine
geometry origin and comes from fixing the hyperplane at infinity
\cite{Samuel}.

The following results can be verified by direct calculation.
\begin{Prop} \label{prop:parallel}
i) The proportionality factors $C_i$ satisfy the equations
\begin{equation}
 \D_jC_i = A_{ij}T_j(C_j-C_i), \; \; i\ne j.
\end{equation}
ii) The transformed lattice is a quadrilateral lattice with
Combescure-transformed functions of the form
\begin{align*}
{\cal C}(A_{ij}) &= A_{ij}\frac{T_jC_j}{C_i},
 \; \; i\ne j, \\
     {\cal C}(\bX_i) &= \bX_i, \\
     {\cal C}(H_i) &= C_iH_i.
\end{align*}
iii) All the quadrilaterals with vertices $\{ \bx, T_i\bx , {\cal C}(\bx) ,
{\cal C}(T_i\bx) \} $ are planar.
\end{Prop}
From the last property of Proposition \ref{prop:parallel}, it
follows that the lattices $\bx$ and $\calC(\bx)$ form a
quadrilateral strip with the $N$ dimensional basis $\bx$ and the
transversal direction given by the Combescure transform $\calC$
(direction $\calC$); see Fig. 8.

 \begin{center}
\leavevmode
 \epsfxsize=5cm
 \epsffile{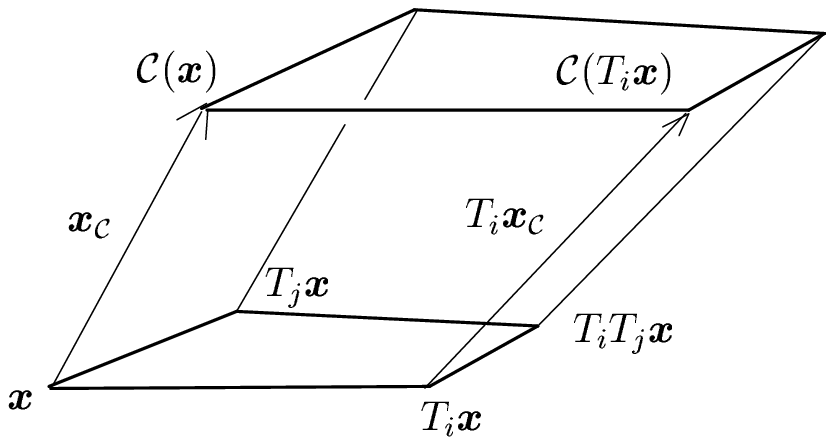}
 \end{center}
Figure 8.

Therefore the recursive application of a Combescure transformation to the 
$N$-dimensional quadrilateral lattice $\bx$ can be viewed as generating
a new dimension (say, the $N+1$st) of the lattice. The corresponding
data are simply:
\begin{align*}
H_{N+1} &= 1 \; , \\
     \bX_{N+1} &= \bx_\calC \; ,
\end{align*}
up to an arbitrary function of $n_{N+1}$, always present in the definition
of $H$ and $\bX$ (see~\cite{MQL}).

We observe that the transversal vector $\bx_\calC$, given by
\begin{equation}
\label{def:X-C}
\bx_\calC = {\calC}(\bx) - \bx,
\end{equation}
satisfies the equations
\begin{equation} \label{eq:D-iX-C}
 \D_i\bx_\calC = (T_i\sigma_i) \D_i\bx =
(T_iv_i^*) \bX_i,
\end{equation}
where  the functions $\sigma_i$ and $v^*_i$, $i=1,...,N$ are given by
\begin{equation} \label{def:sigma-v*}
\sigma_i = C_i - 1, \; \; \;
v_i^*= (C_i - 1)H_i.
\end{equation}
The following facts are easy to verify.
\begin{Cor}
i) Functions $v_i^*$ satisfy the adjoint linear system
(\ref{eq:lin-H}). \\
ii) Functions $\sigma_i$ satisfy the equation
\begin{equation} \label{eq:sigma}
\D_j\sigma_i = \frac{\D_j H_i}{H_i} T_j(\sigma_j-\sigma_i), \; \;
i\neq j .
\end{equation}
iii) In the notation of Theorem \ref{th:vect-Darb},
the vector $\bx_\calC$ can be
rewritten as
\begin{equation}
 \bx_\calC = \bOm[\bX,v^*];
\end{equation}
i. e., the function $\bx_\calC:\ZZ^N \ra \RR^M$ is a solution of
the Laplace equation
\begin{equation} \label{eq:Lapl-X-C}
\D_i\D_j\bx_\calC = \left(T_i\frac{\D_jv^*_i}{v^*_i} \right)
\D_i\bx_\calC +
\left(T_j \frac{\D_iv^*_j}{v^*_j} \right) \D_j\bx_\calC.
\end{equation}
iv) The lattice $\bx_\calC$ is also a Combescure transform
of $\bx$.
\end{Cor}

From the above considerations we can extract the following construction of 
the
Combescure transform, which will be used in the next Sections.

\begin{Prop} \label{prop:Combescure}
In order to construct a Combescure transform of the lattice $\bx$ we\\
i) find a scalar solution $v^*_i$ of the adjoint linear problem
\[ \D_j v^*_i = (T_j v^*_j) Q_{ji} \; ;
\]
ii) the Combescure transform of $\bx$ is then given by
\begin{equation}
 \calC(\bx) = \bx + \bOm[\bX, v^*] = \bOm[\bX, H + v^* ].
\end{equation}
\end{Prop}

Given any scalar solution $\phi$ of the Laplace equation
(\ref{eq:Laplace}), we define its Combescure transformed 
function $\phi_\calC$ in terms of
$\phi$ in the same way in which $\bx_\calC$ follows from $\bx$:
\begin{equation} \label{eq:phi-C}
\D_i\phi_\calC = (T_i\sigma_i)\D_i\phi.
\end{equation}
Equivalently, since $\phi$ defines a scalar solution $v_i$,
$i=1,...,N$ of the linear problem (\ref{eq:lin-X}) via
\begin{equation} \label{def:v-phi}
\D_i\phi = (T_iH_i)v_i,
\end{equation}
we have
\begin{equation}
 \phi_\calC = \bOm [ v,v^*].
\end{equation}

\subsection{Combescure congruences}

Let us consider an important example of congruence obtained
from a quadrilateral lattice and its Combescure-transformed lattice.

From Proposition \ref{prop:parallel} iii) it follows that, given a
pair of parallel lattices, the lines
passing through $\bx$ and ${\cal C}(\bx)$ define a congruence
which we call Combescure
congruence. 

The focal lattices of this congruence can be found in the following way.
Given a real function $t:\ZZ^N \ra \RR$,
define a new lattice $\by$ with points on
the lines of the congruence
\begin{equation} \by = \bx + t\bx_\calC;
\end{equation}
the tangent vectors of the new lattice are given by
\begin{equation}
\D_i\by = \left( 1 + T_i\left( \sigma_i t \right) \right)\D_i\bx    +
(\D_it) \: \bx_\calC.
\end{equation}
When
\begin{equation}
 t = -\frac{1}{\sigma_i}\; ,
\end{equation}
then the line of the $i$-th tangent vector $\D_i\by$ is the line
of the congruence and therefore the lattice
\begin{equation} \label{eq:foc-latt-Comb}
\by_i = \bx - \frac{1}{\sigma_i}\bx_\calC
\end{equation}
is the $i$-th focal lattice of the Combescure congruence.

\begin{Cor}
All the lattices $\bx$, ${\cal C}(\bx)$, $\by_i$, $i=1,...,N$, are
conjugate to the same (Combescure) congruence.
\end{Cor}

The Combescure congruences will be used extensively throughout
the paper due to the following result.
\begin{Prop} \label{prop:Comb-conj}
Any congruence conjugate and transversal to a quadrilateral lattice
$\bx$ (i.e. not tangent to the lattice in the corresponding points)
comes from a Combescure transform ${\cal C}(\bx)$.
\end{Prop}
\begin{Proof}
Geometrically, the construction of such lattice $\calC(\bx)$
is as follows. Mark on the line $\gl ({\boldsymbol 0})$, ${\boldsymbol 0} 
\in\ZZ^N$, of the congruence a point $\calC(\bx({\boldsymbol 0}))$ different
from $\bx({\boldsymbol 0})$.
The point $T_i\calC(\bx({\boldsymbol 0}))$ is the intersection of the line
$T_i\gl ({\boldsymbol 0})$ with the line passing through 
$\calC(\bx({\boldsymbol 0}))$ and parallel to the line $\langle 
\bx({\boldsymbol 0}), T_i \bx({\boldsymbol 0}) \rangle$. The compatibilty
of this construction, i.e., $T_iT_j\calC(\bx) = T_jT_i\calC(\bx)$, follows from 
the fact that $T_iT_j\calC(\bx)$ is the intersection point of $T_iT_j\gl$
with the plane $\langle \calC(\bx), T_i\calC(\bx), T_j\calC(\bx) \rangle$.

Since this Proposition is one of the most important in our paper,
we give an alternative algebraic proof.
A congruence $\gl$ conjugate to $\bx$ can be described by giving
the vector-function $\bX:\ZZ^N\ra\RR^M$ in the direction of the
line of the congruence which passes through the corresponding point
$\bx$ of the lattice. Our goal is to rescale the direction vector
of the congruence by a function $t$, such that the lattice $\bx + t\bX$
is parallel to $\bx$.

The co-planarity of the neighboring lines of
the congruence implies that, if $\D_i\bX \ne 0$, then $\D_i\bx$ can
be decomposed into a linear combination of $\bX$ and $\D_i\bX$, i.e.:
\begin{equation}
\D_i\bx \in Span\{ \bX , \D_i\bX \}
\end{equation}
This implies that $\D_i\D_j\bx$ is a linear combination of $\bX$, $\D_i\bX$, 
$\D_j\bX$ and$\D_i\D_j\bX$.
But since
\begin{equation}
 \D_i\D_j\bx \in Span\{ \D_i\bx , \D_j\bx \} \subset
Span\{ \bX, \D_i\bX, \D_j\bX \}\; , \quad i\ne j \; , 
\end{equation}
then, there must exist a linear relation between $\bX$, $\D_i\bX$,
$\D_j\bX$ and $\D_i\D_j\bX$, which can be written in the form of
the generalized Laplace equation
\begin{equation}  \label{eq:LaplX}
 \D_i\D_j\bX = (T_iB_{ij})\D_i\bX + (T_jB_{ji})\D_j\bX + C_{(ij)}\bX
 \; , \quad i\ne j \; .
\end{equation}
The compatibility condition between (\ref{eq:LaplX})
implies the existence of the logarithmic
potentials $F_i$ (see also the discussion in \cite{MQL}) such that
\begin{equation}
 B_{ij} = \frac{\D_j F_i}{F_i}\; , \quad i\ne j \; .
\end{equation}
Let us consider functions $\lambda_i : \ZZ^N \ra \RR $
which describe the focal lattices $\by_i$ of the congruence
in terms of the reference lattice $\bx$ and of the direction vectors $\bX$
\begin{equation} \label{eq:foc-yi}
\by_i = \bx - \lambda_i \bX ;
\end{equation}
note that, due to the transversality of the congruence, the 
functions $\lambda_i$
never vanish. Since $\by_i$ are the focal lattices of $\gl$, then
the vectors $\D_i\by_i$ are directed along $\bX$:
\begin{equation}
 \D_i\by_i = \rho_i \bX,
\end{equation}
and this equation can be rewritten, using equation (\ref{eq:foc-yi}), as
\begin{equation} \label{eq:Dix-X}
 \D_i\bx = (T_i\lambda_i )\D_i\bX + \mu_i\bX,
\end{equation}
where $\mu_i = \rho_i + \D_i\lambda_i$.

The application of the partial difference operator $\D_j$ to equation
(\ref{eq:Dix-X}) and the Laplace equation (\ref{eq:Laplace-H}) with
equation (\ref{eq:Dix-X}) give
\begin{multline*}
(T_jT_i\lambda_i)\D_i\D_j\bX + (\D_jT_i\lambda_i)\D_i\bX +
(T_j\mu_i)\D_j\bX + (\D_j\mu_i)\bX =\\
\left( T_i\frac{\D_jH_i}{H_i}\right) \left( (T_i\lambda_i)
\D_i\bX + \mu_i\bX \right) +
\left( T_j\frac{\D_iH_j}{H_j}\right) \left( (T_j\lambda_j)
\D_j\bX + \mu_j\bX \right).
\end{multline*}
Rewriting this equation in the form of the generalized Laplace
equations (\ref{eq:LaplX}) allows to calculate the coefficients
$B_{ij}$:
\begin{equation}
 B_{ij} = \frac{\D_j\left(H_i/\lambda_i\right)}
{H_i/\lambda_i} \; \; \; \; \;\Longrightarrow \; \; \; F_i =
\frac{H_i}{\lambda_i}.
\end{equation}
Comparing  both expressions for $B_{ji}$ one obtains the following
identity
\begin{equation}
\frac{T_iH_j}{H_j} = \frac{(T_i\lambda_i)(T_i\lambda_j)}{\lambda_j
(T_i\lambda_i - T_i\lambda_j)}
\left(\frac{\lambda_j + \mu_i}{T_i\lambda_i} - 1
\right). \label{eq:iden1}
\end{equation}
Since $C_{(ij)}$ should be symmetric with respect to the change of
indices (see \cite{MQL}), then, using equation (\ref{eq:iden1}), one 
arrives to
\begin{equation}
\frac{\mu_i}{T_i\lambda_i}\left(T_i \frac{\mu_j}{T_j\lambda_j} -1 \right)
= \frac{\mu_j}{T_j\lambda_j}\left( T_j \frac{\mu_i}{T_i\lambda_i} -1 \right),
\end{equation}
which implies the existence of
a potential function $t:\ZZ^n \ra \RR$ such that
\begin{equation}
 \frac{T_i t}{t} = \left( 1 -\frac{\mu_i}{T_i\lambda_i}
  \right)^{-1}.
\end{equation}
Now, we can scale the direction vector $\bX$ of the congruence
multiplying it by the potential $t$ and check that
\begin{equation}
\D_i(t\bX) = \left( T_i\frac{t}{\lambda_i} \right)\D_i\bx,
\end{equation}
which asserts that
the lattice with points given by $\bx+t\bX$ is a Combescure transform
of $\bx$. We only remark that an arbitrary scalar constant in the potential $t$ 
corresponds to the freedom in choosing the initial point 
$\calC(\bx({\boldsymbol 0}))$.
\end{Proof}

\section{L\'evy transformations and their adjoint}

In this Section we are interested in the relations between two
quadrilateral lattices in which one of
the lattices
is a focal lattice of the congruence conjugate to the other. 
In the continuous context, these transformations are called L\'evy 
transformations~\cite{Levy}
and are studied in detail in~\cite{Eisenhart-TS,Lane}.
We remark
that, in the limiting case when also the second lattice (net) is focal,
we arrive to the Laplace transformations considered in 
Section~\ref{sec:Laplace}.

\subsection{Adjoint L\'evy transformations}

\label{sec:adj-Levy}

\begin{Def} \label{def:adj-Levy} The $i$-th adjoint L\'evy transform
$\cL^*_i(\bx)$ of the quadrilateral lattice
$\bx$ is the $i$-th focal lattice of a congruence conjugate to $\bx$.
\end{Def}

\begin{center}\leavevmode\epsfxsize=6cm\epsffile{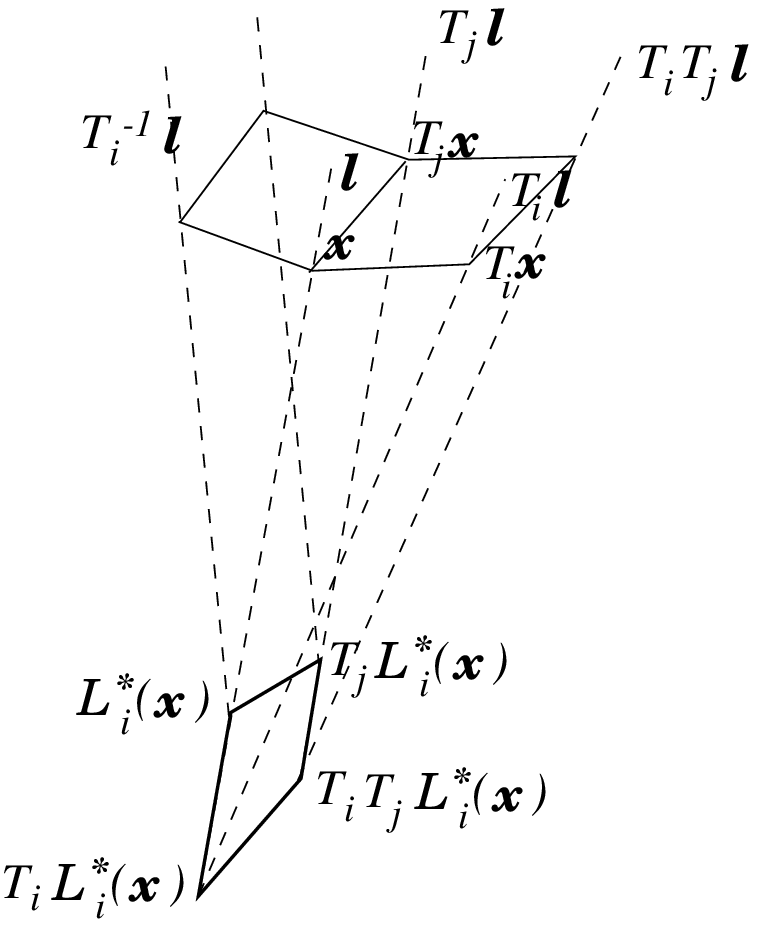}\end{center}
Figure 9.

\begin{Rem}
Adjoint L\'evy transformations are usually called in soliton theory
{\it adjoint elementary Darboux transformations} \cite{ms,OevelSchief,KoSchief}.
\end{Rem}

Assuming that we deal with a generic case, i. e. the congruence conjugate to
$\bx$ is transversal to it, we construct this congruence
via a Combescure transformation vector $\bx_\calC$
of the lattice $\bx$.
Combining Propositions \ref{prop:Combescure} and  \ref{prop:Comb-conj}
with formula (\ref{eq:foc-latt-Comb}) for the focal lattices
of the Combescure congruence, we obtain
\begin{Prop}
i) The adjoint L\'evy transform of the lattice $\bx$ is given by
\begin{equation}  \label{eq:adj-Levy1}
\cL_i^*(\bx) = \bx - \frac{1}{\sigma_i}\bx_\calC \; ,
\end{equation}
where the functions $\sigma_i$ are solutions of the equation
(\ref{eq:sigma}). \\
ii) The Lam\'e coefficients of the new lattice are of the form
\begin{equation}
\cL_i^*(H_i) = T_i^{-1}\left( H_i \frac{\D_i\sigma_i}{T_i\sigma_i} \right) \; ,
\quad \cL_i^*(H_j) = H_j\left( 1 - \frac{\sigma_j}{\sigma_i} \right).
\end{equation}
\end{Prop}

Since $\D_i\cL_i^*(\bx)$ is, by definition, proportional to $\bx_\calC$,
it is easy to check that
\begin{equation} \label{eq:adj-Levy2}
\cL_i^*(\bx) = \bx + \frac{1/\sigma_i}{\D_i(1/\sigma_i)}\D_i\cL_i^*(\bx).
\end{equation}
At this point we can also verify the result which we will use in the
next section.

\begin{Lem} \label{lem:1/sigma}
The function $\displaystyle\frac{1}{\sigma_i}$ satisfies the point
equation of the lattice $\cL_i^*(\bx)$.
\end{Lem}

It is convenient to reformulate our results
in the notation of Theorem \ref{th:vect-Darb}. Using the
functions $v_i^*$ defined
in (\ref{def:sigma-v*}), we have the following algebraic formulation
of the adjoint L\'evy transformation.
\begin{Prop} \label{prop:adj-Levy}
To construct the adjoint L\'evy transform $\cL^*_i(\bx)$ of the
quadrilateral lattice $\bx$: \\ i) find a scalar solution $v_i^*$
of the adjoint linear problem
\[ \D_j v^*_i = (T_j v^*_j) Q_{ji},
\]
which defines the direction vectors $\bx_\calC  =\bOm[\bX,v^*]$ of
a congruence conjugate to $\bx$. \\ 
ii) Its $i$-th focal
lattice is the adjoint L\'evy transform:
\begin{equation}
\cL^*_i(\bx) = \bx - \frac{H_i}{v^*_i}\bOm[\bX,v^*].
\end{equation}
iii) The Lam\'e coefficients and the tangent vectors of the new lattice are of 
the form
\begin{align}
\cL^*_i(H_i) & = -T_i^{-1}\left( v_i^* \D_i \left( \frac{H_i}{v_i^*}
\right) \right) \; ,\\
\cL^*_i(H_j) & = H_j - \frac{v_j^*}{v_i^*}H_i \; , \\
\cL^*_i(\bX_i) & =  \frac{1}{v_i^*}\bOm[\bX,v^*] \; ,\\
\cL^*_i(\bX_j) & = \bX_j - \frac{Q_{ji}}{v_i^*}\bOm[\bX,v^*] \; .
\end{align}
\end{Prop}

Let us observe that the lattices $\bx$ and $\bx + \bx_\calC$ form a 
quadrilateral strip with the $N$ dimensional basis $\bx$ and one transversal 
direction $\bx_\calC$.
The adjoint L\'{e}vy  transformation $\cL_i^*$ of the lattice $\bx$
can be interpreted as the Laplace transformation $\cL_{\calC i}$ of the
strip. 

We also remark that Proposition \ref{prop:conj-foc-conj}
can be formulated in the following way.
\begin{Prop} \label{prop:2adLevy-conj}
Two lattices which have been obtained by the $i$-th adjoint L\'evy
transformation of
the same quadrilateral lattice are conjugate to the same congruence.
\end{Prop}

\subsection{L\'evy transformations}

\label{sec:Levy}

\begin{Def} \label{def:Levy}
The $i$-th L\'evy transform $\cL_i(\bx)$ of the quadrilateral lattice
$\bx$ is a quadrilateral lattice
conjugate to the $i$-th tangent congruence of $\bx$.
\end{Def}
\begin{center}
\leavevmode
\epsfxsize=9cm
\epsffile{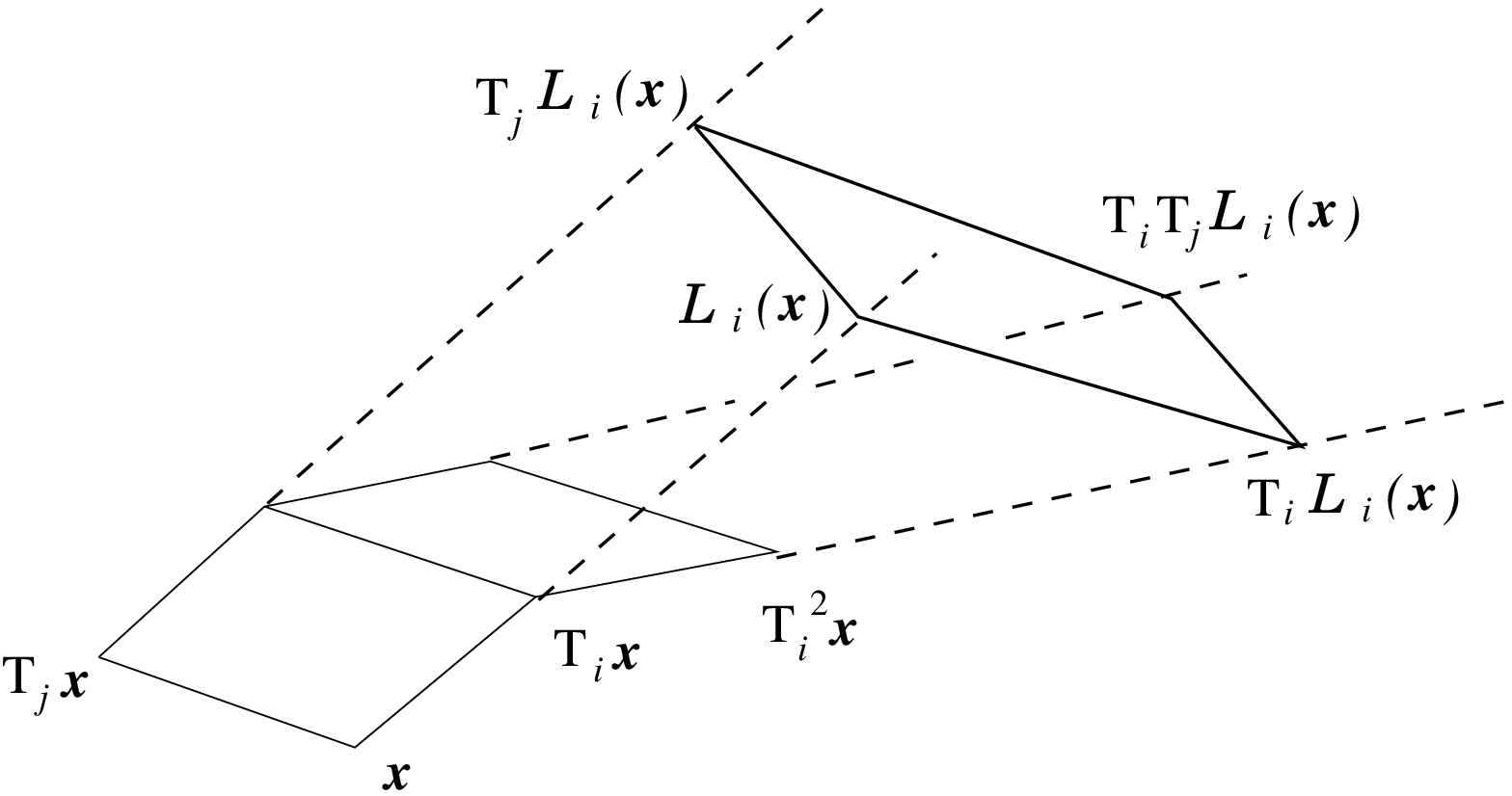}
\end{center}
Figure 10.

\begin{Rem} In the soliton theory, L\'evy transformations
of multi-conjugate systems are usually  called {\it elementary Darboux
transformations} \cite{ms,OevelSchief,KoSchief}.
\end{Rem}

It is evident from Definitions \ref{def:Levy} and \ref{def:adj-Levy},
that the L\'evy transform is in a sense the inverse of the adjoint L\'evy
transform. Therefore, in the notation of this 
Section, formula (\ref{eq:adj-Levy2})
can be rewritten as
\begin{equation}
\bx = \cL_i(\bx) + \frac{1/\sigma_i}{\D_i\left( 1/\sigma_i\right)} \D_i\bx.
\end{equation}
Finally, making use of Lemma \ref{lem:1/sigma}, we may formulate the
following result.

\begin{Prop} \label{prop:Levy}
i) The L\'evy transform $\cL_i(\bx)$ of the quadrilateral
lattice $\bx$ is given by
\begin{equation} \label{eq:Levy-transf}
\cL_i(\bx) = \bx - \frac{\phi}{\D_i\phi}\D_i\bx,
\end{equation}
where the function $\phi:\ZZ^N \ra \RR$ is a solution of the Laplace
equation (\ref{eq:Laplace}) of the lattice $\bx$.\\ 
ii) The Lam\'{e}
coefficients of the new lattice read
\begin{equation}
\cL_i(H_i) = (T_iH_i)\frac{\phi}{\D_i\phi} \; , \quad
\cL_i(H_j) = H_j - \frac{\phi}{\D_i\phi}\D_iH_j \; .
\end{equation}
\end{Prop}
Formula (\ref{eq:Levy-transf}), presented in the form coming from the
$\bar\partial$ approach,
was first written in \cite{BoKo}.

The geometric meaning of the function $\phi$ entering into
formula (\ref{eq:Levy-transf}) can be explained as follows. Given
an additional scalar solution $\phi:\ZZ^N \rightarrow \RR$ of
the Laplace equation (\ref{eq:Laplace}), we define a new
quadrilateral lattice $\tilde{\bx}: \ZZ^N \rightarrow \RR^{M+1}$ as
\begin{equation} \tilde\bx : \ZZ^N \rightarrow
\begin{pmatrix}  \bx \\ \phi \end{pmatrix}.
\end{equation}
The point $\cL_i(\bx)$ is the intersection point of the line $\tilde\bx +
t\D_i\tilde\bx$ with its projection $\bx + t\D_i\bx$ on the $\RR^M$ space,
therefore for the intersection parameter $t_0$ we have
\begin{equation}
\begin{pmatrix}  \bx \\ \phi \end{pmatrix}
+ t_0 \begin{pmatrix}  \D_i\bx \\ \D_i\phi \end{pmatrix}
= \begin{pmatrix}  \cL_i(\bx) \\ 0 \end{pmatrix},
\end{equation}
which implies formula (\ref{eq:Levy-transf}).

Let us observe that the direction of the transversal vector
$\tilde\bx - \bx$ is fixed; this implies that the quadrilaterals
with vertices $\bx$, $T_i\bx$, $\tilde\bx$, $T_i\tilde\bx$ are
planar. Then both lattices form a quadrilateral strip with $N$
dimensional basis and one transversal direction $\cL$. The L\'evy
transformation $\cL_i$ of the lattice $\bx$ can be interpreted as
the Laplace transformation $\cL_{i\cL}$ of this strip. Therefore
the L\'evy transformed lattice $\cL_i(\bx)$ is quadrilateral.

As we mentioned in the Section \ref{sec:Combescure}, given a
solution $\phi$ of the Laplace equation (\ref{eq:Laplace}), we
have automatically, via the formula (\ref{def:v-phi}), the 
solution $v_i$ of the linear problem (\ref{eq:lin-X}). Therefore
we may conclude this Section with the
\begin{Cor}To construct a L\'evy transform of the lattice $\bx$: \\
i) find a scalar solution $v_i$  of the linear problem
(\ref{eq:lin-X}); i. e.,
\[ \D_j v_i = (T_jQ_{ij})v_j.
\]
ii) The L\'evy transform is then given by
\begin{equation}
\cL_i(\bx) = \bx - \frac{\bOm[v,H]}{v_i} \bX_i.
\end{equation}
iii) The Lam\'e coefficients and the tangent vectors of the new lattice are of 
the form
\begin{align}
\cL_i(H_i) & = \frac{1}{v_i}\bOm[v,H] \; ,\\
\cL_i(H_j) & = H_j - \frac{Q_{ij}}{v_i}\bOm[v,H] \; , \\
\cL_i(\bX_i) & = -\D_i\bX_i + \frac{\D_iv_i}{v_i}\bX_i \; ,\\
\cL_i(\bX_j) & = \bX_j - \frac{v_j}{v_i}\bX_i \; .
\end{align}
\end{Cor}

\section{Radial transformations} \label{sec:radial}
Given a quadrilateral lattice $\bx$ and a point $\bp\in\RR^M$,
consider lines passing through that point and the points of the
lattice. The conditions of Definition \ref{def:int-congr} are
obviously satisfied. In this way we obtain a special type of
congruence which we call radial congruence. Such congruence is
of a very degenerate type -- its focal lattices consist of the
point $\bp$ only.

Without loss of generality we may assume that the point $\bp$ is the
coordinate center, and we define the radial congruence $\gr(\bx)$ of $\bx$ with
respect to that point.

\begin{Def}
The radial (or projective)
transform $\cP(\bx)$ of the quadrilateral lattice $\bx$ is a
quadrilateral lattice conjugate to the radial congruence $\gr(\bx)$ of $\bx$.
\end{Def}

\begin{center}
\leavevmode
\epsfxsize=5cm
\epsffile{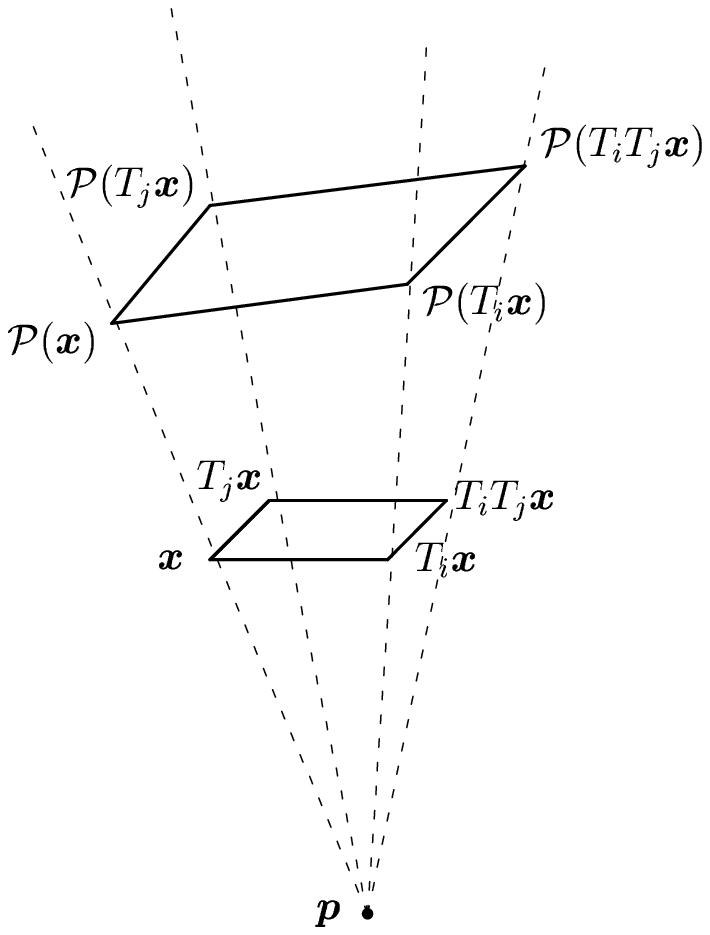}
\end{center}
Figure 11.

\begin{Prop} \label{prop:radial}
i) The radial transform $\cP(\bx)$ is given by
\begin{equation} \label{eq:radial}
\cP(\bx) = \frac{1}{\phi}\bx,
\end{equation}
where $\phi:\ZZ^N\ra \RR$ is a solution of the Laplace equation
(\ref{eq:Laplace}) of the lattice $\bx$. \\ ii) The Lam\'e
coefficients of the new lattice read
\begin{equation}
\cP(H_i) = \frac{H_i}{\phi}.
\end{equation}
\end{Prop}

\begin{Proof}
We first notice that the transformed lattice should consist of the
points of the form given by (\ref{eq:radial}), where $\phi$ must
be such that the new lattice is quadrilateral. For an arbitrary
$\phi$ the new lattice $\tilde\bx =
\frac{1}{\phi}\bx$
satisfies equation
\begin{equation}
\D_i\D_j\tilde\bx = (T_i \tilde A_{ij})\D_i\tilde\bx + (T_j \tilde A_{ji})
\D_j\tilde\bx + \tilde C_{ij}\tilde\bx, \; \; i\not= j,
\end{equation}
with the coefficients
\begin{align} \label{eq:radial-A}
\tilde{A}_{ij}  &=  (T_j\phi)^{-1}(A_{ij}\phi - \D_j\phi ) =
\frac{\D_i\left(H_i/\phi\right)}{H_i/\phi},
 \; \;  \; \; i\neq j,\\   \label{eq:radial-C}
\tilde{C}_{ij}  &=  (T_iT_j\phi)^{-1}
(-\D_i\D_j\phi + (T_iA_{ij})\D_i\phi + 
(T_jA_{ji})\D_j\phi ).
\end{align}
Formula (\ref{eq:radial-C}) precises the form of $\phi$, whereas
(\ref{eq:radial-A}) implies the form of the new Lam\'e
coefficients.
\end{Proof}

\section{Fundamental transformations of the MQL}
The transformations studied in this Section were introduced, in the continuous 
context, by Jonas~\cite{Jonas} as the most general
transformations of conjugate nets on a surface satisfying the permutability
property. Eisenhart, who discovered these transformations independently, but a 
little bit later, called them fundamental transformations~\cite{Eisenhart-TS}. 
The content of Proposition~\ref{prop:fundamental} and
Corollary~\ref{cor:fundamental} can also be found
in~\cite{KoSchief2}.

\label{sec:fundamental}

\subsection{Fundamental transformations}

In the previous Sections we considered transformations between
multidimensional quadrilateral lattices conjugate to the same
congruence. We studied four particular cases:
\begin{enumerate}
\item both lattices are focal lattices of the congruence
(Laplace transformation),
\item one of the lattices is a focal lattice (L\'evy transformation and its
adjoint),
\item parallel lattices (Combescure transformation),
\item lattices conjugate to a radial congruence (radial transformation).
\end{enumerate}
In this Section we study the most general transformation between
multidimensional quadrilateral lattices conjugate to the same
congruence, which contains the above ones as particular reductions.

\begin{Def} \label{def:fundamental}
Two quadrilateral lattices are related
by the fundamental transformation when they are conjugate to the same
congruence, which is called the congruence of the transformation.
\end{Def}
Consider a generic case, when the congruence of the
transformation
can be constructed via a Combescure transformation vector $\bx_\calC$
of the lattice $\bx$.
Since the same congruence should be constructed also
via a Combescure transformation vector $\cF(\bx)_\calC$ of the lattice
$\cF(\bx)$, we have
\begin{equation}  \label{eq:FxC-xC}
\cF(\bx)_\calC = \frac{1}{\theta} \: \bx_\calC;
\end{equation}
i. e., both vectors are related by a radial transformation, where,
by Proposition \ref{prop:radial}, the function $\theta$ satisfies
the point equation of the lattice $\bx_\calC$.

The transformed lattice $\cF(\bx)$ is therefore necessarily of the
form
\begin{equation}  \label{eq:Fx}
\cF(\bx) = \bx - \phi \: \cF(\bx)_\calC =
\bx - \frac{\phi}{\theta} \bx_\calC,
\end{equation}
where the function $\phi$ is to be determined.

The first derivatives of $\cF(\bx)$ are reducible, due to equations
(\ref{def:X-C}), (\ref{eq:FxC-xC}) and (\ref{eq:Fx}), to the form
\begin{equation}
\D_i\cF(\bx) = \left( \frac{T_i\theta}{T_i\sigma_i} - T_i\phi \right)
\D_i\cF(\bx)_\calC + \left( \frac{\D_i\theta}{T_i\sigma_i} - 
\D_i\phi \right)
\cF(\bx)_\calC.
\end{equation}
From these expressions it follows that $\cF(\bx)_\calC$ is a
Combescure transformation vector of $\cF(\bx)$ if and only if
$\theta$ and $\phi$ satisfy
\begin{equation}
\D_i\theta = (T_i\sigma_i)\D_i\phi.
\end{equation}
The above equations imply that $\phi$ is a solution of the point
equation of the lattice $\bx$, whereas $\theta = \phi_\calC$ is
the Combescure transformed function of $\phi$.

\begin{Prop} \label{prop:fundamental}
i) The fundamental transform $\cF(\bx)$ of the quadrilateral
lattice $\bx$ is given by
\begin{equation} \label{eq:fund-transf}
\cF(\bx) = \bx - \frac{\phi}{\phi_\calC}\bx_\calC,
\end{equation}
where \\ i) $\phi:\ZZ^N \ra \RR$ is a solution of the Laplace
equation (\ref{eq:Laplace}) of the lattice $\bx$, \\ ii)
$\bx_\calC$ is the vector of the Combescure transformation of
$\bx$,  \\ 
iii) $\phi_\calC:\ZZ^N \ra \RR$ is the corresponding
Combescure transformed function of $\phi$.
\end{Prop}

\begin{Cor} \label{cor:fundamental}
In the notation of Theorem \ref{th:vect-Darb},
the fundamental transformation can be written in the form
\begin{equation}\label{eq:x-fund-g}
 \cF(\bx) = \bx - \bOm[\bX, v^*]\frac{\bOm[v,H]}{\bOm[v,v^*]},
\end{equation} 
where $v_i$, and $v_i^*$, $i=1,...,N$, are solutions of the linear
problem (\ref{eq:lin-X}) and its adjoint (\ref{eq:lin-H}). 
The Lam\'e coefficients and the tangent vectors are transformed in 
the following way
\begin{align} 
 \cF(H_i) &= H_i - v_i^* \frac{\bOm[v,H]}{\bOm[v,v^*]} \; , 
 \label{eq:X-fund-g}\\
 \cF(\bX_i) &= \bX_i - \bOm[\bX, v^*]\frac{v_i}{\bOm[v,v^*]} 
 \label{eq:H-fund-g}\; ,
\end{align}
and the corresponding transformation of the fields $Q_{ij}$ reads
\begin{equation} \label{eq:Q-fund-g}
\cF(Q_{ij}) = Q_{ij} - \frac{v_j^* v_i}{\bOm[v,v^*]}.
\end{equation}
\end{Cor}

The geometric meaning of the formula (\ref{eq:fund-transf}) can be
explained as follows. Given an additional scalar solution
$\phi:\ZZ^N \rightarrow \RR$ of the Laplace equation
(\ref{eq:Laplace}), we define, like in the case of the Levy
transformation, a new quadrilateral lattice $\tilde{\bx}:\ZZ^N
\rightarrow \RR^{M+1}$ as
\begin{equation} \tilde\bx: \ZZ^N \rightarrow
\begin{pmatrix}  \bx \\ \phi \end{pmatrix}.
\end{equation}
We construct then a Combescure transform of the lattice $\tilde\bx$; i. e.,
we find the corresponding vector $\tilde\bx_\calC$
\begin{equation} \tilde\bx_\calC =
\bOm \left[ \begin{pmatrix}  \bX \\ v \end{pmatrix}, v^*
\right] = \begin{pmatrix} \bOm[\bX,v^*] \\ \bOm[v,v^*]
\end{pmatrix} = \begin{pmatrix} \bx_\calC \\ \phi_\calC
\end{pmatrix}.
\end{equation}
The point $\cF(\bx)$ is the intersection point of the line $\tilde\bx +
t \tilde\bx_\calC$ with its projection $\bx + t\bx_\calC$ on the $\RR^M$
space, therefore for the intersection parameter $t_0$ we have
\begin{equation}
\begin{pmatrix}  \bx \\ \phi \end{pmatrix} + t_0
\begin{pmatrix}  \bx_\calC \\ \phi_\calC \end{pmatrix}
= \begin{pmatrix}  \cF(\bx) \\ 0 \end{pmatrix}.
\end{equation}

 \begin{center}\leavevmode
 \epsfxsize=5cm
 \epsffile{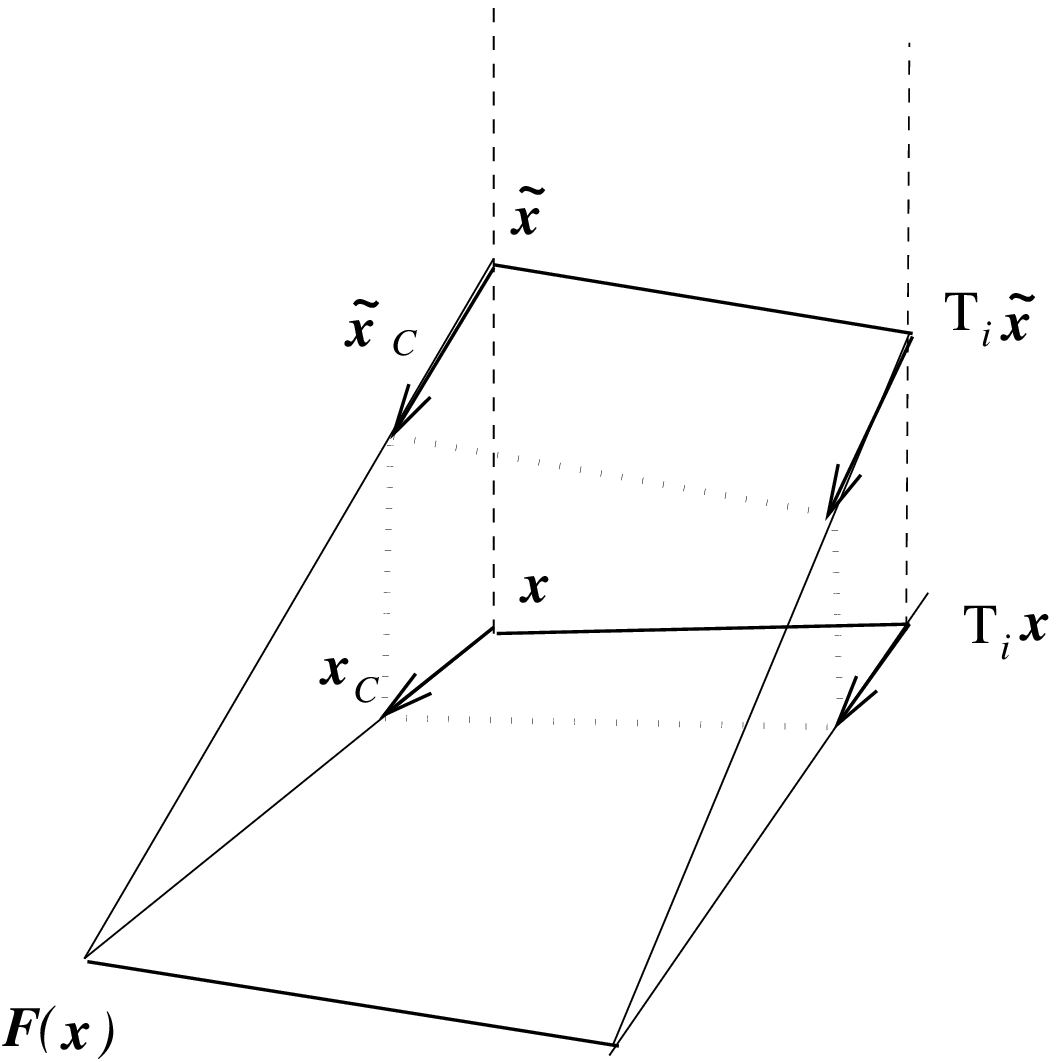}
 \end{center}
Figure 12.

Let us observe that the quadrilaterals with vertices $\bx$, $\tilde\bx$,
$\bx+\bx_\calC$, $\tilde\bx+\tilde\bx_\calC$ are planar.
All the lattices form a
quadrilateral strip with the $N$ dimensional basis and two transversal 
directions
$\cL$ and $\calC$.
The fundamental transformation $\cF$ of the lattice $\bx$
can be interpreted as the Laplace transformation $\cL_{\calC \cL}$ of the
strip, see Fig.~12. Therefore the new lattice $\cF(\bx)$ is quadrilateral.

Given a quadrilateral lattice $\bx$ and its fundamental
transform $\cF(\bx)$ conjugate to the congruence $\gl$, we are
automatically given also $N$ focal lattices $\by_i$ of the congruence.
Obviously, $\by_i$ is the $i$-th adjoint L\'evy transform of both lattices
$\bx$ and $\cF(\bx)$; moreover the lattices $\bx$ and
$\cF(\bx)$ are two different $i$-th L\'evy transforms of $\by_i$. This implies
that the fundamental transformation can be considered as the superposition of
an adjoint L\'evy and a L\'evy transformations.
\begin{Cor}
In order to construct a fundamental transform $\cF(\bx)$ of the
quadrilateral lattice $\bx$ we may proceed in the following way:\\
i) construct a congruence $\gl$ conjugate to $\bx$,
 \\ ii) find the
$i$-th focal lattice $\by_i=\cL^*_i(\bx)$ of the congruence $\gl$,
\\ iii) construct its $i$-th L\'evy transform
\begin{equation}
 \cL_i(\by_i) = \cL_i(\cL^*_i(\bx)) = \cF(\bx).
\end{equation}
\end{Cor}

 \begin{center}
\leavevmode
 \epsfxsize=8cm
 \epsffile{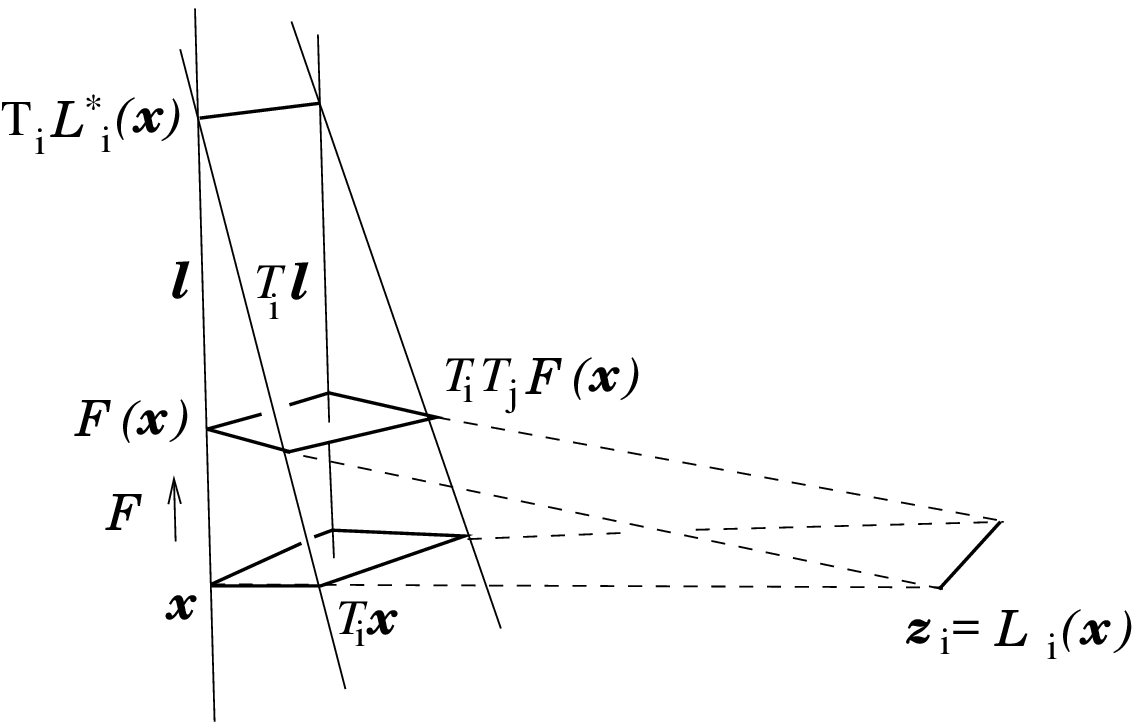}
 \end{center}
Figure 13.

Let us observe also that the transformation $\cF(\bx)$ builds, from the 
lattice
$\bx$, a
quadrilateral strip with basis $\bx$ and  transversal  direction $\cF$.
If we define the
lattice $\bz_i$ as the $\cL_{i\cF}$-th Laplace transform of this strip,
then $\bz_i$ is the $i$-th L\'evy transform of both lattices
$\bx$ and $\cF(\bx)$, while the lattices $\bx$ and
$\cF(\bx)$ are different $i$-th adjoint L\'evy transforms of $\by_i$.
This observation,
together with Proposition \ref{prop:2adLevy-conj}, provides a third way to
construct the fundamental transform $\cF(\bx)$.

\begin{Cor}
In order to construct a fundamental transform $\cF(\bx)$ of the
quadrilateral lattice $\bx$, we may proceed in the following way:\\
i) we find the $i$-th L\'evy transform $\bz_i=\cL_i(\bx)$ of $\bx$; \\
ii) we construct a congruence conjugate to $\bz_i$; \\ 
iii) we find the
$i$-th focal lattice of the congruence
\begin{equation}
\cL^*_i(\bz_i) = \cL_i^*(\cL_i(\bx)) = \cF(\bx).
\end{equation}
\end{Cor}
\begin{Rem}
The fundamental transformation, superposition
of L\'evy and
adjoint L\'evy
transformations, is usually called,
in the soliton theory, {\em binary
Darboux transformation} \cite{ms,OevelSchief,KoSchief}.
\end{Rem}

We end this Section remarking that, from the previous observations, it is 
possible 
to interpret the transformation $\bx \to \cF(\bx)$ as a generic addition of a new 
dimension (the $(N+1)$-st) to the original lattice $\bx$. We will discuss this 
interesting aspect of the fundamental transformations in 
Section~\ref{sec:vect-fund}. 

\subsection{Superposition of fundamental transformations}
\label{sec:superpositions-fund}

In this Section we consider vectorial fundamental transformations,
which are nothing else but superpositions of the fundamental
transformations. Generalizing the procedure of the previous Section,
we consider $K\geq 1$ solutions $\phi^k$, $k=1,...,K$ of the Laplace 
equation of the lattice $\bx$, which we arrange in the $K$ component vector 
$\bphi = (\phi^1 , ... , \phi^K)^t$; this allows to
introduce the quadrilateral lattice 
$\tilde{\bx} = \left( \begin{array}{c} \bx \\ \bphi \end{array} \right)$ 
in the space $\RR^{M+K}$. We also consider $K$
Combescure transformation vectors $\bx_{\calC,k}$; also 
the Combescure transformation vectors
$\bx_{\calC,k}$ can be extended (this procedure involves $K$ arbitrary 
constants) to the Combescure transformations
vectors $\tilde{\bx}_{\calC,k} = \left( \begin{array}{c}
\bx_{\calC,k} \\
\bphi_{\calC,k} \end{array} \right)$ of the lattice $\tilde{\bx}$,
where the $K$ component vector 
$\bphi_{\calC,k} = (\phi^1_{\calC,k}, ..., \phi^K_{\calC,k})^t$ consists of the
Combescure transformed functions $\phi^l_{\calC,k}$
of $\phi^l$; each of the vectors $\tilde\bx_{\calC,k}$ defines
a Combescure transform of the lattice $\tilde\bx$. The $K$
vectors $\tilde\bx_{\calC,k}$ define the $K$-dimensional subspace
\begin{equation}
\tilde\bx + \sum_{k=1}^{K} \tilde\bx_{\calC,k} t^k =
\tilde\bx + (\tilde\bx_{\calC,1}, ... ,\tilde\bx_{\calC,N})
\begin{pmatrix} t^1 \\ \vdots \\ t^K \end{pmatrix} =
\tilde\bx + \tilde\bx_{\calC} \bt = \begin{pmatrix}
\bx \\ \bphi \end{pmatrix} +
\begin{pmatrix}
 \bx_\calC \\ \bphi_\calC \end{pmatrix}
\bt.
\end{equation}
The intersection point of this subspace with $\RR^M$ (in general, a 
$K$-dimensional and an $M$-dimensional subspaces of the $(M+K)$-dimensional  
space 
intersect in a single point) defines the new lattice $\cT(\bx)$
\begin{equation}
\begin{pmatrix}
 \cT(\bx) \\ {\boldsymbol 0} \end{pmatrix} =
\begin{pmatrix} \bx \\ \bphi \end{pmatrix} +
\begin{pmatrix} \bx_\calC \\ \bphi_\calC \end{pmatrix} \bt_0.
\end{equation}
The corresponding values of the parameters $t^k_0$ can be found
from the lower part of the above equation
\begin{equation}
 {\boldsymbol 0} = \bphi + \bphi_\calC \bt_0,
\end{equation}
and then inserted into the upper part, giving
\begin{equation}
\cT(\bx) = \bx - \bx_\calC \bphi_\calC^{-1} \bphi.
\end{equation}
In the notation of Theorem \ref{th:vect-Darb} we have
\begin{equation*} \bphi  = \bOm[\bv,H], \quad
\bphi_\calC = \bOm[\bv , \bv^*], \quad
\bx_\calC = \bOm[\bX ,\bv^* ]
\end{equation*}
and
\begin{equation} \label{eq:x-sup}
\cT(\bx) = \bOm[\bX ,H ] - \bOm[\bX ,\bv^* ] \bOm[\bv , \bv^*]^{-1}
\bOm[\bv,H].
\end{equation}
One can prove that the new lattice $\cT(\bx)$ is also a
quadrilateral one. This is a consequence of Theorem
\ref{th:vect-Darb} and the proof can be found in Section
\ref{sec:vect-fund}. In that Section it will also be shown that the
vectorial fundamental transformation is the superposition of $K$
fundamental transformations.

In this Section we consider only the simplest case $K=2$,
emphasizing the geometric meaning of all the steps involved in the
construction.

\begin{Prop} \label{prop:fund-sup-g}
i) The two component vectorial fundamental transformation is
equivalent to the superposition of two fundamental
transformations:\\ 1) the transformation $\cF_1$ of the lattice $\bx$, with
parameters $\phi^1$ and $\bx_{\calC,1}$:
\begin{equation}
\cF_1(\bx) = \bx - \frac{\phi^1}{\phi_{\calC,1}^1}\bx_{\calC,1},
\end{equation}
2) the transformation $\cF_2$ of the lattice $\cF_1(\bx)$ with
parameters $\phi^{2\prime}$, $\bx_{\calC,2}^\prime$:
\begin{equation}
\cT(\bx) = \cF_2 (\cF_1 (\bx)) = \cF_1(\bx) -
\frac{\phi^{2\prime}}{\phi^{2\prime}_{\calC,2}}
\bx_{\calC,2}^\prime,
\end{equation}
where $\phi^{2\prime}$, $\bx_{\calC,2}^\prime$ are
nothing but the parameters $\phi^2$ and $\bx_{\calC,2}$
transformed by the first transformation
\begin{align*}
\phi^{2\prime} &= \cF_1(\phi^2) = \phi^2 -
\frac{\phi^1}{\phi_{\calC,1}^1}\phi^2_{\calC,1},\\
\bx_{\calC,2}^\prime &= \cF_1(\bx_{\calC,2}) = \bx_{\calC,2} -
\frac{\phi^1_{\calC,2}}{\phi_{\calC,1}^1}\bx_{\calC,1},
\end{align*}
and, correspondingly,
\begin{equation}
 \phi^{2\prime}_{\calC,2} =
\cF_1(\phi^2_{\calC,2}) = \phi^2_{\calC,2}
- \frac{\phi^1_{\calC,2}}{\phi_{\calC,1}^1}\phi^2_{\calC,1}.
\end{equation}
ii) The result of the superposition of $\cF_1$ and $\cF_2$ is independent of 
the order.
\end{Prop}
\begin{Proof}
The proof is by direct calculation; we only remark that, by
construction, $\phi^{2\prime}$ is a solution of the Laplace
equation of the lattice $\cF_1(\bx)$, and $\bx_{\calC,2}^\prime$ is
a vector of the Combescure transformation of the same lattice.

One can look at the above superposition of the fundamental
transformations as follows:\\ a) the fundamental transformation of the
lattice $\Big(\begin{smallmatrix}
\bx \\ 0 \\ \phi^2 \end{smallmatrix}\Big) $ using the solution
$\phi^1$ of the Laplace equation and the Combescure
transformation vector $\Big(\begin{smallmatrix}\bx_{\calC,1} \\ 0
\\ \phi^2_{\calC,1} \end{smallmatrix}\Big)$, which gives
$\Big(\begin{smallmatrix}
\cF_1(\bx) \\ 0 \\ \phi^{2\prime} \end{smallmatrix}\Big)$;\\
b) the simultaneous transformation of the Combescure vector
$\Big(\begin{smallmatrix} \bx_{\calC,2} \\ 0 \\
\phi^2_{\calC,2}
\end{smallmatrix} \Big)$,
which gives the Combescure transformation vector
$\Big(\begin{smallmatrix}
\bx_{\calC,2}^\prime \\ 0
\\ \phi^{2\prime}_{\calC,2} \end{smallmatrix} \Big)$ of the lattice obtained in 
point a);\\
c) the combination of the lattice in $\RR^{M+1}$ constructed in point a) with
the Combescure transformation vector constructed in point b) gives
the lattice $\cT(\bx)$ in $\RR^M$.
\end{Proof}
\begin{Cor} \label{cor:plan-fund}
The points $\bx$, $\cF_1(\bx)$, $\cF_2(\bx)$ and $\cT(\bx) = \cF_1(\cF_2(\bx)) 
=\cF_2(\cF_1(\bx))$ are coplanar. 
\end{Cor}

\section{Are the fundamental transformations really fundamental?}

\label{sec:limits}
The main goal of this Section is to show explicitly that all the
transformations discussed in the previous Sections are special
cases of the fundamental transformations. Since focal lattices can be viewed as 
limiting cases of generic lattices conjugate to the congruence, this statement 
is rather obvious, from a geometrical point of view. Nevertheless, due to the 
fact that the Combescure transformation vector $\bx_\calC$ is not suited well 
to describe tangent congruences, the consequent subtleties associated with 
the analytic limits require a detailed study.

\subsection{Reduction to the Combescure and radial transformations}
We first illustrate the straightforward reduction from the fundamental 
transformations to the Combescure and radial transformations.

To obtain the Combescure transformation from the fundamental one we
put $v_i=0$, $i=1,...,N$, in the Corollary \ref{prop:fundamental}.
This implies that both $\phi$ and $\phi_\calC$  are
constants. The constant $\frac{\phi}{\phi_\calC}$ can
always be absorbed by the corresponding rescaling of $v_i^*$.

In looking for the reduction of the fundamental transformation to
the radial one, we may notice that, in the radial transformation, the
Combescure vector $\bx_\calC$ of the congruence must be
proportional to the lattice vector $\bx$. This gives $v_i^* = H_i$,
$\bx_\calC = \bx$ and, therefore, $\phi_\calC$ is a solution of
the Laplace equation of the points of the lattice $\bx$. This
implies that $\phi_\calC - \phi$ must be a constant $c$:
\begin{equation}
 \cF(\bx) \ra \frac{c}{\phi+c}\bx,
\end{equation}
and this formula is obviously is equivalent to formula (\ref{eq:radial}).

\subsection{Singular limit to the adjoint L\'evy transformation}
\label{sec:limit-adj-Levy}

From Sections \ref{sec:adj-Levy} and \ref{sec:fundamental} it
follows that the adjoint L\'evy transformation $\cL_i^*(\bx)$ can be
viewed as the limiting case of the fundamental transformation
$\cF(\bx)$ in which the transformed lattice becomes the $i$-th
focal lattice of the associated congruence.

As it was shown in Section \ref{sec:fundamental},
the construction of $\cF(\bx)$ is the following sequence
of three geometric processes: \\
i) the extension of the lattice $\bx\subset\RR^M$ to the lattice
$\left( \begin{array}{c} \bx \\ \phi \end{array}
\right)\subset \RR^{M+1}$; \\
ii) the Combescure transformation
\[ \calC\left( \begin{array}{c} \bx \\ \phi \end{array} \right)    =
\left( \begin{array}{c} \bx + \bx_\calC \\ \phi + \phi_\calC
\end{array} \right)  ,
\]
which gives the quadrilateral strip with $N$-dimensional basis $\bx$ and two
transversal directions, called $\cL$
and $\calC$;\\
iii) the Laplace transformation $\cL_{\calC \cL}$ of the strip.

\begin{center}
\leavevmode
\epsfxsize=7cm
\epsffile{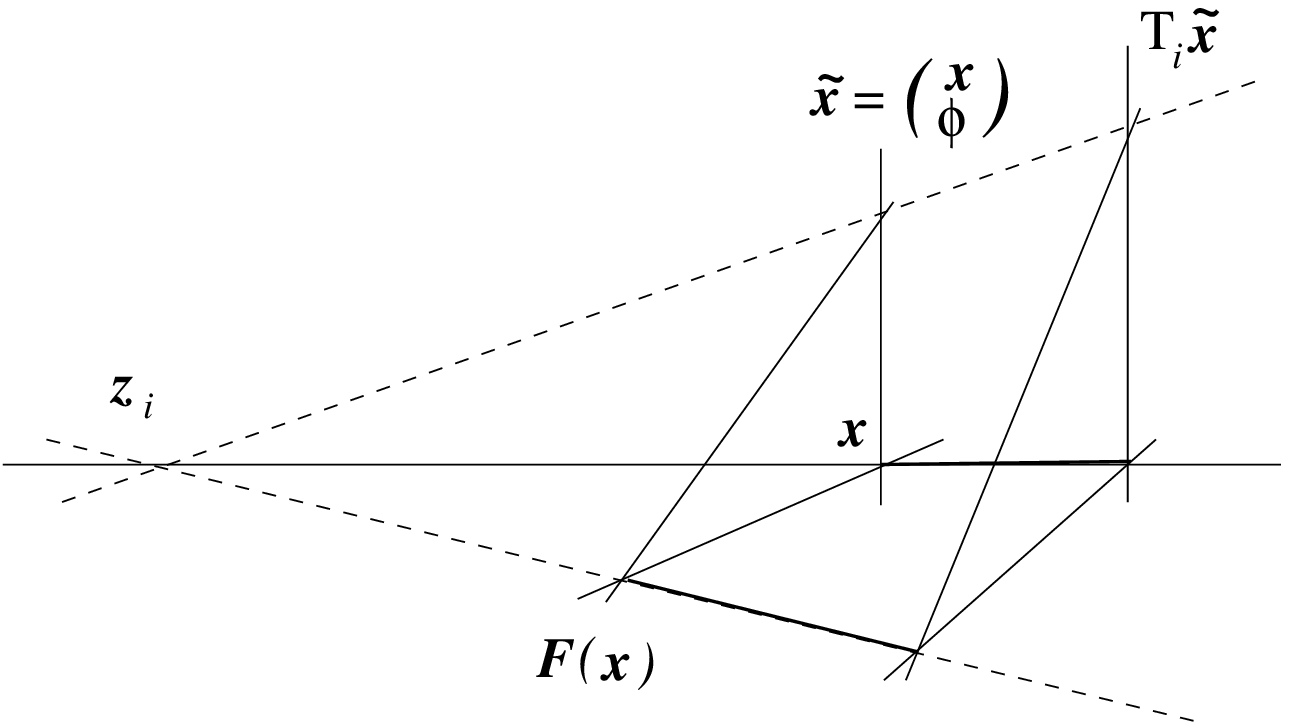}
\end{center}
Figure 14.

In order to investigate the nature of the limit $\cF(\bx) \ra
\cL_i^*(\bx)$, it is convenient to study the properties of
$\phi$ when $\bx$ and $\cF(\bx)$ are given. If $\phi$ is
given in the initial point, then $T_i\phi$ is obtained from the
intersection point $\left(\begin{array}{c} T_i\bx \\ T_i\phi
\end{array} \right)$ of the line passing through
$\left(\begin{array}{c} T_i\bx \\ 0 \end{array} \right)$ in the
$(M+1)$-th direction with the line passing through the points
$\left(\begin{array}{c} \bz_i \\ 0 \end{array} \right)$      and
$\left(\begin{array}{c} \bx \\ \phi \end{array} \right)$, where
$\bz_i\in\RR^M$ was defined in Section \ref{sec:fundamental} as the
intersection of the $i$-th tangent line of the lattice $\bx$ with
the corresponding tangent line of $\cF(\bx)$.

By construction, the vector $\bx - \bz_i$  is proportional to $\D_i\bx$:
\begin{equation}
\bx-\bz_i = \nu \D_i\bx, \; \; \nu\in \RR;
\end{equation}
consequently
\begin{equation}
T_i\phi = \frac{1+\nu}{\nu}\phi.
\end{equation}
In the limit in which $T_i\cF(\bx) \ra T_i\cL_i^*(\bx)$ we have also
$\bz_i \ra \bx$ and $\nu \ra 0$. Therefore
\begin{equation} \label{eq:nu-limit}
T_i\phi \simeq \nu^{-1}\phi, \; \; |\nu | \ll 1.
\end{equation}
We remark that, in formula (\ref{eq:nu-limit}), the lattice
function $\nu$  in the uniform limit $\cF(\bx)\ra
\cL_i^*(\bx)$ is of order $\eps$, $|\eps| \ll 1$. This suggests the
following ansatz for the asymptotics of $\phi$:
\begin{equation} 
\phi = \eps^{-n_i}\alpha(1 + O(\eps) );
\label{eq:lim-ansatz} 
\end{equation}
substituting (\ref{eq:lim-ansatz}) into the Laplace equations
(\ref{eq:Laplace-H}) we obtain
\begin{equation*}
 \D_j\alpha = \frac{\D_jH_i}{H_i}\alpha \; ,  \quad
\D_j\D_k\alpha = \frac{\D_jH_k}{H_k}\D_j\alpha + \frac{\D_kH_j}{H_j}
\D_k\alpha \; , \quad i\ne j \ne k \ne i \; ,
\end{equation*}
which imply that $\alpha = H_i$. From similar considerations we also obtain 
that
\begin{equation}
\phi_\calC = \eps^{-n_i}v_i^*(1 + O(\eps)) ;
\end{equation}
for completeness we write down also the asymptotics of $v_i$:
\begin{align*}
 v_i &= \eps^{-n_i}(\eps^{-1} + Q_{ii} + O(\eps)), \\
v_j &= \eps^{-n_i}Q_{ji}(1 + O(\eps)),
\end{align*}
where
\begin{equation*}
\D_jQ_{ii} = (T_jQ_{ij}) Q_{ji}.
\end{equation*}
Therefore, in the limit $\eps\ra 0$, the asymptotics of the lattice points, 
the Lam\'e coefficients and the tangent vectors read
\begin{align*}
\cF(\bx) &= \bx - \frac{H_i}{v_i^*}\bOm[\bX,v^*] + O(\eps) = \cL_i^*(\bx)+ 
O(\eps) \; ,\\
\cF(\bX_i) &= -\eps^{-1}\frac{\bOm[\bX, v^*]}{v_i^*} + O(1), \\
\cF(\bX_j) &= \bX_j -\frac{\bOm[\bX, v^*]}{v_i^*}Q_{ji} + O(\eps), \\
\cF(H_i) &= \eps T_i^{-1}\left( v_i^* \D_i \left( \frac{H_i}{v_i^*}
\right) \right)  + O(\eps^2), \\
\cF(H_j) &= H_j - \frac{v_j^*}{v_i^*}H_i + O(\eps) \; 
\end{align*}
and agree (up to possible $\eps$-scalings) 
with the formulas of Section \ref{sec:adj-Levy}.

\subsection{Singular limit to the L\'evy transformation}
\label{sec:limit-Levy}
In the limit when the fundamental transformation $\cF(\bx)$ reduces to the 
L\'evy
transformation $\cL_i(\bx)$, the congruence of the transformation becomes the
$i$-th tangent congruence of the lattice $\bx$; i. e., $\bx_\calC$ becomes
proportional to $\D_i\bx$.

On the other hand, iterating equation (\ref{def:X-C}), we obtain the
formal series
\begin{equation}
\bx_\calC = -(T_i\sigma_i)\D_i\bx - (T_i^2\sigma_i)T_i\D_i\bx  - ...
\end{equation}
which, in the above limit, becomes asymptotic in some small parameter $\eps$.
This suggests the following ansatz:
\begin{equation}
 \sigma_i(\bn) \sim \eps^{n_i -1}\beta(\bn)(1 + O(\eps)),
\end{equation}
which gives
\begin{equation}
 \bx_\calC \sim -\eps^{n_i}(T_i\beta)\D_i\bx = - \eps^{n_i}T_i(\beta H_i)
\bX_i.  \label{eq:x-c-lim2}
\end{equation}
Applying the difference operator $\D_j$ to equation (\ref{eq:x-c-lim2})
and using equations (\ref{def:X-C}) and (\ref{eq:MQL-Q}) we infer that
\begin{equation}
\beta = \frac{1}{H_i} \; , \quad \sigma_j \sim
\eps^{n_i}\frac{Q_{ij}}{H_j}\; ;
\end{equation}
i. e.,
\begin{align*}
\bx_\calC &= -\eps^{n_i}(\bX_i + O(\eps)), \\
\sigma_i &=  \eps^{n_i-1}\left( \frac{1}{H_i} + O(\eps) \right), \\
\sigma_j &=  - \eps^{n_i}\left( \frac{Q_{ij}}{H_j} + O(\eps) \right)\; ,
\end{align*}
which allow to calculate the asymptotics of the other relevant objects:
\begin{align*}
v_i^* &=  \eps^{n_i -1}( 1 - \eps Q_{ii} + O(\eps)) \\ v_j^* &=
-\eps^{n_i}(Q_{ij} + O(\eps)), \\
\phi_\calC &= -\eps^{n_i}(v_i + O(\eps)) \; . 
\end{align*}
Therefore, in the limit $\eps\ra 0$, the asymptotics of the lattice points, 
of the Lam\'e coefficients and of the tangent vectors read
\begin{align*}
\cF(\bx) &= \bx - \frac{\phi}{v_i}\bX_i + O(\eps) =\cL_i(\bx) + O(\eps) \; , \\
\cF(\bX_i) &= -\eps \left( \D_i \bX_i - \frac{\D_i v_i}{v_i} \bX_i  \right) +
 O(\eps^2), \\
\cF(\bX_j) & =  \bX_j - \frac{v_j}{v_i}\bX_i + O(\eps), \\
\cF(H_i) & = \frac{1}{\eps}\frac{\phi}{v_i} + O(1), \\
\cF(H_j) &= H_j - \frac{Q_{ij}}{v_i}\phi + O(\eps) \; ,
\end{align*}
and agree with the formulas of Section \ref{sec:Levy}.

\subsection{Singular limit to the Laplace transformations}
\label{sec:limit-Laplace}

The Laplace transformation can be considered as the special limit
of the fundamental transformation such that both lattices
are focal lattices of the congruence of the transformation.
Therefore it can be obtained combining the asymptotics presented in the 
previous Sections~\ref{sec:limit-adj-Levy} and \ref{sec:limit-Levy}. The 
corresponding asymptotics read as follows:
\begin{align*}
\cF(\bx) &= \bx -\frac{H_j}{Q_{ij}}\bX_i + O(\eps) = \cL_{ij}(\bx) + 
O(\eps)\; , \\
\cF(H_i) &= -\frac{1}{\eps} \frac{H_j}{Q_{ij}} + O(1) \; , \\
\cF(H_j) &=  \eps T_j^{-1} \left( Q_{ij} \D_j \left(
\frac{H_j}{Q_{ij}} \right) \right)
+ O(\eps^2) \; , \\
\cF(H_k) &= H_k - \frac{Q_{ik}}{Q_{ij}} H_j + O(\eps) \; , \\ 
\cF(\bX_i) & = \eps \left( \D_i\bX_i - \frac{\D_iQ_{ij}}{Q_{ij}} \bX_i \right)
+ O(\eps^2) \; , \\
\cF(\bX_j) & = -\frac{1}{\eps} \frac{\bX_i}{Q_{ij}}  + O(1)  \; , \\
\cF(\bX_k) & = \bX_k - \frac{Q_{kj}}{Q_{ij}} \bX_i +O(\eps)  \; .
\end{align*}

\section{Connection with vectorial Darboux transformations and permutability 
theorems}
\label{sec:vect-fund}

\subsection{Fundamental transformations from the vectorial formalism}

The main goal of this Section is to show that the fundamental transformations 
and, therefore, all the particular
transformations discussed in the previous Sections, are special
cases of the vectorial transformation described in Theorem
\ref{th:vect-Darb} and introduced in \cite{MDS}.

Consider the following splitting of the vector space $\WW$ of
Theorem \ref{th:vect-Darb}:
\begin{equation} \WW = \EE\oplus\VV\oplus\FF  \;, \; \; \;
 \WW^* = \EE^*\oplus\VV^*\oplus\FF^* \; \; \; \; ;
\end{equation}
if
\begin{equation}
 \bY_i = (\bX_i, \bv_i, 0 )^T, \; \; \;
\bY_i^* = ( 0 , \bv_i^*, \bX^* ) \; ,
\end{equation}
then, the corresponding potential matrix is of the form
\begin{equation}
\bOm[\bY,\bY^*] = \begin{pmatrix}
\II_\EE & \bOm[\bX,\bv^*] & \bOm[\bX, \bX^*] \\
0       & \bOm[\bv,\bv^*] & \bOm[\bv, \bX^*] \\
0       &      0          &   \II_\FF
\end{pmatrix}
\end{equation}
and its inverse is:
\begin{equation}
\bOm[\bY,\bY^*]^{-1} =
\begin{pmatrix}
\II_\EE & -\bOm[\bX,\bv^*]  \bOm[\bv,\bv^*]^{-1} &
-\bOm[\bX, \bX^*] +\bOm[\bX,\bv^*]\bOm[\bv,\bv^*]^{-1}\bOm[\bv, \bX^*]  \\
0       & \bOm[\bv,\bv^*]^{-1} & -  \bOm[\bv,\bv^*]^{-1}\bOm[\bv, \bX^*] \\
0       &      0          &   \II_\FF
\end{pmatrix}.
\end{equation}
This implies that
\begin{align*}
\hat{\bY}_i & = \begin{pmatrix}
\hat{\bX}_i \\ \hat{\bv}_i \\ 0
\end{pmatrix} =
\begin{pmatrix}
\bX_i -\bOm[\bX,\bv^*]  \bOm[\bv,\bv^*]^{-1}\bv_i \\
\bOm[\bv,\bv^*]^{-1}\bv_i \\ 0
\end{pmatrix},\\
\hat{\bY}_i^* &= ( 0, \hat{\bv}_i^*, \hat{\bX}_i^* )
= ( \; 0 , \;   \bv^*_i \bOm[\bv,\bv^*]^{-1}, \;
\bX_i^* - \bv_i^* \bOm[\bv,\bv^*]^{-1}\bOm[\bv, \bX^*]  \; )
\end{align*}
and
\begin{equation} \label{eq:Q-fund}
\hat{Q}_{ij} = Q_{ij} - \bv_j^* \bOm[\bv,\bv^*]^{-1}\bv_i.
\end{equation}
Theorem \ref{th:vect-Darb} implies, in particular, that,
up to a constant operator,
\begin{equation} \label{eq:x-fund}
\bOm[\hat{\bX},\hat{\bX}^*] =
\bOm[\bX, \bX^*] - \bOm[\bX,\bv^*]\bOm[\bv,\bv^*]^{-1}\bOm[\bv, \bX^*].
\end{equation}
The fundamental transformation can be obtained in the simplest
case, by putting $\FF = \VV =\RR$, $\EE = \RR^M$, $\bw = (0,0,1)^T$
and choosing the projection operator on the space $\EE$ along
$\VV\oplus\FF$. Then $\bX_i^* = H_i$, the scaled tangent vectors
are just $\bX_i$ and $\bx=\bOm[\bX,H]$; the transformation
data $\bv_i$ and $\bv_i^*$ are scalar functions. The transformed
lattice points and the transformed functions $Q_{ij}$ are given then by 
formulas
(\ref{eq:x-fund}) and (\ref{eq:Q-fund}), which coincide with
(\ref{eq:x-fund-g}) and (\ref{eq:Q-fund-g}).

We recall that, in Section~\ref{sec:superpositions-fund},  
the geometric meaning of equation (\ref{eq:x-fund}) was given
in the case in which
$\FF = \RR$,  $\VV =\RR^K$, $\EE = \RR^M$.

\subsection{Permutability of the fundamental transformations}
\label{sec:permutability}

Let assume that the transformation datum space split as
$\VV=\VV_1\oplus\VV_2$, so that we write
$\bOm[\bv,\bv^*]=
\begin{pmatrix} m_{11}&m_{12}\\m_{21}&m_{22}
\end{pmatrix}$ with $m_{ij}=\bOm[\bv_{(i)},\bv_{(j)}^*]:
\VV_j\ra\VV_i$. 
Correspondingly, we have the following decompositions
\begin{equation} \label{eq:split}
\begin{aligned}
\bv_i&=\begin{pmatrix}\bv_{(1),i}\\ \bv_{(2),i}\end{pmatrix},\\
\bv_i^*&=(\bv_{(1),i}^*, \bv_{(2),i}^*),\\
\bOm[\bX,\bv^*]&=(M_{(1)},M_{(2)}),\quad M_{(i)}=\bOm[\bX,\bv_{(i)}^*],\\
\bOm[\bv,\bX^*]&=\begin{pmatrix} M_{(1)}^*\\ M_{(2)}^*\end{pmatrix},
\quad \quad\;\;M_{(i)}^*=\bOm[\bv_{(i)},\bX^*]
\end{aligned}
\end{equation}
If $m_{22}\in\hbox{\rm GL}(\VV_2)$, we have the factorizations
\begin{equation} \label{eq:fact}
\begin{aligned}
\bOm[\bv,\bv^*]&=
\begin{pmatrix}1& m_{12}m_{22}^{-1}\\0&1\end{pmatrix}
\begin{pmatrix} m_{11}-m_{12}m_{22}^{-1}m_{21}&0\\
m_{21}&m_{22}\end{pmatrix} \\ &=\begin{pmatrix}
m_{11}-m_{12}m_{22}^{-1}m_{21}& m_{12}\\0 &m_{22}
\end{pmatrix}\begin{pmatrix}1&0\\ m_{22}^{-1}m_{21}&1
\end{pmatrix}.
\end{aligned}
\end{equation}
Using the formulae (\ref{eq:split}) and (\ref{eq:fact}),
together with (\ref{eq:x-fund}), we obtain
\begin{equation} \label{eq:darboux2}
\begin{aligned}
\hat{Q}_{ij}=&Q_{ij}-\langle\bv_{(2),j}^*,m_{22}^{-1}\bv_{(2),i}
\rangle\\
&-\langle\bv_{(1),j}^*-\bv_{(2),j}^*m_{22}^{-1}m_{21},
(m_{11}-m_{12}m_{22}^{-1}m_{21})^{-1}
(\bv_{(1),j}-m_{12}m_{22}^{-1}\bv_{(2),j})\rangle, \\
\hat{\bX}_i=&\bX_i-M_{(2)}m_{22}^{-1}\bv_{(2),i}\\
&-(M_{(1)}-M_{(2)}m_{22}^{-1}m_{21})
(m_{11}-m_{12}m_{22}^{-1}m_{21})^{-1}
(\bv_{(1),i}-m_{12}m_{22}^{-1}\bv_{(2),i}),  \\
\hat{\bX}_i^*=&\bX_i^*-\bv_{(2),i}^*m_{22}^{-1}M_{(2)}^*\\
&-(\bv_{(1),i}-m_{12}m_{22}^{-1}\bv_{(2),i})
(m_{11}-m_{12}m_{22}^{-1}m_{21})^{-1}
(M_{(1)}^*-m_{22}^{-1}m_{21}M_{(2)}^*).
\end{aligned}
\end{equation}

As we shall see, these formulae coincide with those coming from performing
first a fundamental transformation  with the transformation data $(\VV_2, 
\bv_{(2)}, \bv^*_{(2)})$: 
\begin{align*}
Q_{ij}^\prime&=Q_{ij}-\langle\bv_{(2),j}^*,m_{22}^{-1}\bv_{(2),i}\rangle,\\
{\bX}_i^\prime&=\bX_i-M_{(2)}m_{22}^{-1}\bv_{(2),i},\\
({\bX}_i^*)^\prime&=\bX_i^*-\bv_{(2),i}^*m_{22}^{-1}M_{(2)}^*,
\end{align*}
and then transforming with the data $(\VV_1,\bv_{(1)}^\prime,
\bv^*_{(1)})^\prime)$, where $\bv_{(1)}^\prime,\bv^{*\prime}_{(1)}$
are the data $\bv_{(1)},\bv^*_{(1)}$ after the first fundamental transform 
indicated by $^\prime$.
Therefore the resulting functions are:
\begin{align*}
Q_{ij}^{\prime\prime}&=Q_{ij}^\prime-
\langle(\bv_{(1),j}^*)^\prime,M(\bv^\prime,(\bv^*)^\prime)^{-1}
\bv_{(1),i}^\prime\rangle,\\
\bX_i^{\prime\prime}&=\bX_i^\prime-M(\bX^\prime,(\bv^*)^\prime)M(\bv^\prime,
(\bv^*)^\prime)^{-1}\bv^\prime,\\
(\bX_i^*)^{\prime\prime}&=(\bX_i^*)^\prime-(\bv_{(1),i}^*)^\prime
M(\bv^\prime,(\bv^*)^\prime)^{-1}M(\bv^\prime,(\bX^*)^\prime).
\end{align*}
To show this, it is important to use the relations (\ref{eq:x-fund}) to realize 
that
\begin{align*}
\bOm(\bX^\prime,(\bv^*)^\prime)&=M_{(1)}-M_{(2)}m_{22}^{-1}m_{21},\\
\bOm(\bv^\prime,(\bX^*)^\prime)&=M_{(1)}^*-m_{(12)}m_{22}^{-1}M_{(2)}^*,\\
\bOm(\bv^\prime,(\bv^*)^\prime)&=m_{11}-m_{12}m_{22}^{-1}m_{21},
\end{align*}
so that the above equations for the second fundamental transformation
are just (\ref{eq:darboux2}):
\begin{align*}
Q_{ij}^{\prime\prime}&=\hat{Q}_{ij},\\
\bX_i^{\prime\prime}&=\hat{\bX}_i,\\
(\bX_i^*)^{\prime\prime}&=\hat{\bX}_i^*.
\end{align*}
\begin{Prop} \label{prop:permutability}
The vectorial Darboux transformation (\ref{eq:darboux2}) 
with the transformation data $(\VV_1 \oplus \VV_2, 
\Big(\begin{smallmatrix} \bv_{(1)}\\
\bv_{(2)}\end{smallmatrix}\Big)$, $(\bv_{(1)}^*,\bv_{(2)}^*))$
coincides with the following composition of fundamental
transformations:
\begin{enumerate}
\item First transform with data $(\VV_2,\bv_{(2)},\bv_{(2)}^*)$,
 and denote the transformation by $^\prime$.
\item On the result of this transformation apply a second one with data
$(\VV_1,\bv_{(1)}^\prime,(\bv_{(1)}^*)^\prime)$.
\end{enumerate}
\end{Prop}
\begin{Cor}\label{cor:sup-1}
Assuming that $m_{11}\in\hbox{\rm GL}(\VV_1)$ and following the above steps, 
it is easy to show that this composition does not depend on the order
of the two transformations.
\end{Cor}
\begin{Cor}\label{cor:sup-2}
Applying the mathematical induction to Proposition \ref{prop:permutability},
it is possible to show that, assuming a general splitting $\VV = 
\oplus_{i=1}^K
\VV_i$ of the transformation space, the final result does not depend
on the order in which the $K$ transformations are made.
\end{Cor}

\subsection{Fundamental transformations as integrable discretization}

In Section~\ref{sec:superpositions-fund} we have observed that the fundamental 
transformation $\cF$ can be interpreted as generating a new dimension 
(the $(N+1)$-st) of the lattice $\bx$; more precisely, a single fundamental 
transformation can be interpreted as an elementary translation in this new 
dimension. Moreover the Combescure vector $\bx_\calC$ of the transformation 
can be viewed as the corresponding normalized tangent vector $\bX_{N+1}$. 
Obviously, in order to have an $(N+1)$ dimensional quadrilateral lattice, we 
have to apply recursively fundamental transformations.

The application of two fundamental transformations $\cF_1$ and $\cF_2$ to the 
quadrilateral lattice $\bx$ can be viewed as one step in the generation of 
two new 
dimensions; the permutability theorem (Proposition~\ref{prop:fund-sup-g}) 
guarantees that these translations commute. Moreover, the elementary 
quadrilateral
\begin{equation} \label{eq:quad-sup}
\{ \bx , \cF_1(\bx), \cF_2(\bx) , \cF_1(\cF_2(\bx)) = \cF_2(\cF_1(\bx)) \}
\end{equation}
is planar (see Corollary~\ref{cor:plan-fund}), which makes the theory 
self-consistent.

The statements about the permutability of the fundamental transformations 
$\cF_1$ and $\cF_2$, and about the planarity of the elementary quadrilateral
(\ref{eq:quad-sup}) are also valid  in the limiting case in which $\bx$ 
represents a submanifold parametrized by conjugate coordinates (see Fig.~15).
Actually this last result, that was known to Jonas~\cite{Jonas} and
Eisenhart~\cite{Eisenhart-TS}, is 
very significant in the light of our modern knowledges concerning integrable 
discretizations of integrable PDE's.

\begin{center}
\leavevmode
\epsfxsize=7cm
\epsffile{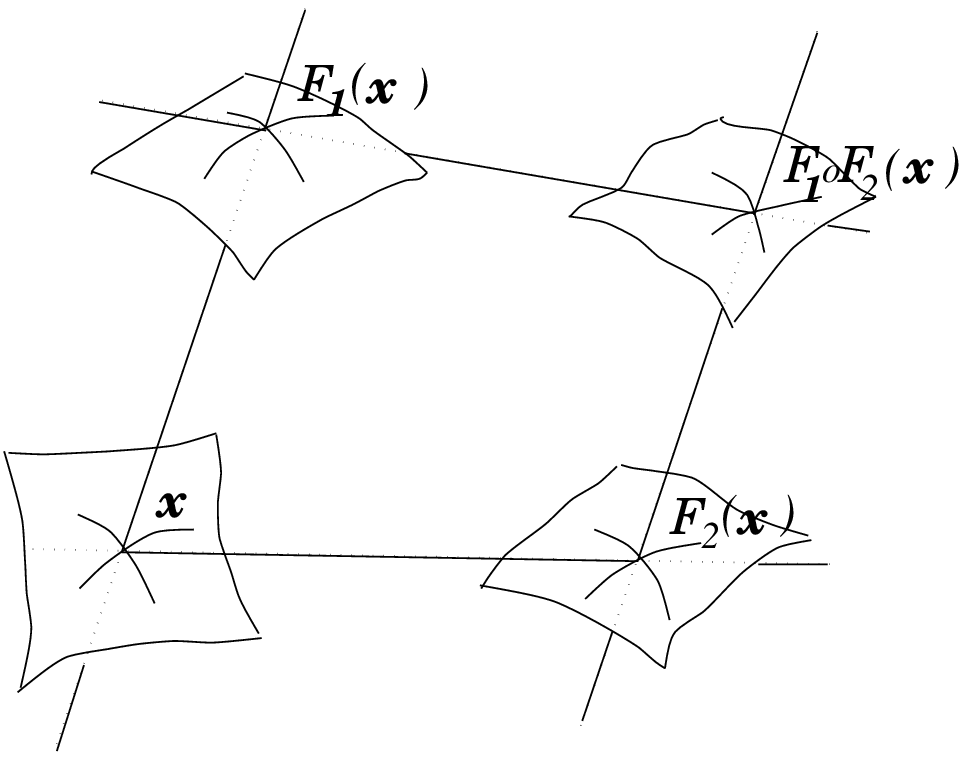}
\end{center}
Figure 15.

Since most (if not all) the known integrable PDE's with geometric meaning 
describe essentially reductions and/or iso-conjugate deformations of the 
multiconjugate systems, then the following, universally accepted nowadays, 
empirical rules:
\begin{enumerate}
{\em \item the Darboux-type transformations of integrable PDE's generate their 
natural integrable discrete versions~\cite{LeBen,NimmoSchief},
\item if the original PDE's have a geometrical meaning, then these 
transformations provide the natural discretization of the corresponding 
geometric notions~\cite{BP1,Tsarev,CDS,BS,KoSchief2}}
\end{enumerate}
find their natural explanation in the light of the planarity of the elementary 
quadrilateral (\ref{eq:quad-sup}).

For discrete integrable systems there is obviously no essential
difference between "finite transformations" and new dimensions. 
This shows once more that, from the point of view of the theory of 
integrable systems, the
discrete ones  are  more basic. 

We finally remark that all the basic transformations we considered here: 
the L\'evy, adjoint L\'evy and fundamental transformations, can be considered 
as Laplace transformations of quadrilateral strips.
This observation shows that, although the Laplace transformations 
are of a very special type, they can be considered as the basic objects of 
the theory of transformations of lattices. This interpretation provides, for
example, a very transparent geometric meaning to the additional solution 
$\phi$ of the Laplace equation entering into the L\'evy transformation.

This formulation in terms of the Laplace transformations 
remains also valid in the limit from the "quadrilateral lattice $\bx$"
to the "conjugate net $\bx$" but, since the intermediate 
steps of the transformation involve "differential-difference" nets, 
it was unknown to the geometers
who studied conjugate nets only. 

\section{$\bar{\partial}$ formalism and transformations}

\label{sec:D-bar}
\subsection{The $\bar\partial$ dressing for the Darboux and MQL equations}
The central role of the $\bar\partial$ problem in 
the study of integrable multidimensional systems was established in [1]; 
soon after 
that, the $\bar\partial$ problem was incorporated successfully in the 
dressing method, 
giving rise to the $\bar\partial$ dressing method \cite{ZakMa}, which is 
a very general and convenient inverse method, based on the theory of 
complex analysis, introduced to construct:  

\noindent
i) integrable nonlinear 
systems of partial differential equations, together with large 
classes of solutions; 

\noindent
ii) the finite transformations (of B\"acklund and Darboux type) between 
different solutions of these integrable systems;

\noindent
iii) the integrable discrete analogues of these integrable systems.

The Darboux equations (\ref{eq:Darboux-H}) and their integrable discrete 
analogues, the MQL equations (\ref{eq:MQL-H}), provide a very precious 
illustrative 
example of the 
power and elegance of the $\bar\partial$ dressing method. The goal of 
this section is to 
reconsider the main results  of the previous sections, investigated so 
far from geometric 
and algebraic points of view, 
in the framework of the $\bar\partial$ formalism. More precisely, we shall 
present the 
$\bar\partial$ formulation of the radial, Combescure and fundamental 
transformations,  
together with their limiting cases: the L\'evy, adjoint L\'evy and Laplace 
transformations; 
we shall also discuss the  
permutability theorem and the essential equivalence between integrable  
discretizations of integrable PDE's and finite transformations of them. 
We shall 
find that the main results of the previous sections have a very 
elementary 
interpretation in the framework of the $\bar\partial$ formalism.
Although the $\bar\partial$ formalism associated with the Darboux and MQL 
equations is scalar, we have decided to consider its matrix generalization 
because we expect that the matrix analogue of the Darboux and MQL equations will 
find a geometric meaning.

Let us consider the following nonlocal $\bar\partial$ problem \cite{ZakMa}:
\begin{equation} \label{eq:D-bar}
\partial_{\bar\lambda}\chi (\lambda )=
\partial_{\bar\lambda}\eta (\lambda )+
\int_{\CC}\chi (\lambda')R(\lambda',\lambda )d\lambda'\wedge d\bar\lambda' ,
\quad
\lambda , \lambda'\in \CC, 
\end{equation}
for square $D\times D$ matrices,  
where $R(\lambda',\lambda )$ is a given $\bar\partial$ datum, which 
decreases 
fastly enough at $\infty$ in $\lambda$ and $\lambda'$, and the function    
$\eta (\lambda )$, the {\it normalization} of the unknown 
$\chi (\lambda )$, is a given function of $\lambda$ and $\bar\lambda$, which 
describes, in particular, the polar behaviour of $\chi (\lambda )$ in $\CC$ 
and its behaviour at 
$\infty$: $\chi -\eta\to 0$ as $\lambda\to\infty$. 
Therefore the $\bar\partial$ problem (\ref{eq:D-bar}) is  equivalent to the 
following Fredholm 
integral equation of the second type:
\begin{equation} \label{eq:Fredholm}
\chi (\lambda )=\eta (\lambda )+
{1\over 2\pi i}\int_{\CC}{d\lambda'\wedge d\bar\lambda'\over 
\lambda'-\lambda}
\int_\CC\chi (\lambda'')R(\lambda'',\lambda')d\lambda''\wedge d\bar\lambda''.
\end{equation}
We remark that the dependence of $\chi (\lambda )$ and $R(\lambda ,\lambda')$ 
on 
$\bar\lambda$ and $\bar\lambda'$ will be sistematically omitted, 
for notational convenience, 
throughout this section. Furthermore it will be assumed that the $\bar\partial$ 
problem (\ref{eq:D-bar}) be 
uniquely solvable; i.e., if $\xi (\lambda )$ solves the homogeneous version of 
the $\bar\partial$ problem (\ref{eq:D-bar}) and $\xi (\lambda )\to 0$ 
as $|\lambda |\to\infty$, then $\xi (\lambda )=0$.

The dependence of $R(\lambda',\lambda )$ (and, consequently, of 
$\chi (\lambda )$) on the 
continuous $\bu \in R^{N}$ and discrete $\bn\in Z^{N}$ space 
coordinates  
is  assigned, respectively, through the following compatible equations:
\begin{equation} \label{eq:evol-R-cont}
\partial_iR(\lambda ,\lambda')=K_i(\lambda )R(\lambda ,\lambda')-
R(\lambda ,\lambda')K_i(\lambda'),\quad i=1,..,N,
\end{equation}
\begin{equation} \label{eq:evol-R-dis}
T_iR(\lambda ,\lambda')=(1+K_i(\lambda ))      
R(\lambda ,\lambda'){(1+K_i(\lambda'))}^{-1},\quad i=1,..,N,
\end{equation}
where $\partial_i=\partial /\partial u_i,~i=1,..,N$ and  
$K_i(\lambda ),~i=1,..,N$ are given commuting matrices  
constant in $\bu$ and $\bn$; in the following, for simplicity, the 
matrices $K_i(\lambda )$ will be assumed to be diagonal. If we are 
interested in the construction of continuous (discrete) systems we 
concentrate on (\ref{eq:evol-R-cont}) 
(on (\ref{eq:evol-R-dis})) only; but, in general, both dependences can be 
considered at the same 
time.  
Equations (\ref{eq:evol-R-cont}) and (\ref{eq:evol-R-dis}) admit the general 
solution
\begin{equation*}
R(\lambda ,\lambda';{\bu},{\bn})=
G(\lambda )R_0(\lambda ,\lambda')(G(\lambda ))^{-1},
\end{equation*}
where
\begin{equation*}
G(\lambda )=\exp (\sum\limits_{i=1}^{N}u_iK_i(\lambda ))
\prod\limits_{j=1}^{N}{(1+K_j(\lambda )})^{n_j}.
\end{equation*}
We finally assume that $R_0(\lambda ,\lambda')$ be identically zero in 
both variables in a neighbourhood of the following points: the poles 
($\lambda_i$) and the zeroes of $\det (1+K_i(\lambda ))$, $i=1,..,N$ and the 
poles of $\eta (\lambda )$. This restriction ensures the analyticity of 
$\chi -\eta $ at these points \cite{BogdanovManakov,BoKo}.  
We briefly recall that, in the $\bar\partial$ dressing method, a crucial role 
is played by
the {\it long derivatives} $\cD_{u_i}$, $\cD_{n_i}$, $i=1,..,N$, defined, 
respectively, 
by
\begin{align} \label{eq:long-cont}
\cD_{u_i}\chi (\lambda )&:=\partial_i\chi (\lambda ) +
\chi (\lambda )K_i(\lambda ),\quad i=1,..,N \\ \label{eq:long-dis}
\cD_{n_i}\chi (\lambda )& :=\Delta_i\chi (\lambda ) +
(T_i\chi (\lambda ))K_i(\lambda ),\quad i=1,..,N 
\end{align}
which are the generators of the Zakharov-Manakov ring of operators 
\cite{ZakMa}; i.e., any linear 
combination, with coefficients depending only on $\bu$ and $\bn$, of the 
operators
\begin{equation}
\prod_k \cD^{l_k}_{u_k},\quad \prod_k \cD^{l_k}_{n_k},\quad
l_k\in N
\end{equation}
transforms solutions of (\ref{eq:D-bar}) into solutions of 
(\ref{eq:D-bar}) (corresponding, in general, to different 
normalizations). For instance:
\begin{equation} \label{eq:F1} 
\partial_{\bar\lambda}(\cD_{n_i}\chi (\lambda ))=(T_i\chi (\lambda ))
\partial_{\bar\lambda}K_i(\lambda )+\cD_{n_i}(\partial_{\bar\lambda}\eta )+
\int_\CC d\lambda'\wedge d\bar\lambda'(\cD_{n_i}\chi (\lambda' ))
R(\lambda',\lambda ),
\end{equation}                    
\begin{eqnarray} \nonumber
\partial_{\bar\lambda}(\cD_{n_i}\cD_{n_j}\chi (\lambda )) & = &
(T_i\cD_{n_j}\chi (\lambda ))\partial_{\bar\lambda}K_i(\lambda )+
(T_j\cD_{n_i}\chi (\lambda ))\partial_{\bar\lambda}K_j(\lambda )+
\cD_{n_i}\cD_{n_j}(\partial_{\bar\lambda}\eta )+ \\
& +&  \int_\CC d\lambda'\wedge d\bar\lambda'(\cD_{n_i}\cD_{n_j}\chi (\lambda'))
R(\lambda',\lambda ), \label{eq:F2}
\end{eqnarray}                     
The goal of the method is to use this ring of operators to construct a set 
of solutions $\{\xi (\lambda )\}$ of (\ref{eq:D-bar}) such that 
$\xi (\lambda )\to 0$ as $\lambda\to\infty$ 
and use uniqueness to infer the set of equations: $\{\xi (\lambda )=0\}$, 
which are equivalent to the integrable nonlinear system.  

A given choice of the rational functions $K_i(\lambda )$ gives rise to 
solutions of a   
particular integrable nonlinear system; for instance, the Darboux and MQL 
equations (\ref{eq:Darboux-H}) and (\ref{eq:MQL-A}) 
correspond to the following choice \cite{ZakMa,BoKo} 
(see Proposition \ref{prop:Lapl-D} below):
\begin{equation} \label{def:K-Db}
K_i(\lambda ):={\alpha_i\over \lambda -\lambda_i},\quad
i=1,..,N,
\end{equation}
where $\alpha_i$ are the constant diagonal matrices.
 
Different normalizations are associated instead with 
different solutions of such nonlinear system. As it was observed in 
\cite{BogdanovManakov}, the richness of this 
mechanism of constructing solutions is typical of multidimensional problems 
since, in the case 
of the   {\it local} $\bar\partial$ problem, arising in 1+1 dimensions, 
different 
normalizations are all gauge equivalent. 
In this paper we shall limit our considerations to {\bf bounded} 
(in $\lambda$ and 
$\bar\lambda$) normalizations, which give rise to bounded  
(in $\lambda$ and $\bar\lambda$) solutions of the 
$\bar\partial$ problem (\ref{eq:D-bar}).     

We first recall the basic results concerning the 
$\bar\partial$ - integrability of the Darboux 
and MQL equations, obtained, respectively, in \cite{ZakMa} and \cite{BoKo}.  
\begin{Prop} \label{prop:Lapl-D}
Let $\varphi (\lambda )$  be the 
solution of (\ref{eq:D-bar}) corresponding to the 
{\it canonical} normalization $\eta =1$. Then 
the complex function
\begin{equation} \label{eq:psi-DB}
\psi (\lambda )=\varphi (\lambda )G(\lambda )
\end{equation}
solves the continuous and discrete Laplace equations:
\begin{equation} \label{eq:Laplace-DB}
L_{ij}[H]\psi (\lambda )=\Lambda_{ij}[H]\psi (\lambda )=0,
\; \; i,j=1,..,N, \; \; i\ne j
\end{equation}
where
\begin{align} \label{def:L-c-D}
L_{ij}[H]& :=\partial_i\partial_j
-(\partial_jH_i)H^{-1}_i\partial_i
-(\partial_iH_j)H^{-1}_j\partial_j ,
\\ \label{def:L-d-D}
\Lambda_{ij}[H]& :=\Delta_i\Delta_j-
T_i((\Delta_jH_i)H^{-1}_i)\Delta_i-
T_j((\Delta_iH_j)H^{-1}_j)\Delta_j
\end{align}
and the set of functions $H_i,\; i=1,..,N$, defined by  
\begin{equation} \label{def:H-DB}
H_i:=\varphi (\lambda_i)G_i
\end{equation}
\begin{equation} \label{def:G-D}
G_i:=\exp (\sum\limits_{k=1,k\ne i}^Nu_kK_k(\lambda_i))
\prod\limits_{k=1,k\ne i}^N{(1+K_k(\lambda_i ))}^{n_k},
\end{equation}
solve the matrix analogues of the Darboux (\ref{eq:Darboux-H}) and MQL equations 
(\ref{eq:MQL-H}).
\end{Prop}
\begin{Proof}
In the philosophy of the $\bar\partial$ method, one shows that the 
solutions  $\tilde L_{ij}\varphi (\lambda )$, 
$\tilde \Lambda_{ij}\varphi (\lambda )$ of the homogeneous version of the 
$\bar\partial$ problem (\ref{eq:D-bar}) go to zero as 
$\lambda\to\infty$, where:
\[
\tilde L_{ij}\varphi (\lambda ):=\cD_{u_i}\cD_{u_j}\varphi (\lambda )-
(\cD_{u_j}\varphi (\lambda_i))\varphi (\lambda_i)^{-1}
\cD_{u_i}\varphi(\lambda)- 
(\cD_{u_i}\varphi (\lambda_j))\varphi (\lambda_j)^{-1}
\cD_{u_j}\varphi(\lambda),
\] 
\begin{equation} \label{eq:lambda-DB}
\tilde \Lambda_{ij}\varphi (\lambda ):=\cD_{n_i}\cD_{n_j}\varphi (\lambda )-
T_i((\cD_{n_j}\varphi (\lambda_i))\varphi (\lambda_i)^{-1})
\cD_{n_i}\varphi(\lambda ) -
\end{equation}
\[
T_j((\cD_{n_i}\varphi (\lambda_j))\varphi (\lambda_j)^{-1})
\cD_{n_j}\varphi(\lambda ).
\]
Therefore uniqueness implies that 
\[
\tilde L_{ij}\varphi (\lambda )= 
\tilde\Lambda_{ij}\varphi (\lambda )=0
\] 
or, equivalently, equations (\ref{eq:Laplace-DB}). Finally, evaluating equation 
(\ref{eq:lambda-DB}) at 
$\lambda =\lambda_k$, $k\ne i\ne j\ne k$ and using (\ref{def:H-DB}), we obtain 
the Darboux 
and MQL equations respectively.
\end{Proof}

The above function $\psi (\lambda)$ allows one to construct the 
$D\times M$ matrix solution ${\bx}$:
\[ {\bx}({\bu},{\bn})=
\int \psi (\lambda )h(\lambda )d\lambda\wedge d\bar\lambda ,
\]
of the Laplace equations (\ref{eq:Laplace-DB}),  
where $h(\lambda )$ is an arbitrary localized $D\times M$ matrix function of 
$\lambda$ and $\bar\lambda$ (but independent of the coordinates).  
If the $\bar\partial$ problem (\ref{eq:D-bar}) 
is scalar, i.e.: $D=1$, ${\bx}$ is 
an $M$-dimensional vector solution of the Laplace equations. Therefore,   
keeping ${\bn}$ fixed, ${\bx}$ describes 
an $N$ dimensional manifold in $\RR^M$, parametrized by the conjugate 
coordinates ${\bu}$ 
(a conjugate net). Different values of ${\bn}$ can therefore be interpreted 
as defining an 
$N$ - dimensional (quadrilateral) sequence of conjugate nets. In the second 
interpretation we priviledge, 
instead, the discrete aspect of the problem: keeping ${\bu}$ fixed, 
${\bx}$ describes 
an $N$ dimensional quadrilateral lattice in $\RR^M$, while the continuous 
coordinates ${\bu}$ 
describe ``iso-conjugate" deformations of this lattice.

We finally remark that equation (\ref{eq:F1}) can be viewed as the continuous 
limit 
$\epsilon\to 0$ of (\ref{eq:F2}), in which: 
$\epsilon n_i\to u_i$ and $T_i\sim 1+\epsilon\partial_i$ (replacing 
$\alpha_i$ by $\epsilon\alpha_i$).

Exploiting completely the possible normalizations of the $\bar\partial$ 
problem, one obtains 
more solutions of the 
Laplace equations, together with the relations between them. The 
radial (or projective) 
and the Combescure transformations can be obtained in this way.

\subsection{Radial transformations}
\begin{Prop}  \label{prop:radial-D}
Let $\varphi_{\cP} (\lambda )$ be the solution of (\ref{eq:D-bar}) 
corresponding 
to the normalization $\eta =\phi^{-1}$, where $\phi$ is any solution of the 
continuous and 
discrete Laplace equations (\ref{eq:Laplace-DB}). Define the function   
\begin{equation} \label{def:psi-P}
\psi_{\cP} (\lambda ):=
\varphi_{\cP} (\lambda )G(\lambda );
\end{equation}
then:

\noindent
i) $\psi_{\cP} (\lambda )$ is related to the function $\psi (\lambda )$, 
defined in (\ref{eq:psi-DB}), through the radial (gauge) transformation:
\begin{equation}
\psi_{\cP}(\lambda )=\phi^{-1}\psi (\lambda );
\end{equation}

\noindent
ii) $\psi_{\cP} (\lambda )$ solves the Laplace equations 
\begin{equation}
L_{ij}[{\cP}(H)]\psi_{\cP} (\lambda )=
\Lambda_{ij}[{\cP}(H)]\psi_{\cP} (\lambda )=0, \quad i,j=1,..,N, i\ne j,
\end{equation}
where the functions 
\begin{equation}
{\cP}(H_i)=\varphi_{\cP} (\lambda_i)G_i=\phi^{-1}H_i
\end{equation}
solve the matrix Darboux and MQL equations.
\end{Prop}
\begin{Proof}
The proof goes as in Proposition \ref{prop:Lapl-D}. The uniqueness of the
$\bar\partial$ problem implies the following equations: 
\[
\varphi_{\cP} (\lambda )-\phi^{-1}\varphi (\lambda )=0,
\]  
\[\tilde L_{ij}\varphi_{\cP} (\lambda )+  
\phi^{-1}( L_{ij}[H]\phi )\phi^{-1}\varphi (\lambda )=0,
\]
\[\tilde \Lambda_{ij}\varphi_{\cP} (\lambda )+
(T_iT_j\phi^{-1})(\Lambda_{ij}[H]\phi )\phi^{-1}\varphi (\lambda )=0,
\]
equivalent, respectively, to (\ref{eq:evol-R-cont}) and (\ref{eq:evol-R-dis}).

Therefore the $D\times M$ matrix 
\begin{equation} \label{eq:rad-D}
{\cP}({\bx})=
\int_\CC\psi_{\cP} (\lambda )h(\lambda )d\lambda\wedge d\bar\lambda ,
\end{equation}
satisfies the equations
\[
L_{ij}[{\cP}(H)]{\cP}({\bx})=
\Lambda_{ij}[{\cP}(H)]{\cP}({\bx})=0, \quad i,j=1,..,N, i\ne j
\]
\[
{\cP}({\bx})=\phi^{-1}{\bx}
\]
and, if the $\bar\partial$ problem (\ref{eq:D-bar}) is scalar ($D=1$), it 
defines  
the radial transform ${\cP}({\bx})$ of ${\bx}$ (see Section \ref{sec:radial}).
\end{Proof}

\subsection{Combescure Transformations}
We first introduce the basic, localized in $\lambda$ and $\bar\lambda$, 
solutions of the 
$\bar\partial$ problem, corresponding to the simple pole normalization 
$\eta ={(\lambda -\mu )}^{-1}$. These solutions were first used in a 
multidimensional context 
in \cite{BogdanovManakov} and used extensively in \cite{BK-Comb}. The following 
proposition can be found in 
\cite{BoKo}.
\begin{Prop} \label{prop:Comb-D}
Let $\varphi (\lambda ,\mu )$ be the solution of (\ref{eq:D-bar}) 
corresponding to the simple pole normalization  
$\eta ={(\lambda -\mu )}^{-1},~\mu\ne \lambda_i,~i=1,..,N$. Define 
the function
\begin{equation} \label{eq:norm-D}
\psi (\lambda ,\mu ):=G(\mu )^{-1}\varphi (\lambda ,\mu )G(\lambda ),
\end{equation}
then:

\noindent
i) $\psi (\lambda ,\mu )$ solves the Laplace equations
\begin{equation} \label{eq:LC-DB}
L_{ij}[H(\mu )]\psi (\lambda ,\mu )=
\Lambda_{ij}[H(\mu )]\psi (\lambda ,\mu )=0,
\end{equation}
and the functions 
\[
H_i(\mu )=
G(\mu )^{-1}\varphi (\lambda_1 ,\mu )G_i
\]
solve the Darboux and MQL equations;

\noindent
ii) $\psi (\lambda ,\mu )$ is a Combescure transform of $\psi (\lambda )$,
i.e., the following formulas hold:
\begin{equation} \label{eq:Comb-D}
\partial_i\psi (\lambda ,\mu )=C_i(\mu )\partial_i\psi (\lambda ), \quad
\Delta_i\psi (\lambda ,\mu )=(T_iC_i(\mu ))\Delta_i\psi (\lambda ),
\end{equation}
where:
\begin{equation} \label{def:c-Comb-D}
C_i(\mu )=H_i(\mu )H^{-1}_i
\end{equation}
and:
\begin{equation} \label{eq:E1}
\partial_iH_j(\mu )=C_i(\mu )\partial_iH_j, \quad
\Delta_iH_j(\mu_j)=(T_iC_i(\mu ))\Delta_iH_j, \quad i\ne j
\end{equation}
\[
\partial_iC_j(\mu )+
(C_j(\mu )-C_i(\mu ))(\partial_iH_j){H_j}^{-1}=0,
\]
\begin{equation} \label{eq:E3}
\Delta_iC_j(\mu )+
(T_iC_j(\mu )-T_iC_i(\mu ))(\Delta_iH_j){H_j}^{-1}= 0.
\end{equation}
\end{Prop}
\begin{Proof} 
The uniqueness of the $\bar\partial$ problem implies the following 
equations: 
\[
\tilde \Lambda'_{ij}\varphi (\lambda ,\mu )=0,
\]
\begin{equation} \label{eq:E2}
\cD'_{n_i}\varphi (\lambda ,\mu)-
T_i(\varphi (\lambda_i ,\mu )(\varphi (\lambda_i))^{-1})
\cD_{n_i}\varphi (\lambda )=0
\end{equation}
and their continuous analogues, equivalent, respectively, to (\ref{eq:LC-DB}) 
and (\ref{eq:Comb-D}), 
where $\tilde \Lambda'_{ij}$ is obtained from 
$\tilde \Lambda_{ij}$ replacing $\cD_{n_i}$ by 
\[
\cD_{n_i}^\prime f:=\cD_{n_i}f-{\alpha_i\over \mu -\lambda_i}f, \quad i=1,..,N.
\]
Equations (\ref{eq:E1}) follow by multiplying equation (\ref{eq:E2}) 
by $(1+K_j(\lambda ))^{-1}$ and then setting $\lambda =
\lambda_j$; equations (\ref{eq:E3}) are direct consequences of (\ref{eq:E1}) 
and (\ref{def:c-Comb-D}).  
\end{Proof}

\begin{Rem}
The formula (\ref{eq:Comb-D}) suggests that one could start with the 
solution of (\ref{eq:D-bar}) 
normalized by $\eta =(\lambda -\mu )^{-1}G(\mu)^{-1}$,  
avoiding in this way the introduction of the generalized operators $\cD'_{u_i}$,  
$\cD'_{n_i}$ and simplifying the proof. This is actually a key observation 
in the following construction of more general solutions, bounded in $\lambda$, 
of the Laplace equations.
\end{Rem}

The canonical and simple pole normalizations allow one to construct the 
prototype examples of, respectively, bounded and localized solutions of the 
Laplace equations. This is due to the fact that the corresponding 
normalizations: 
$\eta =1$ and $\eta =(\lambda -\mu )^{-1}G(\mu )^{-1}$
satisfy the equations
\[
\cD_{n_i}(\partial_{\bar\lambda}\eta ) =\cD_{u_i}(\partial_{\bar\lambda}\eta 
)=0,
\]
implying that the forcings of equations (\ref{eq:F1}), ((\ref{eq:F2}) do not 
depend on $\eta$. 
Observing that:
\[
\cD_{n_i}f=\cD_{u_i}f=0, \; \; i=1,..,N \quad \Longleftrightarrow \quad 
f=\gamma (\lambda )G(\lambda )^{-1}, 
\] 
where $\gamma (\lambda )$ is an arbitrary function of $\lambda ,\bar\lambda$, 
but constant in the coordinates, we infer that a general, bounded in 
$\lambda ,\bar\lambda$, solution of the Laplace equations is obtained 
considering the solution $\Phi (\lambda )$ of the $\bar\partial$ 
problem (\ref{eq:D-bar}) corresponding to the normalization
\begin{equation}
\eta =a+{i\over 2}\int_\CC{d\mu'\wedge d\bar\mu'\over \lambda -\mu'}
\gamma (\mu')G(\mu')^{-1} \quad \Rightarrow \quad 
\partial_{\bar\lambda}\eta =
\gamma (\lambda )G(\lambda )^{-1}, \label{eq:norm-C}
\end{equation}
where $\gamma$ is any localized function of $\lambda , 
\bar\lambda$, constant in the coordinates and $a$ is any constant (in
$\lambda$ and in the coordinates) matrix. The general solution 
$\Psi (\lambda ) =\Phi (\lambda ) G(\lambda )$ of the Laplace 
equations reduces to the solutions $\psi (\lambda)$ and 
$\psi (\lambda ,\mu )$, 
corresponding to the canonical and simple pole normalizations, through the 
following obvious 
specifications:
\begin{align*}
a=1, \; \gamma (\lambda )=0 \quad & \Rightarrow \quad 
\Psi (\lambda )=\psi (\lambda ), \\
a=0, \; \gamma (\lambda )=\delta (\lambda -\mu ) \quad & \Rightarrow \quad
\Psi (\lambda )=\psi (\lambda ,\mu ).
\end{align*}  
\begin{Prop} \label{prop:gen-norm-D}
Let $\Phi (\lambda )$ be the solution of (\ref{eq:D-bar}) 
corresponding to the normalization (\ref{eq:norm-C}). Define the function 
$\Psi (\lambda )$ in the usual way:
\[
\Psi (\lambda )=\Phi (\lambda )G(\lambda );
\]
then:

\noindent
i) $\Psi (\lambda )$ solves the Laplace equations
\begin{equation} \label{eq:LL-DB}
L_{ij}[H]\Psi (\lambda )=
\Lambda_{ij}[H]\Psi (\lambda )=0, \; \; i,j=1,..,N,~i\ne j
\end{equation}
and the functions 
\[
H_{i}=\Phi (\lambda_i)G_i
\]
solve the Darboux and MQL equations.

\noindent
ii) If $\Psi^{(l)}(\lambda )=\Phi^{(l)}(\lambda )G(\lambda ), l=1,2$ are two 
different 
solutions of (\ref{eq:LL-DB}) corresponding to the different normalizations 
$a^{(l)}$, $\gamma^{(l)}(\lambda )$, $l=1,2$, then these solutions are related 
by the
Combescure transformation, i.e.: 
\begin{equation} \label{eq:norm-Comb-D}
\partial_i\Psi^{(2)}(\lambda )=
C^{(2,1)}_i\partial_i\Psi^{(1)}(\lambda ), \quad
\Delta_i\Psi^{(2)}(\lambda )=
(T_iC^{(2,1)}_i)\Delta_i\Psi^{(1)}(\lambda ), \quad i=1,..,N
\end{equation}
where the functions 
\begin{equation} \label{eq:C-norm-Comb-D}
C^{(2,1)}_i={H^{(2)}_i}(H^{(1)}_i)^{-1}
\end{equation}
\[
H^{(l)}_i=\Phi^{(l)}(\lambda_i)G_i, \quad l=1,2
\]
satisfy the equations
\[
\partial_iH^{(2)}_j=C^{(2,1)}_i\partial_iH^{(1)}_j,~~~
\Delta_iH^{(2)}_j=(T_iC^{(2,1)}_i)\Delta_iH^{(1)}_j,~~~i\ne j
\]
\[
\partial_iC^{(2,1)}_j+
(C^{(2,1)}_j-C^{(2,1)}_i)(\partial_i{H}^{(1)}_j)({H}^{(1)}_j)^{-1}=0,
\]
\[
\Delta_iC^{(2,1)}_j+ (T_iC^{(2,1)}_j-T_iC^{(2,1)}_i)(\Delta_i{H}^{(1)}_j)
({H}^{(1)}_j)^{-1}= 0.
\]
    
\noindent
iii) The following relations hold:
\begin{equation} \label{eq:EE-DB}
\cD_{n_i}\Phi (\lambda )=\Phi (\lambda_i)\alpha_i\varphi (\lambda ,\lambda_i),
\quad \cD_{u_i}\Phi (\lambda )=(T_i\Phi (\lambda_i))\alpha_i\varphi 
(\lambda ,\lambda_i), \quad
i=1,..,N.
\end{equation}

\noindent
iv) If $\lambda_0\ne \lambda_i,\; i=1,..,N$ is an additional complex parameter 
associated with the additional coordinates $u_0$ and $n_0$:
\[
\partial_{u_0}R(\lambda ,\lambda')={\alpha_0\over \lambda
-\lambda_0}R(\lambda ,\lambda')-
R(\lambda ,\lambda'){\alpha_0\over \lambda'
-\lambda_0},~~~i=1,..,N,
\]
\begin{equation} \label{eq:T0-DB}
T_0R(\lambda ,\lambda')=(1+{\alpha_0\over \lambda
-\lambda_0})      
R(\lambda ,\lambda')(1+{\alpha_0\over \lambda'
-\lambda_0})^{-1},~~~i=1,..,N,
\end{equation}
where $\alpha_0$ is a diagonal matrix and $R_0(\lambda',\lambda )$ is zero 
in a neighborough of $\lambda =\lambda_0$ and $\lambda' =\lambda_0$,  
then $\varphi (\lambda ,\lambda_0)$ and $\Phi (\lambda )$ are connected through 
the analogues 
of equations (\ref{eq:EE-DB}):
\[
\cD_{n_0}\Phi (\lambda )=\Phi (\lambda_0)\alpha_0\varphi (\lambda ,\lambda_0),
\quad \cD_{u_0}\Phi (\lambda )=(T_0\Phi (\lambda_0))\alpha_0\varphi 
(\lambda ,\lambda_0),
\]
equivalent to equations  
\begin{equation} \label{eq:T00-DB}
\cD_{u_0}\Psi (\lambda )=
\Psi (\lambda_0)\alpha_0\psi (\lambda ,\lambda_0), \quad
\cD_{n_0}\Psi (\lambda )=
(T_0\Psi (\lambda_0))\alpha_0\psi (\lambda ,\lambda_0).
\end{equation}
\end{Prop}
\begin{Proof}
As before, the uniqueness of the $\bar\partial$ problem implies equations
\[
\tilde \Lambda_{ij}[\Phi (\lambda )]=0,
\]
\[
\cD_{n_i}\Phi^{(2)}(\lambda )-
T_i(\Phi^{(2)}(\lambda_i)(\Phi^{(1)}(\lambda_i))^{-1})
\cD_{n_i}\Phi^{(1)}(\lambda )=0,
\]
\[
\cD_{n_i}\Phi (\lambda )-
(T_i\Phi (\lambda_i )\alpha_i\varphi (\lambda ,\lambda_i)=0
\]
and their continuous analogues, equivalent, respectively, to equations 
(\ref{eq:LL-DB}), (\ref{eq:norm-Comb-D}) and (\ref{eq:EE-DB}). 
The rest of the 
proof is as in the previous Propositions.
\end{Proof}
\begin{Rem}
We remark that the localized solutions $\Phi$ of (\ref{eq:D-bar}),
corresponding to the normalization (\ref{eq:norm-C}) with $a=0$, can be
obtained integrating the simple pole solutions with an arbitrary measure:
\[
\Phi(\lambda) ={i\over 2\pi }\int_\CC {d\mu\wedge d\bar\mu
\gamma (\mu)G(\mu)^{-1}\varphi(\lambda,\mu)} \; .
\]
This formula establishes a contact with the class of Combescure related
solutions of the Laplace equation obtained in~\cite{BK-Comb,BK-mKP}.
\end{Rem}

\begin{Rem} 
The Combescure solutions introduced in this Proposition  form a linear 
space. For instance, the solution $\Psi (\lambda )$, corresponding to the 
normalization
\[
1+{i\over 2}\int_\CC{d\lambda^{\prime}\wedge d\bar\lambda^{\prime}\over 
\lambda -\lambda^{\prime}}
\gamma (\lambda^{\prime})G(\lambda^{\prime})^{-1},
\]
is the linear combination
\begin{equation} \label{eq:PsiC}
\Psi (\lambda )=\psi (\lambda )+\psi_{\cal C}(\lambda )
\end{equation}
of the solution $\psi (\lambda )$, corresponding to the canonical normalization, 
and of the 
solution $\psi_{\cal C}(\lambda )$, corresponding to the normalization 
\[
{i\over 2}\int_\CC{d\lambda^{\prime}\wedge d\bar\lambda^{\prime}\over 
\lambda -\lambda^{\prime}}
\gamma (\lambda^{\prime})G(\lambda^{\prime})^{-1}.
\]

Therefore the $D\times M$ matrix solutions 
\[ {\bf x}^{(l)}({\bf u},{\bf n})=\int_\CC\Psi^{(l)}(\lambda )h(\lambda )
d\lambda\wedge d\bar\lambda ,~~~l=1,2
\]
of the Laplace equations 
\[
L_{ij}[H^{(l)}]{\bx}^{(l)}=
\Lambda_{ij}[H^{(l)}]{\bx}^{(l)}=0,~~~~l=1,2~~~~~i,j=1,..,N,~i\ne j
\]
satisfy the Combescure relations
\[
\partial_i{\bx}^{(2)}=
C^{(2,1)}_i\partial_i{\bx}^{(1)},~~~~~
\Delta_i{\bx}^{(2)}=(T_iC^{(2,1)}_i)\Delta_i{\bx}^{(1)},~~~i=1,..,N
\]
At last, from the equation (\ref{eq:PsiC}) we have the relation
\[
{\calC}({\bx})={\bx}+{\bx}_{\calC},
\]
where
\[
{\calC}({\bf x})=\int_\CC\Psi (\lambda )h(\lambda )
d\lambda\wedge d\bar\lambda ,
\] 
\[
{\bx}_{\calC}=\int_\CC\psi_{\cal C}(\lambda )h(\lambda )
d\lambda\wedge d\bar\lambda .
\] 
In the scalar case $D=1$, the $M$-dimensional vectors 
${\bf x}^{(l)},~~l=1,2$, ${\cal C}({\bf x})$, ${\bf x}$ and 
${\bf x}_{\cal C}$ are related by 
the Combescure transformation formulas of Section \ref{sec:Combescure}.
\end{Rem}

\subsection{Fundamental Transformations and their Composition}

So far we have used only different normalizations of the $\bar\partial$ problem. 
In order to generate more solutions of the Laplace equation, this mechanism must 
be combined 
with a more classical one, discovered long ago \cite{Cal} in the context of 1+1 
dimensional problems.  
\begin{Prop} \label{prop:fund-D-bar}
Let us consider the (by assumption uniquely solvable) 
$\bar\partial$ problem
\begin{equation} \label{eq:fund-D-bar}
\partial_{\bar\lambda}\tilde\chi (\lambda )=
\partial_{\bar\lambda}\tilde\eta (\lambda )+
 \int_{\CC}\tilde\chi (\lambda')\tilde R(\lambda',\lambda )d\lambda'\wedge 
d\bar\lambda' ,
\quad \lambda , \lambda'\in \CC, 
\end{equation}
where the $\bar\partial$ datum $\tilde R(\lambda',\lambda )$ is related to 
$R(\lambda',\lambda )$ through the transformation

\begin{equation} \label{eq:fund-R-D}
\tilde R(\lambda',\lambda )=g(\lambda')R(\lambda',\lambda )g(\lambda )^{-1}, 
\quad
\end{equation}
where $g(\lambda)$ is a diagonal matrix (more generally -- commuting with 
$K_i(\lambda)$) and independent of $\bn$, $\bu$, and $R_0(\lambda',\lambda )$ is 
assumed to be zero in a neighbourhood of
the zeroes and poles of $\det g(\lambda )$. Then:

\noindent
i) if $\tilde{\eta}$ satisfies the equation
$\cD_{n_i}(\partial_{\bar{\lambda}}\tilde{\eta})=0$, then the corresponding
solutions of (\ref{eq:fund-D-bar}),~(\ref{eq:fund-R-D})  give rise 
to solutions of the Laplace 
equations.

\noindent
ii) If $\chi (\lambda )$ solves the $\bar\partial$ problem (\ref{eq:D-bar}), 
then the 
function $\chi (\lambda )g(\lambda )^{-1}$ solves the $\bar\partial$ problem 
(\ref{eq:fund-D-bar}), 
corresponding to the inhomogeneous term: 
\begin{equation}
\partial_{\bar\lambda}\tilde\eta = 
(\partial_{\bar\lambda}\eta )g(\lambda )^{-1}+
\chi (\lambda )\partial_{\bar\lambda}g(\lambda )^{-1}.
\end{equation}
\end{Prop}
\begin{Proof} Since $\tilde R(\lambda',\lambda )$ satisfies equations 
(\ref{eq:evol-R-cont}) and (\ref{eq:evol-R-dis}), then 
the results of Propositions \ref{prop:Lapl-D}-\ref{prop:gen-norm-D}
apply also to this case. 
ii) follows from taking the $\partial_{\bar\lambda}$ 
derivative of $\chi (\lambda )g(\lambda )^{-1}$ and using (\ref{eq:D-bar}).
\end{Proof}

The matrix function $g(\lambda )$ appearing in this Proposition is usually 
chosen to 
be a rational 
function of $\lambda $ such that $g(\lambda )\to 1$ as $\lambda\to\infty$, in 
order 
to preserve the properties at $\infty$ of the $\bar\partial$ problem. We shall 
show now that the simplest nontrivial example of this type:
\begin{equation} \label{eq:g-ex-D}
g(\lambda )=1+{\beta\over \lambda -\mu},
\end{equation}
corresponds to the Fundamental Transformation of a quadrilateral lattice and 
conjugate net. 
\begin{Prop} \label{prop-fund-D}
Let $\Phi (\lambda )$ and $\tilde{\Phi}(\lambda )$ be the 
solutions of, respectively, the $\bar\partial$ problems (\ref{eq:D-bar}) 
and (\ref{eq:fund-D-bar})
(with $g(\lambda )$ defined in (\ref{eq:g-ex-D})), corresponding to the 
normalizations 
\[
\eta =a+{i\over 2}\int_\CC{d\mu\wedge d\bar\mu\over \lambda -\mu}
\gamma (\mu)G(\mu)^{-1},
~~~\tilde\eta =a+{i\over 2}\int_\CC{d\mu\wedge d\bar\mu\over \lambda -\mu}
\gamma (\mu)g(\mu )^{-1}G(\mu)^{-1}.
\]
Let $\varphi (\lambda ,\mu )$ be the solution of the 
$\bar\partial$ problem (\ref{eq:D-bar}) corresponding to the normalization $\eta 
= 
(\lambda -\mu )^{-1}$. Define the function 
\begin{equation} \label{def:bpsi-D}
\tilde\Psi (\lambda ):=
\tilde\Phi (\lambda )G(\lambda );
\end{equation}
then:

\noindent
i) $\tilde{\Psi} (\lambda )$ satisfies the Laplace equations   
\begin{equation} \label{eq:fun-L-D}
L_{ij}[{\cF}(H)]\tilde{\Psi}(\lambda )=0,~~~~~
\Lambda_{ij}[{\cal F}(H)]\tilde{\Psi}(\lambda )=0,~~~~~~~
i,j=1,..,N,~~i\ne j,
\end{equation}
and the functions
\begin{equation} \label{eq:fun-H-D}
{\cal F}(H_i):=\tilde{\Phi}(\lambda_i)G_i,~~~~i=1,..,N
\end{equation}
satisfy the matrix Darboux and MQL equations.

\noindent
ii) $\tilde{\Psi} (\lambda )$ is the fundamental transform of 
$\Psi (\lambda )$, i.e.: 
\begin{equation} \label{eq:fund-psit-D}
\tilde{\Psi} (\lambda )=[\Psi (\lambda )+
A\psi (\lambda ,\mu )](1+{\beta\over \lambda -\mu})^{-1},
\end{equation}
where the matrix $A$ is defined in the following two ways:
\begin{equation} \label{eq:A-nu-D}
A=-\Psi (\bnu )(\psi (\bnu ,\mu ))^{-1},~~~~A={\tilde\Psi (\mu )}\beta ,
\end{equation}
\[
(\Psi (\bnu ))_{lm}:=\Psi_{lm}({\nu_m}),~~~
(\psi (\bnu ,\mu ))_{lm}:=\psi_{lm}(\nu_m,\mu ),~~~~~~l,m=1,..,D
\]
and $\nu_m, m=1,..,D$ are the zeroes of $\det g(\lambda )$.
\end{Prop}
\begin{Proof} i) is an immediate consequence of part i) of 
Proposition \ref{prop:fund-D-bar}. 
To prove part ii), first 
remark that
\[
\partial_{\bar\lambda}\left( g(\lambda )^{-1} \right)
=\pi\sum\limits_{k=1}^L
(\nu_k-\mu )\delta (\lambda -\nu_k)P_k,
\]
where $P_k, k=1,..,D$ are the usual matrix projectors: 
$(P_k)_{lm}=\delta_{lk}\delta_{km}$. 
Then observe that the matrix $B$, defined by the following generalized equation
\[
[\Phi (\lambda )+B\varphi (\lambda ,\mu )]g(\lambda )^{-1}=0,
\]
is given by
\[
B=-\Phi (\bnu )(\varphi (\bnu ,\mu ))^{-1},
\]
where
\[
(\Phi (\bnu ))_{lm}=\Phi_{lm} (\nu_m ),~~~~~(\varphi (\bnu ,\mu ))_{lm}=
\varphi_{lm}(\nu_m,\mu ),~~~~l,m=1,..,D.
\]
The uniqueness of the $\bar\partial$ problem (\ref{eq:D-bar}) implies that 
\begin{equation} \label{eq:fund-phit-D}
\tilde\Phi (\lambda )-[\Phi (\lambda )+B\varphi (\lambda ,\mu )]
(1+{\beta\over \lambda -\mu})^{-1}=0.
\end{equation}
In addition, since $\tilde\Phi (\lambda )$ is analytic in $\lambda =\mu$, 
it follows that   
$B=\tilde\Phi (\mu )\beta$. Multiplying (\ref{eq:fund-phit-D}) by 
$G(\lambda )$ one obtains equation (\ref{eq:fund-psit-D}), with 
$A=BG(\mu )$.    
\end{Proof}

If the $\bar\partial$ problem is scalar ($D=1$), then
\[
\tilde{\Psi} (\lambda )=[\Psi (\lambda )-
{\Psi (\nu )\over \psi (\nu ,\mu )}\psi (\lambda ,\mu )]
{\lambda -\mu \over \lambda -\nu}, \quad \nu =\mu -\beta ,
\]
and the quadrilateral lattices (and conjugate nets)
\[
{\bx}=\int_\CC \Psi (\lambda )h(\lambda )d\lambda \wedge d\bar\lambda,
\quad {\bx}_\calC(\mu )=\int_\CC \psi (\lambda ,\mu )h(\lambda )d\lambda \wedge 
d\bar\lambda,
\]  
\[
{\cF}({\bx})=
\int_\CC \tilde\Psi (\lambda ){\lambda -\nu\over \lambda -\mu}h(\lambda )
d\lambda \wedge d\bar\lambda 
\]
are related through the Fundamental transformation 
(see Section \ref{sec:fundamental})
\[
{\cF}({\bx})={\bx}-
{\Psi (\nu )\over \psi (\nu ,\mu )}{\bx}_\calC(\mu ).
\]

This result can be generalized in a straightforward way to the case of the 
composition of 
several fundamental transformations. In terms of the $\bar\partial$ datum, 
the sequence 
of transformations reads:
\[
R(\lambda ,\lambda')\to 
R_1(\lambda ,\lambda')=g_1(\lambda )R(\lambda ,\lambda')g_1(\lambda')^{-1}\to
\]
\[
R_{12}(\lambda ,\lambda')=
g_2(\lambda )R_1(\lambda ,\lambda')g_2(\lambda')^{-1}=
g_1(\lambda )g_2(\lambda )R(\lambda ,\lambda')(g_1(\lambda')g_2(\lambda'))^{-1}
\to 
\cdot\cdot 
\]
\[ 
\to R_{12\cdot\cdot L}=\prod\limits_{k=1}^Lg_k(\lambda )R(\lambda ,\lambda')
\prod\limits_{k=1}^L(g_k(\lambda'))^{-1},
\]
where
\begin{equation} \label{def:gi-fund-D}
g_i(\lambda )=1+{\beta_i\over \lambda -\mu_i},~~~~~~i=1,..,L.
\end{equation}
Therefore the sequence of $L$ fundamental transformations $g_i(\lambda )$ is 
equivalent to a 
single transformation in which
\begin{equation} \label{def:g-fund-D}
g(\lambda )=\prod\limits_{k=1}^L(1+{\beta_k\over \lambda -\mu_k}).
\end{equation}
Furthermore the commutation of the diagonal matrices $g_i(\lambda ),~i=1,..,L $ 
implies 
that the   
sequence of fundamental transformations does not depend on the order 
in which it is obtained 
(the famous permutability theorem has therefore a very elementary 
interpretation in the 
$\bar\partial$ formalism). The corresponding transformation in 
configuration space is 
described by the following   

\begin{Prop} \label{prop:fund-config-D}
Let $\Phi (\lambda )$ and $\tilde{\Phi} (\lambda )$ be the 
solutions, respectively, of the $\bar\partial$ problems (\ref{eq:D-bar}) 
and (\ref{eq:fund-D-bar}), (\ref{eq:fund-R-D}), (\ref{def:g-fund-D}), with 
$\mu_k\ne\mu_j$, $k\ne j$,
corresponding to the normalizations
\[
\eta =a+{i\over 2}\int_\CC{d\mu\wedge d\bar\mu\over 
\lambda -\mu}\gamma (\mu)G(\mu)^{-1},
~~~\tilde\eta =a+{i\over 2}\int_\CC{d\mu\wedge d\bar\mu\over 
\lambda -\mu}
\gamma (\mu)g(\mu )^{-1}G(\mu)^{-1}.
\]
Let $\varphi (\lambda ,\mu_k),~k=1,..,L$ be the solutions of the $\bar\partial$ 
problem (\ref{eq:D-bar}) 
corresponding to the normalizations $\eta = (\lambda -\mu_k)^{-1},~k=1,..,L$. 
Define the function $\tilde{\Psi} (\lambda )$ as in (\ref{def:bpsi-D}); then:
 
\noindent
i) the function $\tilde{\Psi} (\lambda )$     
satisfies the Laplace equations
\begin{equation} \label{eq:Lapl-f-D}
L_{ij}[\tilde H]\tilde{\Psi}(\lambda )=0,~~~~~
\Lambda_{ij}[\tilde H]\tilde{\Psi}(\lambda )=0,
\quad i,j=1,..,N,
~~i\ne j
\end{equation}
and
\[
H_i=\Phi (\lambda_i)G_i,~~~~\tilde H_i=\tilde\Phi (\lambda_i)G_i,~~~~~i=1,..,N.
\]

\noindent
ii) The following relation holds : 
\[
\tilde\Psi (\lambda )=[\Psi (\lambda )+
\sum\limits_{k=1}^LA^{(k)}\psi (\lambda ,\mu_k)]
\prod\limits_{k=1}^L(1+{\beta_k\over \lambda -\mu_k})^{-1},
\]
where the $D\times D$ matrices $A^{(k)},~k=1,..,L$ are defined in two 
independent ways; 
through the following linear system of $L$ equations for $D\times D$ matrices:
\[
\sum\limits_{k=1}^MA^{(k)}\psi (\bnu_i ,\mu_k)=-\Psi (\bnu_i),~~~~i=1,..,L,
\]
where
\[
(\Psi (\bnu_i))_{lm}=\Psi_{lm}(\nu_{im}),~~~~~
(\psi (\bnu_i ,\mu_k))_{lm}=\psi_{lm}(\nu_{im},\mu_k),~~~~~l,m=1,..,D,
\]
and $\nu_{im}, m=1,..,D$ are the zeroes of $\det g_i(\lambda )$, and 
through the equations
\[
A^{(k)}=\tilde\Psi (\mu_k)\beta_k\prod\limits_{l=1,l\ne k}^L(1+{\beta\over 
\mu_k-\mu_l}),
~~~~k=1,..,L.
\]
\end{Prop}
\begin{Proof}
The proof is a straightforward generalizations of that of 
Proposition \ref{prop-fund-D}. 
\end{Proof}

In the scalar case, the above equations simplify to 
\[
\tilde\Psi (\lambda )=[\Psi (\lambda )+
\sum\limits_{k=1}^LA^{(k)}\psi (\lambda ,\mu_k)]
\prod\limits_{k=1}^L{\lambda -\mu_k\over \lambda -\nu_k},~~~~\nu_k=\mu_k-
\beta_k,
\]
\[
\sum\limits_{k=1}^L A^{(k)}\psi (\bnu_i ,\mu_k)=-\Psi (\nu_i),~~~~i=1,..,L.
\]

Therefore the $M$ - dimensional vector   
\[
\tilde {\bx} =\int_\CC\tilde\Psi (\lambda )
\prod\limits_{k=1}^L{\lambda -\nu_k\over \lambda -\mu_k}
h(\lambda )d\lambda \wedge d\bar\lambda ,
\]
obtained combining in an arbitrary order $L$ fundamental transformations 
described 
by the Combescure vectors
\[ {\bx}^{(k)}_{\calC}=\int_\CC \psi (\lambda ,\mu_k)h(\lambda )
d\lambda \wedge d\bar\lambda,~~~k=1,..,L
\]
satisfies the following equation 
\[
\tilde {\bx}={\bx}+\sum\limits_{k=1}^L
A^{(k)}\bx^{(k)}_{\calC},
\]
which agrees with equation (\ref{eq:x-sup}).

\subsection{L\'evy, adjoint L\'evy and Laplace transformations}  
As we have seen in Section \ref{sec:limits}, the fundamental transformation 
contains, as 
significant geometric limits, the  L\'evy, adjoint L\'evy and Laplace 
transformations. Here we shall briefly discuss the analytic counterpart of 
these geometric 
limits, limiting our considerations to the scalar case.
\begin{Prop} 
Let $\Phi (\lambda )$, $\tilde\Phi (\lambda )$ and 
$\varphi (\lambda ,\mu )$  be the solutions of the {\bf scalar} 
$\bar\partial$ problems (\ref{eq:D-bar}) 
and (\ref{eq:fund-D-bar}), (\ref{eq:fund-R-D}) considered in Proposition 
\ref{prop-fund-D} and 
therefore connected by the fundamental transformation (\ref{eq:fund-psit-D}). 
Then:

\noindent
1) if $\nu\to\lambda_i$, the fundamental transformation ${\cal F}$ reduces to 
the 
adjoint L\'evy transformation ${\cL}^*_i$:
\begin{equation} \label{eq:lim-aL-DB}
\tilde\Psi (\lambda ) \to 
[\Psi (\lambda )-{{H_i}\over {H}_i(\mu )}\psi (\lambda ,\mu )]
{\lambda -\mu \over \lambda -\lambda_i},
~~~\nu\to\lambda_i.
\end{equation}
\[
\Rightarrow~~~{\cF}({\bx})\to {\bx}-{H_i\over H_i(\mu )}{\bx}(\mu )=
{\cL}^*_i({\bx}),~~~\nu\to\lambda_i.
\]
\noindent
2) If $\mu\to\lambda_j$, then the fundamental transformation ${\cF}$ 
reduces to the L\'evy 
transformation ${\cL}_j$:
\begin{equation} \label{eq:lim-L-DB}
\tilde\Psi (\lambda ) \to 
[\Psi (\lambda )-{\Psi (\nu )\over \Delta_j\Psi (\nu )}\Delta_j\Psi (\lambda )]
{\lambda -\lambda_j \over \lambda -\nu},~~~\mu\to\lambda_j.
\end{equation}
\[
\Rightarrow~~~{\cF}({\bx})\to {\bx}-
{\Psi (\nu )\over \Delta_j\Psi (\nu )}\Delta_j{\bx}={\cL}_j({\bx}),
~~~\mu\to\lambda_j.
\]
\noindent
3) if $\nu\to\lambda_i$ and $\mu\to\lambda_j$, then the Fundamental 
Transformation ${\cF}$ reduces to the Laplace  Transformation 
${\cL}_{ji}$:
\begin{equation} \label{eq:lim-La-DB}
\tilde\Psi (\lambda ) \to 
[\Psi (\lambda )-{H_i\over \Delta_jH_i}\Delta_j\Psi (\lambda )]
{\lambda -\lambda_j \over \lambda -\lambda_i},
~~~\nu\to\lambda_i,~~\mu\to\lambda_j.
\end{equation}
\[
\Rightarrow~~~{\cF}({\bx})\to {\bx}-
{H_i\over \Delta_jH_i}\Delta_j{\bx}={\cL}_{ji}({\bx}),
~~~\nu\to\lambda_i,~~~\mu\to\lambda_j.
\]
\end{Prop}
\begin{Proof} 
We first observe that 
\[ \frac{\Psi(\nu)}{\psi(\nu,\mu)} =
\frac{\Phi(\nu)}{G^{-1}(\mu)\varphi(\nu,\mu)}
\stackrel{\nu\to\lambda_i}{\longrightarrow}
\frac{\Phi(\lambda_i)}{G^{-1}(\mu)\varphi(\lambda_i,\mu)} =
\frac{H_i}{H_i(\mu)} \; ,
\] 
implying equation (\ref{eq:lim-aL-DB}). Equation (\ref{eq:lim-L-DB}) follows
from
\[ \frac{\psi(\lambda,\mu)}{\psi(\nu,\mu)} =
\frac{\varphi(\lambda,\mu)G(\lambda)}{\varphi(\nu,\mu)G(\nu)} 
\stackrel{\mu\to\lambda_i}{\longrightarrow}
\frac{\varphi(\lambda,\lambda_i)G(\lambda)}{\varphi(\nu,\lambda_i)G(\nu)} 
\stackrel{*}{=} 
\frac{(\cD_{n_i}\Phi(\lambda))G(\lambda)}{(\cD_{n_i}\Phi(\nu))G(\nu)} =
\frac{\D_i\Psi(\lambda)}{\D_i\Psi(\nu)} \; ,
\] 
where the equality $\stackrel{*}{=}$ is a consequence of equation
(\ref{eq:EE-DB}). Finally, equation (\ref{eq:lim-La-DB}) follows from  
(\ref{eq:lim-L-DB}) observing that 
\[ \frac{\Psi(\nu)}{\D_j\Psi(\nu)} =
\frac{\Phi(\nu)G(\nu)}{\D_j(\Phi(\nu)G(\nu))} 
\stackrel{\nu\to\lambda_i}{\longrightarrow}
\frac{H_i}{\D_j H_i(\mu)} \; .
\] 
\end{Proof}

\begin{Rem}
i) The L\'evy transformation was first derived, in the 
$\bar\partial$ context, in \cite{BoKo}, in the 
particular case in which $\Psi (\lambda )$ is canonically normalized.

\noindent
ii) We have seen that the limits of the fundamental 
transformation have a very elementary interpretation in the $\bar\partial$ 
formalism as limits 
on the zeroes and poles 
of the corresponding transformation function $g(\lambda )$; this is another 
indication of the
power of this approach. As a consequence of that, the basic identities 
(\ref{eq:Lid-ij})-(\ref{eq:Lid-ijk}) 
associated 
with the Laplace 
transformations have the following elementary interpretation in terms of 
multiplications of 
rational functions:
\[
{\lambda -\lambda_i\over \lambda -\lambda_j}{\lambda -\lambda_j\over \lambda 
-\lambda_i}=1
~~~~~\Rightarrow ~~~~~~{\cal L}_{ij}\circ {\cal L}_{ji}=id,
\]
\[
{\lambda -\lambda_j\over \lambda -\lambda_k}{\lambda -\lambda_i\over \lambda 
-\lambda_j}=
{\lambda -\lambda_i\over \lambda -\lambda_k}
~~~~~\Rightarrow ~~~~~~{\cal L}_{jk}\circ{\cal L}_{ij}={\cal L}_{ik},
\]
\[
{\lambda -\lambda_k\over \lambda -\lambda_i}{\lambda -\lambda_i\over \lambda 
-\lambda_j}=
{\lambda -\lambda_k\over \lambda -\lambda_j}
~~~~~\Rightarrow ~~~~~~{\cal L}_{ki}\circ{\cal L}_{ij}={\cal L}_{kj}.
\]
\end{Rem} 

\subsection{Finite Transformations versus Discretization}
We finally conclude this section with a short discussion, 
in the framework of the 
$\bar\partial$ formalism, on the connections \cite{LeBen} 
between finite 
transformations of integrable continuous  systems and the 
integrable discrete analogues of such  
continuous systems. It is enough to observe that the fundamental 
transformation of a conjugate 
net which, on the $\bar\partial$ level, reads:
\[
\tilde R(\lambda',\lambda )=
(1+{\beta\over \lambda' -\mu})R(\lambda',\lambda )(1+{\beta\over \lambda 
-\mu})^{-1},
\] 
can be formally interpreted as the shift in the additional discrete 
variable  
$n_0$ described in equation (\ref{eq:T0-DB}), after the identifications: $\beta 
=
\alpha_0, \mu =\lambda_0$. It is a simple exercise to verify that the 
fundamental 
transformation (\ref{eq:fund-psit-D}) is equivalent to the relation 
(\ref{eq:T00-DB}). This is another confirmation of the fact that finite 
transformations of 
integrable 
continuous systems provide their natural 
integrable discretization.

\section*{Acknowledgements}
A. D. would like to thank A. Sym for pointing out (see also
~\cite{PrusSym}) the important role of 
the rectilinear
congruences in the theory of integrable geometries (soliton surfaces). 
He also acknowledges
partial support from KBN grant no. 2P03 B 18509.
M. M. acknowledges  partial support from  CICYT  proyecto PB95--0401
and from the exchange agreement between {\em Universit\` a La
Sapienza }  of Rome and  {\em Universidad Complutense} of Madrid.

\end{document}